\documentstyle[aps]{revtex}

\input{epsf}

\begin{document}

\setcounter{page}{0}
\thispagestyle{empty}

\title{Propagation of Gluons 
\\${}$\\From a Non-Perturbative Evolution Equation
\\${}$\\in Axial Gauges
\\${}$}
\bigskip

\author{Klaus Geiger\thanks{The author died in the crash of Swissair
flight 111 on September 2, 1998. This manuscript was slightly edited
based on a version circulated to a few colleagues on the day before
the accident.}}

\address{Physics Department, Brookhaven National Laboratory, Upton, NY 11973}

\medskip

\date{\today}

\maketitle

\begin{abstract}
We derive a non-perturbative evolution equation for the gluon propagator
in axial gauges based on the framework of Wetterich's formulation of the
exact renormalization group. We obtain asymptotic solutions to this equation
in the ultraviolet and infrared limits. 
\end{abstract}
\vskip 1.5in


\newpage

\section{INTRODUCTION AND SUMMARY}
\label{sec:section1}
\bigskip

Non-Abelian gauge theories, and in particular  QCD,
are nowadays fairly  well understood in the short-distance (large-momentum)
regime where asymptotic freedom allows reliable calcuations within
perturbation theory.
On the other end of the scale,
in the long-distance (low-momentum) domain, fundamental unanswered
questions remain, linked intimately to the phenomenon of confinement 
(or the lack of detailed knowledge thereof),
and posing severe infrared problems that present a tough challenge
for developing  adequate non-perturbative methods to perform
practical calculations.
Whereas the non-perturbative effects on QCD Green functions are small 
when all relevant momenta are large compared to the inverse
confinement length, 
the properties of the vacuum, the dynamics of the QCD phase transition,
or the formation of color-neutral hadronic excitations from colored 
quark and gluon fluctuations, are completely dominated by the
non-perturbative infrared physics.
Although lattice simulations provide to date the most rigorous 
non-perturbative studies of QCD, they suffer in one
way or another from finite lattice size effects and violation of 
translational or rotational invariance.
Moreover, the continuum limit of results obtained on a discrete
Euclidean space lattice is a difficult problem itself.
\smallskip

\subsection{Average effective action and non-perturbative evolution equation}

Therefore, it is clear that  non-perturbative methods, formulated
in continuous space and  maintaining  the symmetries of translations 
and rotations, are of fundamental need to complement insight
into the infrared properties of QCD.
Such a method has been developed \cite{wett1,wett2,wett3}
during the last few years and has found 
diverse applications \cite{wett4,wett5,wett6}. 
It embodies the concept of the {\it average effective action}
in continuous Euclidean or Minkowski space
within the renormalization-group framework of quantum field theory.
The basic idea is to study the theory within a volume $\Omega \propto 1/ \kappa^{4}$
and effectively integrate out all quantum fluctuations
that can be localized within that volume, i.e.,
fluctuations with squared momentum $q^2$ larger than $\kappa^2$.
The {\it average effective action} $\overline{\Gamma}_\kappa$ 
is formulated as a functional integral over the microscopic
quantum fields, and can be shown to be equal to the usual
effective action $\Gamma$ for macroscopically averaged fields
\footnote{
In a sense this concept is analogous to a  quasi-particle picture
of quantum fluctuations,
wherein elementary excitations are effectively 
embodied in a quasi-particle with Compton wavelength $r_c\propto 1/\kappa$:
On  distance scales $r > r_c$ the particle appears as 
an elementary object, but as one increases  the resolution to shorter
distances by a larger $\kappa' > \kappa$, excitations with 
wavelengths $\propto 1/\kappa^{\,\prime}$ reveal themselves
as a substructure of the original quasi-particle.
Vice versa, a decrease of resolution by lowering $\kappa$, averages
over fluctuations with longer wavelengths, and
yields a larger quasi-particle.
Loosely speaking,
in the extreme short-distance limit $\kappa \rightarrow \infty$,
the quasi-particle would be,  for instance, a single elementary bare gluon, 
while in the opposite limit of inifinite volume,
$\kappa \rightarrow 0$, the quasi-particle would correspond to our Universe.
The variation of the the scale $\kappa$ therefore controls which, and how much,
physics one includes in the panorama.
}.
The vacuum properties are obtained in the limit $\kappa \rightarrow 0$
where the volume $\Omega \propto 1 / \kappa^{4}$ tends to infinity.
In this paper, however, we are interested in the non-perturbative
infrared behavior of gluons propagating in an unconfined quarkless world.
The volume of such an idealized colored world cannot, of course,
be infinite, since in reality confinement limits it to be
of the size of a hadronic state $\Omega \,\sim 1$ fm$^3$.
 Hence, as we ignore the existence of
the QCD phase transition between the colored and the hadronic world,
we must cut out the long-distance hadronic physics beyond distances
of order 1 fm, and need to restrict $\kappa$ to be larger than
the mass scale of the QCD phase transition:
\begin{equation}
\kappa \;\;\ge  \;\;\kappa_{\rm PT} \;\; \approx  200 \;{\rm MeV}
\label{kappaPT}
\;.
\end{equation}
As we shall see, the introduction of a new scale $\kappa$ into the theory
is intimately related to  the standard renormalization program of QCD,
in which one needs to introduce a mass scale at which 
the  Green functions are  normalized (since they are 
not normalizable at zero momentum, due to the infrared divergence).
\medskip

The dependence of the average effective action $\overline{\Gamma}_\kappa$
on the variation of the scale $\kappa$
is controlled by an exact non-perturbative evolution equation
\cite{wett1,wett2}, 
which is very sensitive to the infrared properties.
It is of the generic form
\begin{equation}
\kappa^2 \frac{\partial}{\partial \kappa^2}\;\overline{\Gamma}_\kappa
\;\;=\;\;
{\cal K}\left[\kappa^2,\overline{\Gamma}^{(2)}_\kappa
\right]
\;
\label{EE}
\end{equation}
where the kernel ${\cal K}$ depends explicitly only on the
(exact) 2-point function $\overline{\Gamma}_\kappa^{(2)}$,
but not on higher-order Green functions (which however 
implicitly enter in determining the 2-point function).
It has been shown \cite{wett2} that $\overline{\Gamma}_\kappa$  approaches
the classical action in the ultraviolet limit $\kappa \rightarrow \infty$
and becomes the usual effective action in the infrared limit $\kappa\rightarrow
0$.
A solution to the evolution  equation (\ref{EE}) therefore interpolates between
the (short-distance) classical action and the (long-distance) effective 
action.
Since $\overline{\Gamma}_\kappa \equiv \sum_n \overline{\Gamma}_\kappa^{(n)}$ 
generates the $\kappa$ dependent one-particle irreducible (1PI) 
Green functions $\overline{\Gamma}_\kappa^{(n)}$
(such as the inverse propagator for $n=2$, or the vertex functions for $n\ge 3$),
the evolution equation for $\overline{\Gamma_\kappa}$ 
is equivalent to an infinite set of corresponding equations for 
the 1PI functions $\overline{\Gamma}_\kappa^{(n)}$, which 
are the differential version of the 
well-known Dyson-Schwinger equations \cite{DSreview}, however
with an additional infrared cut-off given by $\kappa$.
Just as in the case of the infinite number of Dyson-Schwinger equations,
a truncation to a finite number of coupled equations 
is unavoidable, if one wishes to find an explicit, but approximate solution.
\smallskip

\subsection{Evolution of the gluon propagator}

The purpose of this paper is to demonstrate the powerful
potential of the average effective action $\overline{\Gamma}_\kappa$
and its evolution equation
by studying the simplest non-trivial object in QCD without quarks,
namely the gluon propagator.

Since the gluon propagator $\Delta_\kappa$ is related to the 
inverse of the 2-point function $\overline{\Gamma}_\kappa^{(2)}$,
we can obtain from the evolution equation for $\overline{\Gamma}_\kappa$
a corresponding equation for $\Delta_\kappa$,
which determines how the propagator changes as the scale $\kappa$ is lowered
from some large initial value in the ultraviolet all the way into
the deep infrared regime.
Unfortunately, the evolution equation for $\Delta_\kappa$ contains
in addition the unknown 3-gluon and 4-gluon vertex functions
$\overline{\Gamma}_\kappa^{(3)}$ and $\overline{\Gamma}_\kappa^{(4)}$,
which are themselves determined by
similar, but even more complex equations, involving further higher-order
functions $\overline{\Gamma}_\kappa^{(5)}$, $\overline{\Gamma}_\kappa^{(6)}$,
and so forth.
However, by working within the class of axial gauges, 
the evolution equation for the propagator becomes remarkably
simple (at least formally), because the exact propagator 
is just the bare propagator times a renormalization function
${\cal Z}_\kappa$,
\begin{equation}
\Delta_\kappa(q) \;\;=\;\;{\cal Z}_\kappa(q) \; \Delta^{(0)}_\kappa(q)
\label{ED}
\;,
\end{equation}
and the evolution equation (\ref{EE}) translates to an evolution
equation for ${\cal Z}_\kappa$,
\begin{equation}
\kappa^2 \frac{\partial}{\partial \kappa^2}\;{\cal Z}_\kappa^{-1}(q)
\;\;=\;\;
{\cal K}^\prime\left[\kappa^2,\Delta_\kappa,\overline{\Gamma}_\kappa^{(3)},
\overline{\Gamma}_\kappa^{(4)} \right]
\; ,
\label{EZ}
\end{equation}
where the kernel ${\cal K}^\prime$ explicitly depends  on the
exact propagator $\Delta_\kappa$ and the exact
3- and 4-gluon vertex functions
$\overline{\Gamma}_\kappa^{(3)}$ and $\overline{\Gamma}_\kappa^{(4)}$.
In the class of axial gauges, it is furthermore possible
to project out all contributions of 4-gluon vertex functions, so that
the remaining unknown object is the exact 3-gluon vertex.
The latter can be eliminated by exploiting the gauge symmetry properties of QCD, 
in particular the Slavnov-Taylor identities, which
provide a constraint equation between the 3-gluon vertex 
$\overline{\Gamma}_\kappa^{(3)}$ and the propagator $\Delta_\kappa$.
The strategy is then to construct an ansatz for $\overline{\Gamma}_\kappa^{(3)}$ 
in terms of $\Delta_\kappa$ such that this constraint equation is identically
satisfied. As a result, one arrives at
an evolution equation for $\Delta_\kappa$ 
in terms of  the propagator alone, which must be solved as a function of $\kappa$.
The crucial point of success in this program is the choice
for  $\overline{\Gamma}_\kappa^{(3)}$. Although constrained by gauge symmetry,
this choice is hardly unique. In the present paper we construct a particularly
simple ansatz, since our main motivation is to illustrate the concept and
the techniques involved. 
\smallskip

\subsection{Connection of propagator with gluon distribution function}

An important point that one should bear in mind throughout is,
that the gluon propagator $\Delta_\kappa(q)$, in general,  is a gauge-dependent object.
Only in the ultraviolet regime ($q\rightarrow \infty$), where asymptotic
freedom is approached, it reduces to a gauge-independent
form as given by the perturbative one-loop formula \cite{taylor96},
In the infrared domain ($q\rightarrow 0$), on the other hand, confinement should
manifest itself in the behavior of the gluon propagator, but here
the gauge-dependence foils an unambiguous assignment of confinement effects.
Yet, the fact that the propagator is gauge-dependent
does not imply that it does not contain physics; rather, it is that the
physics is obscure and difficult to extract.
\smallskip 

Because of this problem it is desirable to relate the gluon propagator
to gauge-invariant quantities, for example the Wilson loop
or the gluon distribution function of hadrons measured in experiments.
The latter is intimately connected with the spectral density
of gluon modes in the propagator. Therefore
the evolution equation for the propagator can be transcribed,
as we shall show, into
a corresponding evolution equation for the gluon distribution
function. Indeed, 
in the regime where the longitudinal (or energy) component of $q$
is much larger than the invariant $q^2$, one recovers the
famous DGLAP equation \cite{DGLAP}, the perturbative evolution equation
for the gluon distribution function.
Such a physical scenario is realized, for example, certain hard
processes occurring in high-energy
hadron collisions or deeply inelastic lepton hadron scattering
where a hard gluon can be knocked out and initiate
a gluon jet with $q_0 \approx q_z \gg q_\perp \gg q^2$
that evolves by means of fluctuating (real and virtual) gluonic offspring
towards lower and lower momenta.
\smallskip

\subsection{Related literature}

A large body of work concerning  
non-perturbative analyses of the gluon propagator 
exists in the literature \cite{DSreview}, which may be subdivided into 
analytical and lattice studies.
\smallskip

Most  analytical studies were carried out by attempting to solve the
Dyson-Schwinger equation for the gluon propagator in pure SU(3) gauge
theory without quarks, and in various covariant and non-covariant gauges, 
for example in the Landau gauge 
\cite{mandelstam,atkinson0,pennington,smekal,atkinson1},
the temporal and spacelike axial gauge 
\cite{delbourgo1,delbourgo2,schoenmaker,BBZ,atkinson2,west,alekseev0}, 
and the light-cone gauge 
\cite{alekseev1,alekseev2}.
The non-covariant axial and light-cone gauges have the advantage 
that they are ghost-free and involve only the physical gluon degrees of freedom,
whereas in covariant gauges one faces a complex coupling
between gluon and ghost variables. On the other hand, the structure
of the propagator is more complicated in the non-covariant gauges.
In either case, approximate solutions for the gluon propagator
obtained in the literature from the Dyson-Schwinger equation vary widely
\cite{buettner} 
in the infrared behavior of the gluon propagator,
whereas the large-momentum
behavior is dictated by the well-known perturbative result.
Predictions for the dependence of the propagator in the small-momentum limit 
include an infrared enhancement $\propto q^{-4}$ or $\propto q^{-2}(\ln q^2)^{-1}$, 
infrared constant $\propto q^2$, or infrared vanishing $\propto q^4$.
Recall however, that the gluon propagator is a gauge-dependent
object, so that these very different results are not, necessarily,
contradicting each other. 
\smallskip

Lattice studies are at present equally obsure, since here (in addition
to the gauge-dependence) finite lattice size effects make it difficult
to penetrate the deep infrared where the gluon wavelength
becomes close to or larger than the linear lattice length. 
There have been a number of lattice simulations of the gluon propagator
\cite{lattice1,lattice2,lattice3}, 
all of which used a fixed lattice Landau gauge,
and thus are plagued by Gribov ambiguities that can lead to significant
systematic errors.
It is therefore not surprising that fits to the lattice results
to date are not unique and consequently do not allow, 
at present, for a definite conclusion regarding
the infrared behavior of the gluon propagator.
Nevertheless, viewed as a whole, these studies seem to suggest that
the Landau-gauge gluon propagator is finite and non-zero at $q^2=0$,
although a propagator that vanishes at $q^2=0$ has also been claimed
\cite{lattice1} to be consistent.
\smallskip

\subsection{Strategy of procedure}

A roadmap of our approach to arrive at a solution
for the gluon propagator within the framework of the average effective
action may be given by the following list of conceptual steps:
\begin{description}
\item[1.]
We consider the pure SU(3) gauge theory without quarks in Minkowski space,
and from the very beginning  we choose to work in the class of axial gauges.
\item[2.]
We start from the corresponding vacuum persistence amplitude
$Z = \exp(iW)$, which allows us to
separate out the ghost contribution so that in effect we
deal with a ghost-free theory involving solely the
gauge fields.
\item[3.]
The  generating functional $W=-i\ln Z$ is then extended to a
scale-dependent version $W_\kappa$ by including an infrared
regulating source term $\Re_\kappa = {\cal A}_\mu {\cal R}_\kappa^{\mu\nu} {\cal A}_\nu$ 
that is quadratic in the gauge fields ${\cal A}$ and depends on the momentum
scale $\kappa$, such that only quantum fluctuations with momenta $\ge \kappa$
are included and the limit $\kappa\rightarrow 0$ recovers the full theory.
\item[4.]
From $W_\kappa$ we obtain then the corresponding
 scale-dependent effective action ${\Gamma}_\kappa$ 
which generates the one-particle irreducible $n$-point functions
${\Gamma}_\kappa^{(n)}$, such as the inverse propagator, the 3-gluon
and 4-gluon vertex functions.
all of which explicitly  depend on the cut-off scale $\kappa$.
\item[5.]
Subtracting  from ${\Gamma}_\kappa$ the infrared regulator $\Re_\kappa$,
and averaging over all gauge field configurations with in the
effective volume $\Omega \propto 1/\kappa^4$, we arrive at the
average effective action $\overline{\Gamma}_\kappa$.
Differentiation of $\overline{\Gamma}_\kappa$ with respect its to $\kappa$ dependence
leads the to the desired exact evolution equation.
\item[6.]
From the evolution equation for 
$\overline{\Gamma}_\kappa \equiv \sum_n \overline{\Gamma}_\kappa^{(n)}$ 
we then project out the quadratic term
$\overline{\Gamma}_\kappa^{(2)}$ that is related to the inverse
gluon propagator. 
After  decomposing the tensor structure of the inverse propagator,
we obtain a set of coupled equations for  two independent
scalar functions, $a_\kappa$ and $b_\kappa$.
\item[7.]
Next we focus our attention to the light-cone gauge, a special
case of the axial gauges, in which the function $a_\kappa$ drops
out, so that we are left with a single evolution equation
for the dimensionless function ${\cal Z}_\kappa = q^2/b_\kappa$.
Moreover, all 4-gluon vertex contributions can be eliminated,
and consequently only the 3-gluon vertex function survives
in the determination of  ${\cal Z}_\kappa$.
\item[8.]
By constructing a specific ansatz for the 3-gluon vertex function
that obeys the constraint of the Slavnov-Taylor identity for the
gluon propagator $\Delta_\kappa$, we obtain a closed equation for
the $\Delta_\kappa$. The formal solution of this final
evolution equation is simply
$\Delta_\kappa =  {\cal Z}_\kappa \;\Delta^{(0)}_\kappa$,
where $\Delta^{(0)}_\kappa$ is the bare propagator.
\item[9.]
The remaining integration of the final evolution equation 
for ${\cal Z}_\kappa$ must be
done numerically, but in the ultraviolet and infrared limits,
we are able to extract analytical solutions, which depend (aside
from the gluon momentum $q$) on the scale $\kappa$.
In the limit $\kappa \rightarrow 0$ one obtains then 
from ${\cal Z}_{\kappa=0}(q^2)$ the full gluon propagator in the 
light-cone gauge, $\Delta(q) =  {\cal Z}_{0}(q^2) \;\Delta^{(0)}(q)$.
\item[10.]
In its spectral representation,
the gluon propagator can be related to the gauge-independent gluon
distribution function $G(q,\kappa)$ through the renormalization function
${\cal Z}_\kappa(q^2)$, and the evolution equation for ${\cal Z}_\kappa$
can be transcribed into a corresponding evolution equation for $G$.
In the high-momentum limit we recover the perturbative DGLAP
evolution equation, and we find that our solution coincides with the
perturbative result.
\end{description} 
\smallskip

\subsection{Main results}

Although the full solution to the our evolution equation
for the gluon propagator $\Delta_\kappa(q) =  {\cal Z}_{\kappa}(q^2) 
\;\Delta^{(0)}(q)$ in the light-cone gauge requires a numerical analysis,
we are able to arrive at analytical solutions for ${\cal Z}_\kappa$
in the extreme limits 
$q^2/\kappa^2 \rightarrow \infty$ and $q^2/\kappa^2 \rightarrow 0$.
In the former case, the {\it ultraviolet limit}, we obtain: 
\begin{equation}
{\cal Z}^{-1}_\kappa(q^2) \;\;\stackrel{q^2\rightarrow \infty}{\approx} \;\;
1\;-\; \frac{11 g^2 C_G}{48\pi^2} \;\ln\left(\frac{q^2}{\kappa^2}\right)
\;.
\end{equation}
On the other end of the energy scale, in the {\it infrared limit},
the leading behavior turns out to be:
\begin{equation}
{\cal Z}^{-1}_\kappa(q^2) \;\;\stackrel{q^2\rightarrow 0}{\approx} \;\;
 \frac{g^2 C_G}{48\pi^2} \;\;\frac{q^2}{\kappa^2}
\;.
\end{equation}
The corresponding limiting behavior of the actual gluon propagator
then follows as:
\begin{equation}
\Delta_{\kappa}(q) 
\;\;\stackrel{q^2\rightarrow \infty}{\propto}\;\;
\frac{1}{q^2\;\ln(q^2/\kappa^2)}
\;,\;\;\;\;\;\;\;\;\;\;\;\;\;\;\;
\Delta_{\kappa}(q) 
\;\;\stackrel{q^2\rightarrow 0}{\propto}\;\;
\frac{\kappa^2}{q^4}
\;.
\end{equation}
The ultraviolet behavior is consistent with asymptotic
freedom,  corresponding to a screening of
the color charge due to $g_0^2/g^2 = {\cal Z}_\kappa^{-1} < 1$.
The infrared solution would, on the other hand,  correspond to a
linearly rising potential $V(r) \propto r$ as $r\rightarrow \infty$,
in accordance with the phenomenological picture of confinement.
These results are certainly rather qualitative,
firstly, because the inclusion of quark degrees of freedom which we left out 
here, may alter the details of the infrared behavior and, secondly,
because the weakest point of our analysis is the aforementioned {\it ansatz}
for the 3-gluon vertex function, which may not be all that good
in the long-wavelength limit.
But even for our specific ansatz, 
an exact numerical solution of the evolution equation for
the propagator needs to be carried out before more robust conclusions can be drawn.
\smallskip

\subsection{Organization of the paper}

\noindent
The reminder of the paper is structured in accordance with the
above list of procedural steps:

In Sect. II, we recall the necessary basics of the functional formalism,
which we then extend to its scale($\kappa$)-dependent analogue.
The effective action $\Gamma_\kappa$ for this scale-dependent functional formulation,
obtained as usual, is then related to the average effective action
$\overline{\Gamma}_\kappa$, which is the generating functional
for the Green functions in the presence of the cut-off $\kappa$. 
We derive the desired exact evolution equation
for the change of $\overline{\Gamma}_\kappa$ with a variation of $\kappa$.

Sect. III is devoted to applying the formalism to
the evolution of the gluon propagator $\Delta_\kappa^{\mu\nu}$. 
We first derive, from the fundamental evolution equation for $\overline{\Gamma}_\kappa$,
the general equations that govern the $\kappa$-variation of the propagator.
Next we restrict ourselves to the light-cone gauge, and arrive at
a considerably simpler, single evolution equation for the renormalization
function ${\cal Z}_\kappa$, the formal solution of which is equivalent to
the solution of the gluon propagator in the light-cone gauge.

In Sect. IV, we take pragmatical steps to actually solve the evolution equation,
subject to a  necessary  assumption about the form of the  
3-gluon vertex function.
The final master equation for the renormalization function ${\cal Z}_\kappa$
and hence for the propagator $\Delta_{\mu\nu}$, can then be solved in closed
form, and we are able to obtain the above-quoted results in
the ultraviolet and the infrared limits. A phenomenological
formula for the propagator that may be useful for parton model
applications, is constructed by interpolating between the
two extreme limits.

Sect. V applies the results for renormalization function
${\cal Z}_\kappa$ to  illustrate two important
phenomenological connections with experimentally measurable quantities,
namely the QCD running coupling $\alpha_s(q^2)$ and the gluon
distribution function $g_\kappa(q)$.
First, we infer from ${\cal Z}_\kappa$
the running of the coupling $\alpha_s(q^2)$, using standard
renormalization group arguments, and then we
relate ${\cal Z}_\kappa$ via the spectral density $\rho_\kappa$ of the gluon 
propagator, the gluon distribution function $g(q,\kappa)$ and its evolution
equation.

Appendix A summarizes the notation and conventions used in the paper.
Appendix B recalls some basic formulae of the functional formalism in QCD,
and provides a list of relevant Green functions and vertices.
In Appendix C, we discuss the absence of ghosts in axial gauges,
allowing a factorization of the generating functionals for the ghost and the gluon fields.
Appendix D elaborates the details of the general structure of the gluon propagator
in axial gauges, and the simplifications that emerge when specifically
using the light-cone gauge.
Appendix E briefly reviews the connection between the 
gluon propagator and the gluon spectral density, the latter being related
to the experimentally measurable gluon distribution function.

Table 1 provides a summary list of the notation used in this paper.


\setcounter{figure}{0}
\begin{table}[htb]
\epsfxsize=500pt
\centerline{ \epsfbox{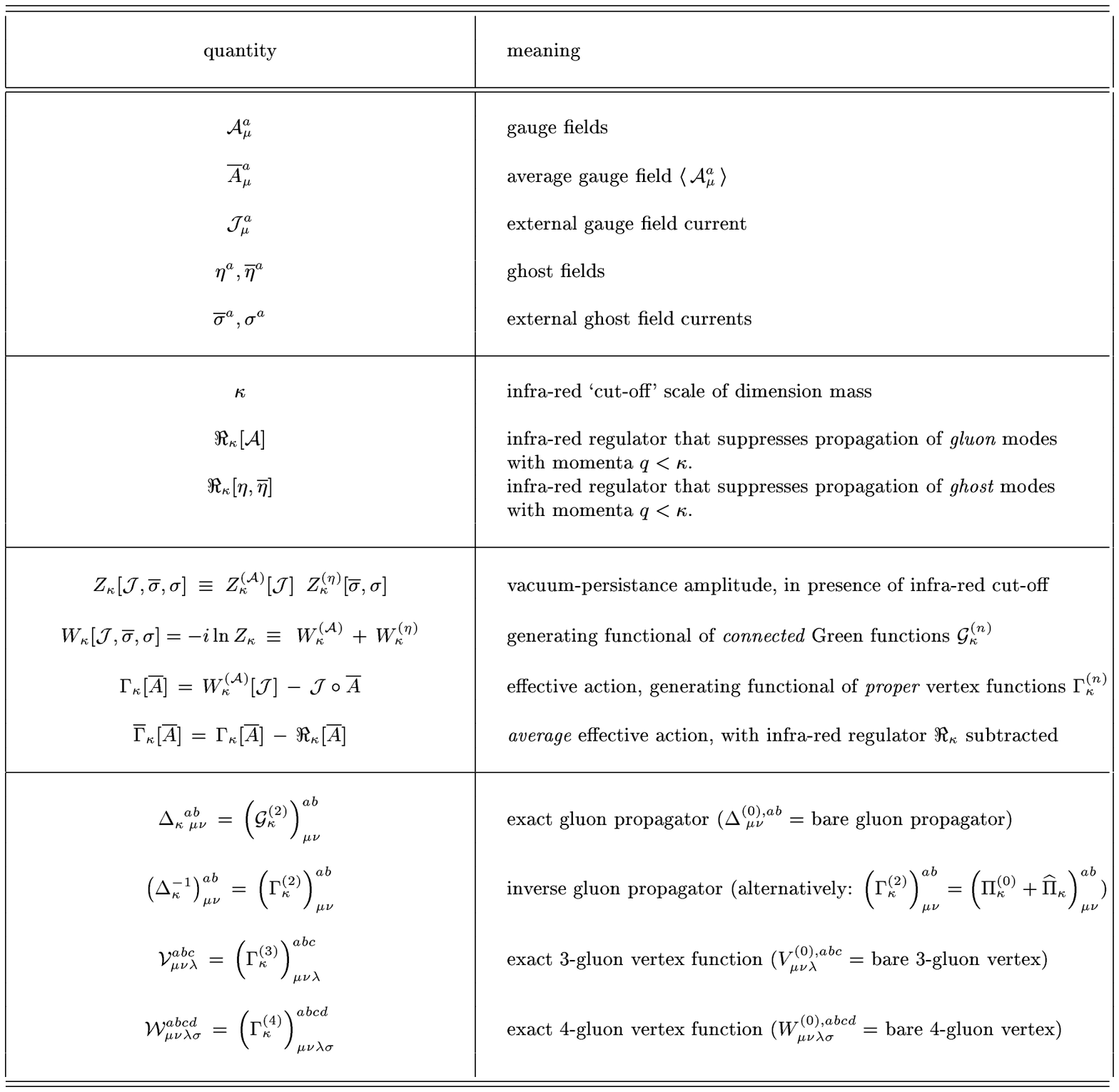} }
\vspace{-4.0cm}
\caption{
List of  basic quantities encountered in the paper. Note that all
quantities with subscript $\kappa$ reduce to the standard
forms when $\kappa = 0$.
Also, note that the separation of $Z_\kappa$ and $W_\kappa$ into
gauge field and ghost field parts holds only in axial gauges.
         }
\end{table}

\newpage

\section{EFFECTIVE AVERAGE ACTION IN NON-COVARIANT GAUGES}
\label{sec:section2}
\bigskip

This Section is devoted to a brief review of
the path-integral formalism for QCD in non-covariant gauges, and
its application to the renormalization group evolution of the 
effective action of QCD, as developed by Reuter and Wetterich 
\cite{wett2}.
We refer to  Appendices A and B, where 
our notational conventions are collected and to Table 1, which
summarizes the notation of basic quantities encountered in the following.

\subsection{QCD path-integral formalism for non-covariant gauges}

We work in Minkowski space 
\footnote{
In order to facilitate the correspondence between
the functional formalism in Euclidean space of 
Reuter and Wetterich \cite{wett2}, and the Minkowski space 
description in the present paper,
we recall the translation rules between Euclidean 
(subscript ``$E$'') and Minkowski (subscript ``$M$'') formulations with
metric $\delta_{\mu\nu} = \mbox{diag}(-,-,-,-)$ 
and $g_{\mu\nu}= \mbox{diag}(+,-,-,-)$, respectively,
\begin{eqnarray}
x^\mu_E &=&(x^0,{\bf x})_E \;\longleftrightarrow\;(ix^0,{\bf x})_M
\;\;\;\;\;\;\;\;\;
\;\;\;\;\;\;\;\;\;
\;\;\;\;\;\;\;\;\;
{\cal A}_{a\;E}^\mu \;=\;({\cal A}_a^0,{\bf {A}}_a)_E \;\longleftrightarrow\;
(-i{\cal A}_a^0,-{\bf {A}}_a)_M
\nonumber \\
D_{ab\;E}^\mu &=&(D_{ab}^0,{\bf D}_{ab})_E \;\longleftrightarrow\;(-iD_{ab}^0,-{\bf D}_{ab})_M
\;\;\;\;\;\;\;\;\;
W[{\cal K}]_E \;\longleftrightarrow\;-iW[{\cal K}]_M
\nonumber
\;.
\end{eqnarray}
Notice that the convention for the four-potential ${\cal A}_\mu$ 
differs from that of an ordinary four-vector $x^\mu$: the
former is defined with common sign, whereas the latter has different
signs of the timelike and spatial components. 
This is chosen for convenience in order
to not have to change the sign of the coupling constant $g$ when 
translating between Euclidean and Minkowski spaces.
}
(as opposed to the Euclidean formulation of Refs. \cite{wett2}), and
consider pure $SU(3)_c$ Yang-Mills theory for $N_c=3$ colors
in the absence of quark degrees of freedom.
Our starting point is the path integral representation 
of the QCD vacuum persistence amplitude 
$Z[{\cal J}] = \langle\, 0 \,| \,0\,\rangle_{{\cal J}}$
in the presence of an external source ${\cal J}$.
Employing the conventions of Appendix B,
we define the generating functional for the connected Green functions  
$W[{\cal J}]$ as usual by 
$Z[{\cal J}] = \exp \left( i \,W[{\cal J}]\right)$, with
\begin{equation}
W[{\cal J}]\;\equiv \; - i \ln Z[{\cal J}]
\;=\;
-i \,{\cal N}\;\ln \left[\frac{}{}
\;\int \,{\cal D}{\cal A} \; \mbox{det} \left(M\right) 
\;\,\delta\left(F^a[{\cal A}]\right)
\;\;\exp \left(\frac{}{}\,i\, S_{\rm YM}\left[{\cal A}\right]
\;+\; S_{\rm ext} \left[{\cal A}\right] \,\right)
\right]
\label{Z1}
\;,
\end{equation}
with the normalization ${\cal N}$ determined by the condition 
$ W[0] = 0$ \cite{cornwall}, and
\begin{equation}
S_{\rm YM} \left[{\cal A}\right] \;=\;
-\frac{1}{4}\;\int d^4x \; {\cal F}_{\mu\nu}{\cal F}^{\mu\nu} 
\;\;\;\;\;\;\;\;\;\;\;\;\;\;\;\;\;
S_{\rm ext} \left[{\cal A}\right] \;=\; 
\int d^4x\;{\cal J}_\mu{\cal A}^\mu 
\label{Z1a}
\;.
\end{equation}
Here ${\cal A}_\mu \equiv T^a\,{\cal A}_\mu^a$ denotes the gauge field,
and the ${\cal F}_{\mu\nu} \equiv T^a\,{\cal F}_{\mu\nu}^a$ the
corresponding field tensor.
The path-integral measure in (\ref{Z1}) is short-notated as  
${\cal D}{\cal A} \equiv \prod_x\prod_\mu\prod_a d {\cal A}_\mu^a(x)$,
and the gauge condition is embodied in
$\delta\left(F^a[{\cal A}]\right)$,
\begin{equation}
F^a [{\cal A}_\mu^b(x)] \;\equiv \;F^a_{{\cal A}}(x)\;=\; 0, \;\;\;\;\;\;\; 
\mbox{for all} \;a, b, \mu
\;.
\label{gaugecond}
\end{equation}
The gauge fixing determines the Jacobian 
$\mbox{det} ( M )$ as the determinant of the Fadeev-Popov matrix
\begin{equation}
M_{ab}(x,y)\;\;=\;\; \frac{\delta F^a_{{\cal A}}(x)}{\delta \omega^b(y)}
\;\;=\;\;
\frac{\delta F^a_{{\cal A}}}{\delta {\cal A}^c_\mu} \;D_\mu^{cb}\;\delta^4(x-y)
\label{Mab}
\;,
\end{equation}
where
$\omega^b$ describes local gauge transformations
$
g[\omega^a]\;\equiv\; \exp\left( - i\,\omega^a(x)\, T_a\right)
$,
under which the gauge fields transform as
$
{\cal A}_\mu^a \;\longrightarrow\;
{\cal A}_\mu^{(\omega)\;a} \;=\;
g[\omega^a]\; {\cal A}_\mu^a \;g^{ -1}[\omega^a]
$,
so that ${\cal F}_{\mu\nu}^a {\cal F}_{\mu\nu}^a$
is gauge invariant.
\bigskip

Because of the practical advantages described before, we choose to work 
with a non-covariant gauge \cite{gaugereview,gaugebook},
for which the gauge condition (\ref{gaugecond}) reads,
\begin{equation}
F^a_{{\cal A}}(x)\;=\; n^\mu\,{\cal A}_\mu^a(x) \;=\;0
\label{gauge1}
\;,
\end{equation}
where
$n^\mu$ is a constant 4-vector, being either space-like ($n^2 < 0$), 
time-like ($n^2 > 0$), or light-like ($n^2=0$).
The particular choice of 
the vector $n^\mu$ is usually dictated by physical considerations 
or computational convenience, and distinguishes
{\it axial gauge} ($n^2 < 0$),
{\it temporal gauge} ($n^2 >  0$), and
{\it light-cone gauge} ($n^2=  0$).
Among these gauges, the light-cone gauge is
most often employed in the literature \cite{gaugereview}.
It is well suited for describing high-energy QCD in the
{\it infinite momentum frame} \cite{infmom}, since Lorentz contraction
and  time-dilation
causes the quantum fluctuations to be concentrated in close proximity 
of the light-cone, the direction of which
naturally suggests the choice of the gauge vector $n^\mu$.
For these reasons we will later adopt the light-cone gauge 
by specifying $n^2 = 0$. For the time being, however, we keep $n^\mu$ general,
so that the considerations apply to the class of non-covariant gauges 
as a whole.
As elaborated in Appendix C, the gauge condition  (\ref{gauge1})
implies for the general case of arbitrary $n^\mu$ 
\begin{equation}
\mbox{det}\left( M\right)\;=\;
\mbox{det}\left(\frac{}{}
 \delta^{ac}\;n^\mu \;[\delta_a^b \partial_\mu \,+\, 
g\,f^{cb}_d\,{\cal A}_\mu^d]\right)
\;\;=\;\;\mbox{det}\left( \delta^{ab}\;n\cdot\partial \right)
\label{detM}
\;,
\end{equation}
because
$ 
\delta F^a_{{\cal A}}/\delta {\cal A}^c_\mu  = \delta^{ac}\;n^\mu\;\delta^4(x-y)
$
and $n\cdot {\cal A}_\mu^d = 0$.
As a consequence, the ghost degrees of freedom decouple, since
$\mbox{det} (M)$ no longer depends on the gauge field ${\cal A}$.
\smallskip
We may cast the generating functional (\ref{Z1}) in a more practical form
by rewriting  the Jacobi determinant $\det (M)$ in terms of
a Gaussian integral over ghost fields $\overline{\eta},\eta$,
\begin{equation}
\mbox{det} \left(M\right) \;=\; 
\int {\cal D}\overline{\eta} {\cal D} \eta \;
\exp \left\{i\;  \frac{}{} \int d^4x \;\overline{\eta}_a(x)\;M^{ab}\;\eta_b(x)\right\}
\int {\cal D}\overline{\eta} {\cal D} \eta \;
\;=\;
\exp \left\{
i\;\int d^4x \;\overline{\eta}_a(x)\;\left( \delta^{ab}\;n^\mu\,\partial_\mu\right)\;\eta_b(x)
\right\}
\;,
\label{detM2}
\end{equation}
and the functional $\delta( F^a_{{\cal A}} )$ as an
 exponential of a gauge-fixing action,
\begin{equation}
{\cal D}{\cal A} \; \,\delta\left(F^a[{\cal A}]\right)
\;=\; {\cal D}{\cal A}
\;\exp \left\{- i \int d^4x \;\frac{1}{2\xi}\;
\left( F^a_{{\cal A}}(x)\right)^2 \right\} 
\;=\; {\cal D}{\cal A}
\;\exp \left\{- \frac{i}{2\xi}\;
\int d^4x \; \left(n \cdot {\cal A}^a(x)\right)^2
\right\}
\label{F2}
\;.
\end{equation}
The gauge parameter $\xi$ allows here, just as in covariant gauges, to
specify a particular gauge within class the of non-covariant gauges, 
e.g. Feynman-type gauges with $\xi =1$, or Landau-type gauges with $\alpha = 0$.
\footnote{
Notice, however, that $\xi$ needs to be kept general at this point and
in the following: it may be fixed only {\it after} the gluon propagator
has been derived explicitly from inverting the terms quadratic
in ${\cal A}$ in (\ref{Z1}).
}
Since $\mbox{det} ( M )$ in (\ref{detM2}) is independent of the gauge fields
${\cal A}$,  it can be pulled out of the functional
integral over the gauge field configurations in (\ref{Z1}),
so that we can factor out the ghost field dependence by rewriting
(\ref{Z1}) as
\begin{equation}
W[{\cal J},\overline{\sigma},\sigma]\;\,=\;\,
- i\, \ln \;\,\left(\frac{}{}
Z^{({\cal A})}[{\cal J}]
\;\;
Z^{(\eta)}[\overline{\sigma},\sigma]
\right)
\label{Z2}
\;.
\end{equation}
Here, and henceforth, we have set arbitrarily the normalization ${\cal N}$ 
appearing in (\ref{Z1}) equal to unity, since it is an
irrelevant constant factor, and introduced the functionals
\begin{equation}
Z^{({\cal A})}[{\cal J}]
\;=\;
\int \,{\cal D}{\cal A} 
\;\exp \left( i\,  \frac{}{} S_{\rm eff}[{\cal A},{\cal J}] \right)
\;\;\;\;\;\;\;\;\;\;\;\;\;
Z^{(\eta)}[\overline{\sigma},\sigma]
\;=\;
 \;\int \,{\cal D}\overline{\eta}{\cal D}\eta\;
\exp\left( i\, \frac{}{} 
S_{\rm eff}[\overline{\eta},\eta,\overline{\sigma},\sigma]
\right) 
\label{Z2a}
\end{equation}
with the combined gauge field action
\begin{eqnarray}
S_{\rm eff}[{\cal A},{\cal J}]&=&
S_{\rm YM}[{\cal A}]\;+\;S_{rm fix}^{(\xi)}[{\cal A}]
\;+\;S_{\rm ext}[{\cal A},{\cal J}]
\nonumber \\
S_{\rm YM}[{\cal A}]&=& -\frac{1}{4}\;\int d^4x \;{\cal F}_{\mu\nu}^a{\cal F}^{\mu\nu}_a 
\label{SA0} \\
S^{(\xi)}_{rm fix}[{\cal A}]&=&
-\frac{1}{2\xi}\;\int d^4x \; \left(n^\mu {\cal A}^a_\mu\right)^2
\nonumber \\
S_{\rm ext}[{\cal A},{\cal J}] &=& \; \int d^4x \;{\cal J}^a_\mu{\cal A}_a^\mu
\label{SA3}
\;,
\end{eqnarray}
and the combined  ghost field action
\begin{eqnarray}
S_{\rm eff}[\overline{\eta},\eta,\overline{\sigma},\sigma]
& = &
S_{\rm ghost}[\overline{\eta},\eta] \;+\;
S_{\rm ext}[\overline{\eta},\eta,\overline{\sigma},\sigma]
\nonumber \\
S_{\rm ghost}[\overline{\eta},\eta] & = &
 \int d^4x \;\overline{\eta}_a\;\left( \delta^{ab}\;n^\mu\,\partial_\mu\right)\;\eta_b
\label{SG0} \\
S_{\rm ext}[\overline{\eta},\eta,\overline{\sigma},\sigma]&=&
\;\int d^4x \;\left(\frac{}{}\overline{\sigma}_a\eta^a\;+\;\sigma_a \overline{\eta}^a\right)
\nonumber
\;.
\end{eqnarray}
\smallskip

\subsection{Generalization to scale-dependent formalism}

On the basis of the generating functional $W[{\cal J}]$ of (\ref{Z2}), one can 
construct a corresponding {\it scale-dependent} functional. Whereas in
(\ref{Z1}) quantum fluctuations with arbitrary momenta are to be included,
the scale-dependent functional should only involve an integration
over modes with momenta larger than some infrared cut-off $\kappa$.
A variation of $\kappa$ describes then the successive integration
over fluctuations corresponding to different length scales
with the aim to recover the full theory in the limit
$\kappa \rightarrow 0$. Following the rationale of Ref. \cite{wett2},
a scale($\kappa$)-dependent generalization $W_\kappa$ of the functional
$W$ in (\ref{Z2}) is defined as
\begin{equation}
W_\kappa[{\cal J},\overline{\sigma},\sigma]\;\equiv\;
W_\kappa^{({\cal A})}[{\cal J}]\;+\;
W_\kappa^{(\eta)}[\overline{\sigma},\sigma]
\;\,=\;\,
-i\, \left\{\frac{}{}
\ln \left(\frac{}{} Z_\kappa^{({\cal A})}[{\cal J}] \right)
\;\;+\;\;
\ln \;\,\left(\frac{}{} Z_\kappa^{(\eta)}[\overline{\sigma},\sigma]\right)
\right\}
\label{Z3}
\;.
\end{equation}
Here the scale-dependent functionals $Z_\kappa$ are related to the 
usual $\kappa$-independent vacuum amplitudes $Z$, eq. (\ref{Z2a}),
by adding invariant infrared cut-offs $\Re_\kappa$ 
for the gauge field ${\cal A}$ and for the ghost fields, respectively,
\begin{eqnarray}
Z_\kappa^{({\cal A})}[{\cal J}]
&=&
\int \,{\cal D}{\cal A} 
\;\exp \left\{i\,  \frac{}{} S_{\rm eff}[{\cal A},{\cal J}]
\right\}
\;\;\exp\left\{\frac{}{} i\,\Re_\kappa[{\cal A}]\right\}\; ,
\label{Z2c}
\\
Z_\kappa^{(\eta)}[\overline{\sigma},\sigma]
&=&
 \;\int \,{\cal D}\overline{\eta}{\cal D}\eta\;
\exp\left\{ i\,\frac{}{} S_{\rm eff}[\overline{\eta},\eta,\overline{\sigma},\sigma]
\right\} 
\;\;\exp\left\{\frac{}{} i\,\Re_\kappa[\overline{\eta},\eta]\right\}
\label{Z2d}
\;,
\end{eqnarray}
with $S_{\rm eff}[{\cal A},{\cal J}]$ and 
$S_{\rm eff}[\overline{\eta},\eta,\overline{\sigma},\sigma]$ defined by
(\ref{SA3}) and (\ref{SG0}), respectively, and the infrared regulators
\begin{eqnarray}
\Re_\kappa [{\cal A}] &=&
- \frac{1}{2} \int d^4x\; {\cal A}^\mu_a \; \left[
{\cal R}_\kappa(\partial^2)\right]^{ab}_{\mu\nu} \;{\cal A}^\nu_b
\label{Delta1}
\;,
\\
\Re_\kappa [\overline{\eta},\eta] &=&
\int d^4x \;\overline{\eta}_a\;
\left[ 
\widetilde{\cal R}_\kappa\left((n\cdot\partial)^2\right)\right]^{ab}
\;\eta_b
\label{Delta2}
\;.
\end{eqnarray}
with
\begin{eqnarray}
& &
\;\;\;\;
\left[{\cal R}_\kappa(\partial^2)\right]^{ab}_{\mu\nu}
\;\;\;\;\;=\;
\;\delta^{ab} \;
R_\kappa(\partial^2) 
\;\left(
g_{\mu\nu} \;-\; \frac{\partial_\mu\partial_\nu}{\partial^2}
\;+\; \frac{1}{\xi} \frac{n_\mu n_\nu}{\partial^2}
\right)
\label{Delta3a}
\\
& &
\left[ \widetilde{\cal R}_\kappa\left((n\cdot\partial)^2\right)\right]^{ab}
\;=\;
\;\;\;\;\;\;\;\delta^{ab} \;\left(
1 + \frac{\widetilde{R}_\kappa\left((n\cdot\partial)^2\right)}{(n\cdot\partial)^2}
\right)
\label{Delta3b}
\;.
\end{eqnarray}
One may wonder about the form
of the infrared regulators $\Re_\kappa$ in (\ref{Delta1}) and (\ref{Delta2}).
These have been constructed,
so that they affect only the gluon and ghost propagators, respectively,
as  $\kappa$-dependent squared mass terms that regularize the infrared poles
in the propagators. 
As we shall see in detail later, by combining
the quadratic pieces of $S_{\rm YM}[{\cal A}]$ with $\Re_\kappa[{\cal A}]$
and similarly the quadratic terms of $S_{\rm ghost}[\overline{\eta},\eta]$ with
$\Re_\kappa[\overline{\eta},\eta]$, the 
inverse gluon propagator $\Delta^{-1}$ and
ghost propagator ${\cal D}^{-1}$, respectively, are modified such that
\begin{eqnarray}
\Delta^{-1} \propto \partial^2  \;\;\;\; & \longrightarrow &
\;\;\;\;
\Delta_\kappa^{-1} \propto \partial^2 R_\kappa(\partial^2) 
\\
& &\nonumber \\
{\cal D}^{-1} \propto n\cdot \partial  \;\;\;\; & \longrightarrow &
\;\;\;\; {\cal D}_\kappa^{-1} \propto n\cdot \partial 
\left( 1 + \frac{\widetilde{R}_\kappa\left((n\cdot \partial)^2\right)
}{(n\cdot \partial)^2}\right)
\;.
\end{eqnarray}
In general,
the functions $R_\kappa$ and  $\widetilde{R}_\kappa$ can be
different, but their specific forms are unconstrained.
One may therefore take the freedom to  choose their analytic form to be the same, 
\begin{equation}
R_\kappa (d^2) \;\;\equiv\;\; R_\kappa\;=\; \widetilde{R}_\kappa
\;,
\end{equation}
but with different arguments $d^2$, namely
the operators $\partial^2$ and $(n\cdot\partial)^2$, respectively. 
The choice of the functional form for $R_\kappa$ specifies the details of how
the fluctuations with eigenvalues of the operators $d^2 =\partial^2$ and
$d^2 = (n\cdot \partial)^2$  larger than 
$\kappa^2$ are integrated out in the computation of the path integral 
(\ref{Z3}). For example, a convenient parametrization 
[ after Fourier transformation to momentum space with $d^2\rightarrow p^{-2}$,
and $p^2 = q^2$ or $p^2=(n\cdot q)^2$ ] is (see Fig. 1),
\footnote{
We shall later use a generalization of this form, which includes an 
additional ultraviolet cut-off 
$\Lambda \gg\kappa$, but
which contains (\ref{Rk}) for $\Lambda \rightarrow \infty$:
$
R_\kappa(p^2)=
 p^2 \,\, \exp\left(-p^2/\kappa^2\right)
\left[ \exp\left(-p^2/\Lambda^2\right)  - \exp\left(-p^2/\kappa^2\right)
\right]^{-1}
$.
}
\begin{equation}
R_\kappa(p^2) \;=\;
p^2 \; \frac{\exp\left(-p^2/\kappa^2\right)}
{1 - \exp\left(-p^2/\kappa^2\right)}
\;,
\label{Rk}
\end{equation}
which has the following
limiting behavior in the ultraviolet and the infrared, respectively:
\begin{equation}
\lim_{p^2/\kappa^2 \rightarrow \infty} \;R_\kappa(p^2) \;=\; 0
\;\;\;\;\;\;\;\;\;\;\;\;
\lim_{p^2/\kappa^2 \rightarrow 0} \;R_\kappa(p^2) \;=\; \kappa^2
\;.
\label{Rklim}
\end{equation}
Hence, the effect of $R_\kappa(q^2)$ is vanishing in the high-momentum limit 
$q \gg \kappa$, but provides an infrared screening as $q\rightarrow 0$.
Moreover, the original functional $W$ of (\ref{Z2}), containing {\it all} 
quantum fluctuations, is recovered from $W_\kappa$ of
(\ref{Z3}) in the limit $\kappa = 0$,
\begin{equation}
W_\kappa[{\cal J},\overline{\sigma},\sigma]\;
\stackrel{\kappa \rightarrow 0}{\longrightarrow} 
\;W[{\cal J},\overline{\sigma},\sigma]
\;.
\end{equation}
\bigskip

\setcounter{figure}{0}
\begin{figure}[htb]
\begin{minipage}[t]{80mm}
\epsfxsize=150pt
\centerline{ \epsfbox{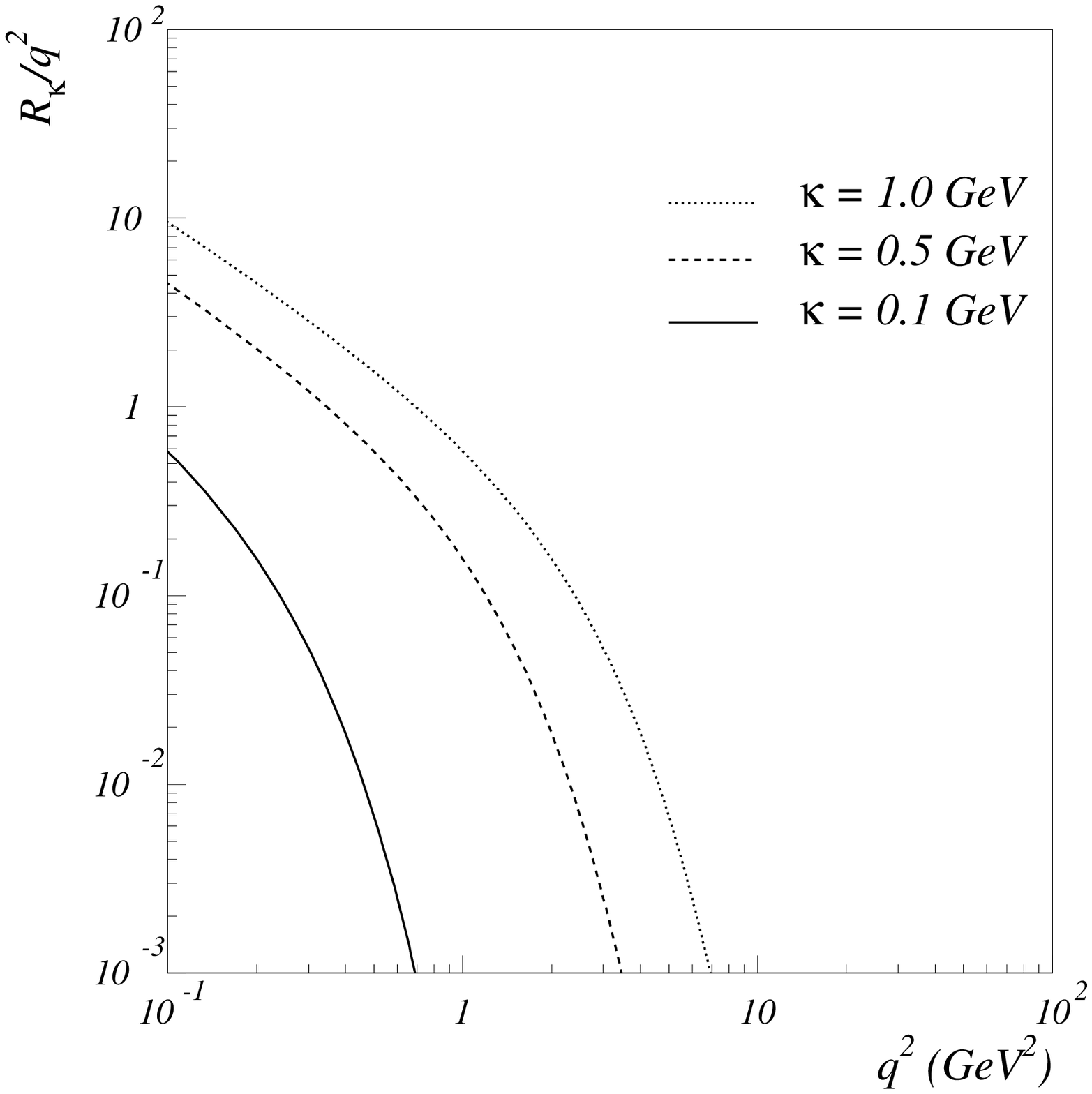} }
\end{minipage}
\begin{minipage}[t]{80mm}
\epsfxsize=150pt
\centerline{ \epsfbox{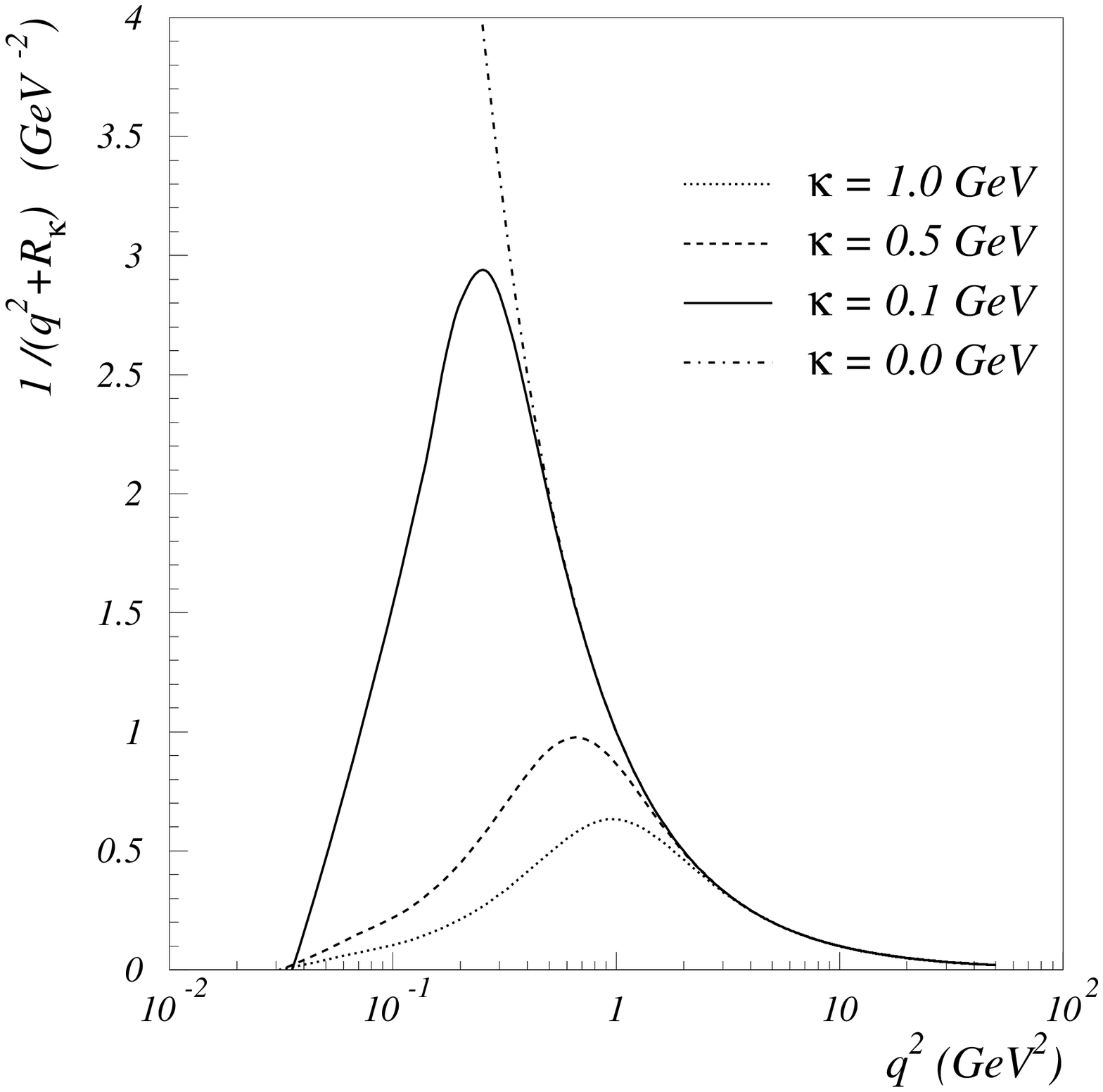} }
\end{minipage}
\caption{
Form of the  infrared regulator $R_\kappa(q^2)$, eq. (\ref{Rklim}),
and its damping effect on the propagator $\Delta_\kappa(q) \propto
1/[q^2+R_\kappa(q^2)]$ at small values of $q^2$. The various curves
illustrate the  different choices of $\kappa$, with $\kappa=0$
corresponding to the case with no infrared cut-off at all.
\label{fig:fig1}
}
\end{figure}
\bigskip
\bigskip
\bigskip
\bigskip

The crux of the above discussion is the convenient decoupling of 
the ghost degrees of freedom 
from the gluon degrees of freedom in (\ref{Z3}) due to the choice of gauge 
(\ref{gauge1}). Since we are interested in the variation
of $W_\kappa$ against $\kappa$ with regard to the physical gluon degrees, 
the first term 
in (\ref{Z3}), $W_\kappa^{(\eta)}$ amounts to an irrelevant constant 
that does not affect the change of $W_\kappa^{({\cal A})}$ and 
therefore may be absorbed in the overall normalization.
In other words,
for the evolution of the physical gluon fields with
changing scale $\kappa$, we can henceforth omit the ghost contribution and
restrict our attention to  $W_\kappa^{({\cal A})}$.
Then one can derive from (\ref{Z3}) - with reference to Appendix B - 
the $\kappa$-dependent generalization $\Gamma_\kappa$  of
the standard effective action $\Gamma \equiv \Gamma_{\kappa=0}$
by introducing the {\it average gauge field}
\begin{equation}
\overline{A}_\mu^a(x)
\;\equiv\;
\langle\,{\cal A}_\mu^a(x)\,\rangle_\kappa
\;\;=\;\;
\left.
\frac{\delta W^{({\cal A})}_\kappa}{\delta {\cal J}_a^\mu(x)}
\right|_{{\cal J}=0}
\;\equiv\;\left({\cal G}_\kappa^{(1)}(x)\right)_{\mu}^{a}
\label{Abar}
\;.
\end{equation}
where the subscript $\kappa$
at $\langle{\cal A}\rangle_\kappa$ indicates that only field modes
that survive the infrared cut-off contribute to the mean value.
The $\kappa$-dependent effective action $\Gamma_\kappa$
is then defined as the Legendre transformation of $W_\kappa^{({\cal A})}$,
\begin{equation}
{\Gamma}_\kappa[\overline{A}] \;=\;
{W}^{({\cal A})}_\kappa[{\cal J}] \;-\; \int d^4x \;{\cal J}_\mu^a \overline{A}_a^\mu
\;,
\label{Gamma00}
\end{equation}
which amounts to a change of variables from $\{ J_\mu \}$, the external source, 
to $\{ \overline{A}_\mu \}$, the average gauge field,
and yields the conjugate of (\ref{Abar}) as
\begin{equation}
\left.
\frac{\delta W^{({\cal A})}_\kappa}{\delta {\cal J}_a^\mu(x)}
\right|_{{\cal J}=0}
\;\equiv\;\left({\cal G}_\kappa^{(1)}(x)\right)_{\mu}^{a}
\;\;=\;\; \overline{A}_\mu^a(x)
\;\;\;\;\;\;\;\;\;
\;\;\;\;\;\;\;\;\;
\left.
 \frac{\delta {\Gamma}_\kappa}{\delta \overline{A}_a^\mu(x)} 
\right|_{\overline{A}=A_0}
\;\equiv\;\left({\Gamma}_\kappa^{(1)}(x)\right)_{\mu}^{a} 
\;\;= \;\,-  {\cal J}_\mu^a(x)
\label{1point}
\;.
\end{equation}
Notice that switching off the external sources ${\cal J}=0$ corresponds to
to the extremum of the effective action at
$\overline{A} = A_0$,  where $A_0$ is the mean
value of the gauge field for which the
effective action achieves its stationary extremum
\footnote{For QCD in the absence of a medium,
$A_0 = 0$, because the vacuum is colorless and Lorentz invariant, which
does not allow a non-vanishing $\langle {\cal A}_\mu^a \rangle$.}
(we take $\overline{A}_0=0$ later).
As summarized in Appendix B, 
repeated functional derivatives of ${W}_\kappa^{({\cal A})}[{\cal J}]$ 
with respect to the sources ${\cal J}$ 
generate the ($\kappa$-dependent) {\it connected} $n$-point Green functions,
and functional differentiation of $\Gamma_\kappa[\overline{A}]$ with respect to the 
average fields $\overline{A}$ yields the  {\it one-particle irreducible} 
$n$-point vertex functions.
In particular, the second functional derivatives determine the 2-point functions
\begin{equation}
\left.
\frac{ -i\,\delta^2 {W}^{({\cal A})}_\kappa}{\delta {\cal J}_a^\mu(x)\delta {\cal J}_b^\nu(y)}
\right|_{{\cal J}=0}
\;\equiv\;
\left({\cal G}_\kappa^{(2)}\right)_{\mu\nu}^{ab} 
\;\;=\;\;
 \Delta_{\kappa\;\mu\nu}^{ab}
\;\;\;\;\;\;\;\;\;\;\;\;
\left.
\frac{\delta^2 {\Gamma}_\kappa}{\delta \overline{A}_a^\mu(x)\delta \overline{A}_b^\nu(y)}
\right|_{\overline{A}=A_0}
\;\equiv\;
\left({\Gamma}_\kappa^{(2)}\right)_{\mu\nu}^{ab} 
\;\;=\;\;
\left( \Delta_{\kappa}^{-1}\right)_{\mu\nu}^{ab}
\;,
\label{2point}
\end{equation}
that is, ${\cal G}_\kappa^{(2)}$ is the exact gluon propagator, 
and ${\Gamma}_\kappa^{(2)}$ is its inverse, with
\begin{equation}
\Delta_{\kappa,\,\mu\nu}^{\;\;ab}(x,y)\;=\;
-\, \frac{\delta \langle\,{\cal A}_\nu^b(y)\,\rangle_\kappa}
{\delta {\cal J}_a^\mu(x)}
\;\;=\;\;
\langle\;{\cal A}_\mu^a(x)\,{\cal A}_\nu^b(y)\;\rangle_\kappa
\label{dAdJ}
\;,
\end{equation}
where again
the contributing field modes ${\cal A}$
are subject to the infrared cut-off at $\kappa$.
Similar relations hold for the higher $n$-point functions (c.f. Appendix B).
\medskip

\subsection{Renormalization issues}

The point of introducing the scale-dependent effective action 
${\Gamma}_\kappa$ satisfying (\ref{Gamma00}) is that
it allows us to vary the scale $\kappa$, say, from some large initial value
corresponding to the perturbative domain down to very small
values in the non-perturbative regime.
In effect, as we change $\kappa$,
more and more gluon fluctuations are included in the effective action,
and at the same time will define the renormalized quantities of the
effective theory, i.e., the gauge field ${\cal A}_\mu$ and the coupling $g$.
As the effective action ${\Gamma}_\kappa$ is a scalar quantity,
the infinities appearing in it {\it must} take the
Lorentz-invariant form of a scalar function 
${\cal Z}$ times $S_{\rm eff}[{\cal A}] =S_{\rm YM}[{\cal A}]+ \Re_\kappa[{\cal A}]$,
i.e.,
$S_{\rm YM}[{\cal A}] = {\cal Z} S_{\rm YM}[{\cal A}_{0}]$, and
$\Re_\kappa[{\cal A}] = {\cal Z} \Re_\kappa[{\cal A}_0]$.
If we define the renormalized field ${\cal A}_\mu$ and 
the renormalized coupling $g$ in terms of the bare, unrenormalized quantitities
${\cal A}_{0\,\mu}$ and $g_0$,
\begin{equation}
{\cal A}_{0\;\mu}^{a} \;=\; Z_{\cal A}^{1/2} \;{\cal A}_\mu^a
\;\;\;\;\;\;\;\;\;\;\;\;
g_{0} \;=\; Z_{g} \;g
\;,
\end{equation}
then the bare ${\cal F}_{0\;\mu\nu}^{\;\;\;a}$ is renormalized by
\begin{eqnarray}
{\cal F}_{0\;\mu\nu}^{\;\;\;a} & = &
\partial_\mu {\cal A}_{0\;\nu}^a \;-\; \partial_\nu {\cal A}_{0\;\mu}^a 
\;+\; g_0\, f^{a}_{bc} \,{\cal A}_{0\;\mu}^b {\cal A}_{0\;\nu}^c
\frac{}{}
\nonumber \\
& = & Z_{\cal A}^{1/2} \;
\left[\frac{}{}
\partial_\mu {\cal A}_{\nu}^a \;-\; \partial_\nu {\cal A}_{\mu}^a 
\;+\; (Z_{\cal A}^{1/2} Z_g) \,g\, f^{a}_{bc} \,{\cal A}_\mu^b {\cal A}_\nu^c
\right]
\end{eqnarray}
and consequently the squared field strength tensor is, 
\begin{eqnarray}
\left({\cal F}_{0\;\mu\nu}^{a}\right)^2 
& = &
Z_{\cal A} \;
\left[\frac{}{}
\left(
\partial_\mu {\cal A}_{\nu}^a \;-\; \partial_\nu {\cal A}_{\mu}^a 
\right)^2
\;+\;
(Z_{\cal A}^{1/2} Z_{g}) \,g\, f_{abc} 
(\partial_\mu {\cal A}_{\nu,a}) {\cal A}^\mu_b {\cal A}^\nu_c
\right.
\nonumber \\
& &
\;\;\;\;\;\;\;\;\;\;\;\;
\;\;\;\;\;\;\;\;\;\;\;\;
\;\;\;\;\;\;\;\;\;\;\;\;\;
\left. \frac{}{}
\;+\;
(Z_{\cal A} Z_{g}^2) \,
g^2\, f_{abc} f_{ab'c'} {\cal A}_{\mu,b} {\cal A}_{\nu, c} 
{\cal A}^\mu_{b'} {\cal A}^\nu_{c'}
\right]
\end{eqnarray}
Clearly, this will only take on a invariant form
of a scalar function ${\cal Z}^{-1}$ times $({\cal F}_{\mu\nu})^2$ provided that
\begin{equation}
Z_{\cal A} \;\;=\;\; Z_{g}^{-2} \;\;\equiv\;\; {\cal Z}^{-1}
\;.
\end{equation}
Indeed, as has been demonstrated originally by Kummer \cite{kummer}
to order $O(g^2)$ and later been proven in general \cite{konetschny77},
this equality of the renormalization factors for the gauge fields, the
3-gluon coupling and the 4-gluon coupling is a unique property
of non-covariant gauges, and therefore holds in the
light-cone gauge employed in this paper.
Similarly, the infrared cut-off for the gauge fields,
$\Re_\kappa[{\cal A}]$ of eq. (\ref{Delta1}) is renormalized by
\begin{eqnarray}
\Re_{\kappa}[{\cal A}_0]
&= &
\frac{1}{2} {\cal A}^{\mu}_{0\;a} \; 
\left[ \frac{}{} g_{\mu\nu} \delta^{ab}
\;R_\kappa(\partial^2)\right] \;{\cal A}^{\nu}_{0\;b}
\;+\;
\frac{1}{2}\left(\frac{1}{\xi} - 1\right) \;
  (n_\mu {\cal A}^{\mu}_{0\;b}) \;
\left[ \frac{(n\cdot \partial)^2 + R_\kappa\left((n\cdot\partial)^2\right)}
{R_\kappa\left((n\cdot\partial)^2\right)}
\;\partial^2\;\delta^{ab} \right]
\; (n_\nu {\cal A}_{0\;b}^{\nu})
\nonumber \\
& = &
\frac{Z_{{\cal A}}}{2} {\cal A}^{\mu}_a \; 
\left[ \frac{}{} \;g_{\mu\nu} \delta^{ab}
\;R_\kappa(\partial^2)\right] \;{\cal A}^{\nu}_b
\;+\;
\frac{Z_{{\cal A}}}{2}\left(\frac{1}{\xi} - 1\right) \;
  (n_\mu {\cal A}^{\mu}_b) \;
\left[ \,
 \frac{(n\cdot \partial)^2 + R_\kappa\left((n\cdot\partial)^2\right)}
{R_\kappa\left((n\cdot\partial)^2\right)}
\;\partial^2\;\delta^{ab} \right]
\; (n_\nu {\cal A}_b^{\nu})
\;.
\end{eqnarray}
Hence, 
\begin{equation}
\left({\cal F}_{0}^{\mu\nu}\right)^2 
 \;\;=\;\; {\cal Z}^{-1}\;\; \left({\cal F}_{\mu\nu}\right)^2
\;,\;\;\;\;\;\;\;\;\;
\Re_\kappa[{\cal A}_0] \;\;=\;\; {\cal Z}^{-1}\;\; \Re_\kappa[{\cal A}]
\;,
\end{equation}
and consequently the forms of $S_{\rm eff}[{\cal A}]$,
$W_\kappa^{(\cal A)}$ and the effective action ${\Gamma}_\kappa[\overline{A}]$ 
are preserved under simple multiplicative renormalization.
Notice that all the physics of the renormalization group is encoded
in a {\it single} scalar renormalization function ${\cal Z}$,
which is a function of the gluon momentum $q$ as well as
of the infrared scale $\kappa$, i.e.,
\begin{equation}
{\cal Z}\;\;\equiv \;\; {\cal Z}_\kappa(q) \;\;=\;\; 
{\cal Z}_\kappa\left(q^2,n\cdot q\right)
\label{Z}
\;,
\end{equation}
where the last equality is true for the class of axial gauges,
for which one can show \cite{kummer} that the $q$-dependence
can only enter in the combination of the two Lorentz invariants
$q^2$ and $(n\cdot q)^2$.
This function ${\cal Z}_\kappa$ will thus be the key 
to the $\kappa$-evolution of the effective action
and the associated gluon propagator.
In particular, we shall exploit the advantageous property of axial
gauges that 
(for specific choices of the gauge vector $n$ and the gauge parameter $\xi$)
the renormalized gluon propagator is simply the renormalization 
function ${\cal Z}_\kappa$ times the bare propagator,
\begin{equation}
\Delta_{\kappa , \,\mu\nu}(q) \;\;=\;\; {\cal Z}_\kappa(q) \;\;
\Delta_{\kappa , \,\mu\nu}^{(0)}(q) 
\;,
\label{RG}
\end{equation}
and similarly, the renormalized running coupling is the bare coupling 
constant multiplied by ${\cal Z}_\kappa^{1/2}$:
\begin{equation}
g \;\equiv \;g(q^2) \;\;=\;\; {\cal Z}^{1/2}_\kappa(q) \;\;g_0
\;.
\label{RGg}
\end{equation}
If we choose the mass scale $\Lambda$ as the point where we normalize
the theory [c.f. (\ref{tdef})], then
\begin{equation}
{\cal Z}_\kappa(\Lambda)\;\;=\;\; 1
\;\;\;\;\;\;\;\;\;\;\;\;\;\;\;
g_0\;\;=\;\; g(\Lambda^2)
\;.
\label{RG0}
\end{equation}
The roadmap for the following is to derive a $\kappa$-evolution 
equation for the effective action, and extract a corresponding
evolution equation for the renormalization function ${\cal Z}_\kappa$,
which then allows us to infer the exact propagator via (\ref{RG})
the running coupling from (\ref{RGg}), subject to the normalization 
condition (\ref{RG0}).
\smallskip

\subsection{The average effective action}

After these preliminaries, we are now in the position 
to derive an {\it average} effective action $\overline{\Gamma}_\kappa$ from
the effective action ${\Gamma}_\kappa$ of (\ref{Gamma00}), as well as an 
{\it exact} evolution equation for this average $\overline{\Gamma}_\kappa$ within
the renormalization group framework. This evolution equation determines how the physics
changes when more and more gluon fluctuations are included
in the functional by successively lowering $\kappa$ towards zero.
The  {\it average effective action} $\overline{\Gamma}_\kappa$ 
is defined \cite{wett1,wett2} as the effective action 
$\Gamma_\kappa$ of (\ref{Gamma00}) minus
the infrared regulator $\Re_\kappa$, eq. (\ref{Delta1}),
\begin{equation}
\overline{\Gamma}_\kappa[\overline{A}] \;=\;
\Gamma_\kappa[\overline{A}] \;\,-\;\, \Re_\kappa[\overline{A}]
\label{Gamma2}
\;,
\end{equation}
and reads in view of (\ref{Gamma00}),
\begin{eqnarray}
\overline{\Gamma}_\kappa\left[\overline{A}\right]
& = &
- i\,\ln \;\,\left[\frac{}{}
\;\int \,{\cal D}{\cal A} \; 
\;\exp \left\{\frac{}{}\,i\, \left( S_{\rm YM}\left[{\cal A}\right]
\,+\,S_{\rm fix}^{(\xi)}\left[{\cal A}\right]
\,+\, \Re_\kappa[{\cal A}]
\,-\, \Re_\kappa[\overline{A}]
\,+\, S_{\rm ext}[{\cal A},{\cal J}]
\,-\, S_{\rm ext}[\overline{A},{\cal J}]
\right) \right\}
\right]
\nonumber \\
& = &
- i\,\ln \;\,\left[\frac{}{}
\;\int \,{\cal D}{\cal A} \; \;\;\exp \left\{\frac{}{}\,i\, 
\int d^4x 
\;
\left(
\frac{}{}
-\frac{1}{2}\;
{\cal A}_a^\mu 
\left(g_{\mu\nu} \partial^2 - \partial_\mu \partial_\nu \right) 
{\cal A}^{\nu\,a}
\; -\;
\frac{1}{2\xi}\;( n^\mu {\cal A}^a_\mu) \; (n^\nu {\cal A}^{\nu\,a})
\right.
\right.
\right.
\label{Gamma10} \\
& &
\;\;\;\;\;\;\;\;\;\;\;\;\;\;\;\;\;\;\;\;\;\;\;\;\;
\;-\;
\frac{1}{2} \,g\, f_a^{bc}
\;
\frac{}{}
\left(\partial_\mu {\cal A}^a_\nu -  \partial_\nu {\cal A}^a_\mu\right)  
\; {\cal A}_b^{\mu} {\cal A}_c^{\nu} 
\;-\;
\frac{1}{4} \,g^2 \,f_a^{ce} f_{b\;e}^{d}
\;
{\cal A}^a_\mu  {\cal A}^b_\nu  {\cal A}_c^{\mu} {\cal A}_d^{\nu}
\nonumber \\
& &
\left.
\left.
\left.
\;\;\;\;\;\;\;\;\;\;\;\;\;\;\;\;\;\;\;\;\;\;\;\;\;
\; -\; \frac{1}{2}\; 
\left({\cal A}_a^\mu\, {\cal R}_{\kappa\;\mu\nu}^{ab}\,{\cal A}^{\nu\,b}
 - \overline{A}_a^\mu\, {\cal R}_{\kappa\;\mu\nu}^{ab}\,\overline{A}^{\nu\,b} \right)
\;+\; {\cal J}^a_\mu ({\cal A}_a^\mu - \overline{A}_a^\mu)
\frac{}{}
\;\right)
\right\}
\;\right]
\nonumber
\;,
\end{eqnarray}
where ${\cal R}_\kappa \equiv {\cal R}_\kappa(\partial^2)$ from (\ref{Delta3a}).
Evaluating the functional integral, and setting $\overline{A} = 0$,
one sees that the classical contribution to the action 
$\overline{S}\equiv \left[S_{\rm YM} + S_{\rm fix}^{(\xi)}\right]_{\overline{A}=0}
=0$ vanishes, so that one arrives at  \cite{ms42}
\begin{equation}
\overline{\Gamma}_\kappa\;\;\equiv\;\;
\overline{\Gamma}_\kappa[\overline{A}=0]\;\;=\;\;
\overline{\Gamma}_\kappa^{(0)} \;+\;\widehat{\overline{\Gamma}}_\kappa
\;,
\label{Gamma11}
\end{equation}
where the $\overline{ \Gamma}^{(0)}_\kappa$ is the `kinetic' part
at $\overline{A}=0$, 
\begin{equation}
\overline{\Gamma}_\kappa^{(0)}
\;=\;
\;\frac{i}{2} \; \int d^4x d^4y 
\;\left( {\Delta}^{(0)}_{\kappa}\right)^{-1} (x-y)
\;{\Delta}_\kappa(y,x),
\;,
\label{Gamma12a}
\end{equation}
and $\widehat{\overline{ \Gamma}}$ is the `interaction' part
at $\overline{A}=0$, 
\begin{eqnarray}
\widehat{\overline{\Gamma}}_\kappa
&=&
\;\;\;\;\;\;\frac{1}{8} \,g^2\;
\;\int d^4x d^4y \int d^4x_1d^4y_1 
W^{(0)}(x,y,x_1,y_1)\;{\Delta}_\kappa(y_1,x_1)\;{\Delta}_\kappa(y,x)
\nonumber
\\
& & 
\;+\;
\frac{i}{12} \,g^2\;
\;\int d^4x d^4y \int\prod_{i=1}^{2} d^4x_id^4y_i 
\;\;V^{(0)}(x,x_1,x_2)\;\;{\Delta}_\kappa(x_1,y_1)\;{\Delta}_\kappa(x_2,y_2)\; 
{\cal V}(y_2,y_1,y) \;{\Delta}_\kappa(y,x)
\nonumber
\\
& & 
\;+\;
\frac{1}{48} \,g^4\;
\;\int d^4x d^4y \int\prod_{i=1}^{3} d^4x_id^4y_i 
\;\;W^{(0)}(x,x_1,x_2,x_3)\;\;{\Delta}_\kappa(x_1,y_1)\;
{\Delta}_\kappa(x_2,y_2)\; {\Delta}_\kappa(x_3,y_3)\;
\nonumber \\
& &
\;\;\;\;\;\;\;\;\;\;\;\;\;\;\;\;\;\;\;\;\;\;\;\;\;
\;\;\;\;\;\;\;\;\;\;\;\;\;\;\;\;\;\;\;\;\;\;\;\;\;
\;\;\;\;\;
\times\;
{\cal W}(y_3,y_2,y_1,y) \;\;{\Delta}_\kappa(y,x)
\nonumber
\\
& & 
\;+\;
\frac{i}{96} \,g^4\;
\;\int d^4x d^4y \int\prod_{i=1}^{2} d^4x_id^4y_id^4z_i 
\;\;W^{(0)}(x,x_1,x_2,x_3)\;\;{\Delta}_\kappa(x_2,z_2)\;{\Delta}_\kappa(x_3,z_3)\; 
\nonumber \\
& &
\;\;\;\;\;\;\;\;\;\;\;\;\;\;\;\;\;\;\;\;\;\;\;\;\;
\;\;\;\;\;\;\;\;\;\;\;\;\;\;\;\;\;\;\;\;\;\;\;\;\;
\;\;\;\;\;\;
\times\;
{\cal V}(z_3,z_2,z_1) \;\;{\Delta}_\kappa(z_1,y_1) \;
{\Delta}_\kappa(x_1,y_2) \;\;{\cal V}(y_1,y_2,y)
\;\;{\Delta}_\kappa(y,x)
\label{Gamma12}
\;.
\end{eqnarray}
Here we  made use of the formulae of Appendix B, in
which ${\Delta}_{\kappa}$
denotes the {\it exact} proagator given by (\ref{2pf}), while
${\cal V}$ and ${\cal W}$
are the {\it exact} proper 3-gluon and 4-gluon vertex functions given by 
(\ref{3pf}) and (\ref{4pf}), respectively.
Correspondingly,
${\Delta}^{(0)}_{\kappa}$ is the {\it bare} propagator,
and  $V^{(0)}$,  $W^{(0)}$ the {\it bare} vertices, explicitly
given by (\ref{R3vertex}) and (\ref{R4vertex}), respectively.
The different contributions in $\widehat{\overline{\Gamma}}_\kappa$
correspond to the diagrams of Fig 2:
the first term is the one-gluon loop,
the second term is the tadpole contribution,
the third term is the 2-gluon loop  with exact 3-vertex,
the fourth term is  the three-loop contribution with exact 4-vertex
and the last term is the three-loop contribution with two exact 3-vertices.
\medskip

Notice that the infrared regulating terms $\Re_\kappa$ in
(\ref{Gamma10}) affect only the contributions that are quadratic
in ${\cal A}$ or $\overline{A}$. Hence, if we write in analogy to
(\ref{Gamma11}) for the effective action $\Gamma_\kappa$ in the presence
of the infrared regulator
\begin{equation}
{\Gamma}_\kappa\;\;\equiv\;\;
{\Gamma}^{(0)}_\kappa \;+\;\widehat{\Gamma}_\kappa
\;,
\label{Gamma11a}
\end{equation}
we have in view of
(\ref{Gamma2}) the following mapping  between 
$\overline{\Gamma}_\kappa$ and  $\Gamma_\kappa$:
\begin{equation}
\Gamma^{(0)} \;\;=\;\; \overline{\Gamma}^{(0)}_\kappa \;+\;\Re_\kappa 
\;\;\;\;\;\;\;\;\;\;\;\;\;\;
\widehat{\Gamma}_\kappa \;\;=\;\; \widehat{\overline{\Gamma}}_\kappa
\label{Gamma11b}
\;.
\end{equation}
\medskip

\setcounter{figure}{1}
\vspace{-2.0cm}
\begin{figure}[htb]
\epsfxsize=500pt
\centerline{ \epsfbox{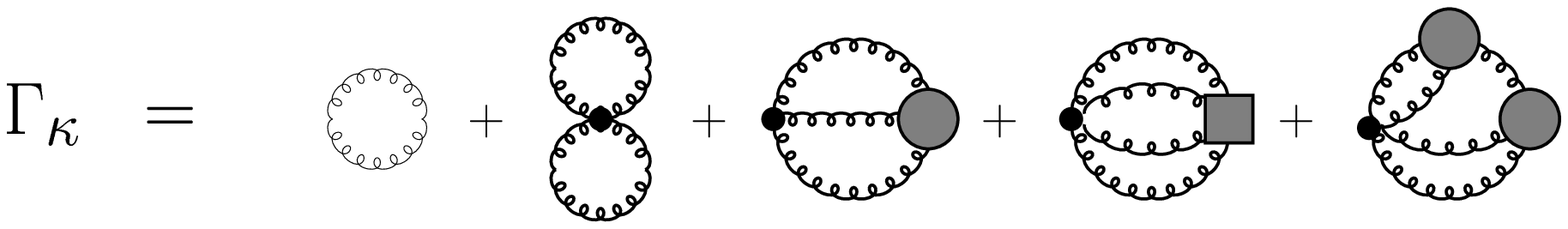} }
\vspace{-18.5cm}
\caption{
Diagrammatic representation of
the effective action 
${\overline{\Gamma}}_\kappa = {\overline{\Gamma}}^{(0)}_\kappa +
\widehat{\overline{\Gamma}}_\kappa$, eqs. (\ref{Gamma11})-(\ref{Gamma12}):
the first term is the `kinetic' term,
the second term is the tadpole contribution,
the third contribution is the 2-gluon loop  with exact 3-vertex,
the fourth term is  the three-loop contribution with exact 4-vertex
and the last diagram is the three-loop contribution with two exact 3-vertices.
The curly lines represent the {\it exact}
gluon propagator in the presence of the
infrared cut-off $\kappa$, the dots are {\it bare} 3-gluon
or 4-gluon vertices vertices, while
the shaded circles (boxes) are {\it exact} 3-gluon and 4-gluon
vertices. 
\label{figure2}
}
\medskip

\end{figure}
\bigskip

\subsection{Evolution equation for the average effective action}

Following Ref. \cite{wett2}, one can derive
an exact evolution equation for the average effective action
$\overline{\Gamma}_\kappa$ defined by (\ref{Gamma2}),
which a type of renormalization group equation 
which governs the scale-dependence of $\overline{\Gamma}_\kappa$
as the infrared cut-off $\kappa$ is
varied. Let us introduce the dimensionless evolution variable,
\begin{equation}
t\;\,\equiv \;\, \ln\left(\frac{\kappa}{\Lambda}\right) \;,\;\;\;\;\;\;\;\;\; 
d t \;\,=\;\, \frac{d\kappa}{\kappa} \;\,=\;\, \frac{1}{2} \;\frac{d\kappa^2}{\kappa^2} 
\;.
\label{tdef}
\end{equation}
where $\Lambda$ is some convenient mass scale 
at which  the theory is normalized (Secs. IV and V), and which may be chosen
to match a specific physics situation, e.g., the total
invariant mass of a  high-energy particle collision, or the large momentum transfer 
in a hard scattering process.  Recalling (\ref{Gamma2}),
and introducing for abbrevation
\begin{equation}
K[{\cal A},\overline{A},{\cal J}] 
\; \equiv \;
\;\exp \left[\;\frac{}{} i \left(
S_{\rm eff}[{\cal A},{\cal J}] + \Re_\kappa[{\cal A}] - {\cal J} \circ \overline{A} 
\right) \right]\, ,
\end{equation}
one obtains for the derivative of $\Gamma_\kappa$,
using (\ref{Z2} -- \ref{SA0}, \ref{Z3}, \ref{Gamma00}),
\begin{eqnarray}
\frac{\partial}{\partial t}\, \Gamma_\kappa[\overline{A}]
\;& = & \;
\frac{\partial}{\partial t}\, W_\kappa^{({\cal A})}[{\cal J}]
\;\;=\;\;
-i \,\frac{\partial}{\partial t}
\;\ln \left[ \int \,{\cal D}{\cal A} 
\;K\left[{\cal A},\overline{A},{\cal J}\right]
\right]
\;\;=\;\;
\left.
\frac{
\int \,{\cal D}{\cal A} 
\; \frac{1}{2} {\cal A}_\mu \,\left(\frac{\partial}{\partial t} {\cal R}_\kappa^{\mu\nu} \right)
\,{\cal A}_\nu\,
\;\;K\left[{\cal A},\overline{A},{\cal J}\right]
}{
\int \,{\cal D}{\cal A} 
\;\;K\left[{\cal A},\overline{A},{\cal J}\right]
}
\right|_{{\cal J}=0}
\nonumber \\
& & \nonumber\\
& = &
\;\;\frac{1}{2} \;\mbox{Tr}
\,\left[\frac{}{}
\; \left(\frac{\partial}{\partial t}\,{\cal R}_\kappa\right)
\;\;
\left(\frac{}{}
{\cal G}_{\kappa}^{(2)}\;+\;\overline{A} \,\circ\, \overline{A}
\right)
\right]
\label{evoleq1a}
\;,
\end{eqnarray}
while for the derivative of the second term on the right-hand side of 
(\ref{Gamma2}) one has
\begin{eqnarray}
\frac{\partial}{\partial t}\,\Re_\kappa[\overline{A}]
& = &
\left.
\frac{1}{
\int \,{\cal D}{\cal A} 
\;\;K\left[{\cal A},\overline{A},{\cal J}\right]
}
\;\;
\left(
\frac{1}{2}\;\frac{\partial}{\partial t}\, 
\;{\cal R}_\kappa^{\mu\nu}
\right)
\;\;
\left(
\frac{}{}
\int \,{\cal D}{\cal A} 
\; {\cal A}_\mu 
\;\;K\left[{\cal A},\overline{A},{\cal J}\right]
\;\;
\int \,{\cal D}{\cal A} 
\; {\cal A}_\nu 
\;\;K\left[{\cal A},\overline{A},{\cal J}\right]
\right)
\;\right|_{{\cal J}=0}
\nonumber \\
& & \nonumber\\
& = &
\;\;\frac{1}{2} \;\mbox{Tr}
\,\left[\frac{}{}
\; \left(\frac{\partial}{\partial t}\,{\cal R}_\kappa\right)
\;\;
\left(\frac{}{} \overline{A}_\mu\,\circ\,\overline{A}_\nu \right)
\right]
\;,
\label{evoleq1b}
\end{eqnarray}
where 
$\mbox{Tr}\left[\ldots \right]$ stands for the trace over
all internal indices, as well as an integration over continuous variables.
Subtracting (\ref{evoleq1b}) from (\ref{evoleq1a}),
utilizing that
$
{\cal G}_{k}^{(2)}= (\Gamma_\kappa^{(2)})^{-1}= (\overline{\Gamma}_\kappa^{(2)} 
+ {\cal R}_\kappa)^{-1}
$,
where
$ \overline{\Gamma}_\kappa^{(2)}$ is the second functional derivative
of $\overline{\Gamma}_\kappa$ with respect to $\overline{A}$, one arrives at
the desired evolution equation for the effective average action (\ref{Gamma2}):
\begin{eqnarray}
\frac{\partial}{\partial t}\, \overline{\Gamma}_\kappa[\overline{A}]
& = &
\;\;\frac{1}{2} \;\mbox{Tr}
\,\left[\frac{}{}
\left(
\frac{\partial}{\partial t} \,{\cal R}_\kappa
\right)
\;\,
\left(\overline{\Gamma}_\kappa^{(2)}\;+\;{\cal R}_\kappa\right)^{-1}
\right]
\;\;\;\equiv\;\;\; {\cal \gamma}_\kappa[\overline{A}]
\label{evoleq2}
\;.
\end{eqnarray}
\bigskip
\bigskip

\section{The evolution equation for the gluon propagator}
\label{sec:section3}
\bigskip

Working henceforth in momentum space, we now take practical steps 
to solve the  evolution equation (\ref{evoleq2}) for the 
gluon propagator. Recall that we defined the exact gluon propagator,
respectively its inverse, as (c.f. (\ref{2point}, \ref{dAdJ})):
\begin{eqnarray}
\Delta_{\kappa,\,\mu\nu}^{\;\;\;ab}(q)
& = &
\;\left({\cal G}_\kappa^{(2)}(q,-q)\right)_{\mu\nu}^{ab}
\;\;=\;\;\langle\;{\cal A}_\mu^a(q)\;{\cal A}_\nu^b(-q)\;\rangle_\kappa 
\;,
\label{Gprop0a}
\\
(\Delta_{\kappa}^{-1})_{\mu\nu}^{ab}(q)
& = &
\;\left({\Gamma}_\kappa^{(2)}(q,-q)\right)_{\mu\nu}^{ab}
\;\;=\;\;
\left(\overline{\Gamma}_\kappa^{(2)}(q,-q)\right)_{\mu\nu}^{ab}
+
{\cal R}_{\kappa,\,\mu\nu}^{\;\;\;ab}(q^2)
\label{Gprop0b}
\end{eqnarray}
Our goal is now to infer from the general evolution equation
(\ref{evoleq2}) for the average effective action $\overline{\Gamma}_\kappa$
a corresponding evolution equation for $\Delta_\kappa^{-1}$,
from which we can then determine the properties of
the propagator  $\Delta_\kappa$ itself.
\bigskip

\subsection{The general case}
\smallskip

We begin by  rewriting (\ref{evoleq2}) as
\begin{eqnarray}
\frac{\partial}{\partial t}\, \overline{\Gamma}_\kappa[\overline{A}]
& = &
\frac{1}{2} \int \frac{d^4 q}{(2\pi)^4 }\;
\;\left( \frac{\partial}{\partial t} \,{\cal R}_\kappa(q^2)
\right)_{\mu\nu}^{ab}
\;\,
\left(\frac{}{}
\left[\overline{\Gamma}_\kappa^{(2)}(q,-q) \,
+\, {\cal R}_\kappa(q^2)\right]^{-1}\right)_{ab}^{\mu\nu}
\;\;\;\equiv\;\;\; {\cal \gamma}_\kappa[\overline{A}]
\label{evoleq2a}
\;.
\end{eqnarray}
As this  is  an {\it exact} equation, any attempt to solve it
in full is certainly out of question, because it would require to 
solve for an infinite number of the vertex functions 
$\overline{\Gamma}_\kappa^{(n)}$ which contribute to both sides
of (\ref{evoleq2a}). On the left-hand side, the $\overline{\Gamma}_\kappa^{(n)}$ 
enter through the series representation of $\overline{\Gamma}_\kappa[\overline{A}]$,
\begin{equation}
\overline{\Gamma}_\kappa[\overline{A}] 
\;=\; \sum_{n=0}^\infty\;\frac{1}{n !}\;
\int \frac{d^4 q_n}{(2\pi)^4} \ldots \frac{d^4 q_1}{(2\pi)^4}\;
\left(\overline{\Gamma}_\kappa^{(n)}(q_1,\ldots q_n)\right)_{\mu_1\ldots\mu_n}^{a_1\ldots a_n}
\;
\overline{A}_{a_1}^{\mu_1}(q_1)\ldots \overline{A}_{a_n}^{\mu_n}(q_n)
\;,
\label{Gamma30}
\end{equation}
while on the right-hand side of (\ref{evoleq2a}), the $\overline{\Gamma}_\kappa^{(n)}$ 
are implicitly encoded in the 2-point function $\overline{\Gamma}_\kappa^{(2)}$.
However, since we are here interested  in the behavior of only the gluon propagator 
$\Delta_\kappa^{\mu\nu} = \langle {\cal A}^\mu{\cal A}^\nu\rangle_\kappa$, 
we do not need to solve (\ref{evoleq2a}) for
the average effective action $\overline{\Gamma}_\kappa[\overline{A}]$ 
as a whole, but only for its contributions
$\overline{\Gamma}_\kappa^{(2)}[\overline{A}]$ 
which are second order in  $\overline{A}$ on the left-hand side of 
(\ref{evoleq2a}), and which are mapped on 
the corresponding quadratic contributions on the right-hand side,
denoted by $\gamma_\kappa^{(2)}[\overline{A}]$,
being the second order term in the series
\begin{equation}
\gamma_\kappa[\overline{A}] 
\;=\; \sum_{n=0}^\infty\;\, \gamma_\kappa^{(n)}
\label{Gamma30a}
\;.
\end{equation}
That is,
instead of (\ref{evoleq2a}) for the full $\overline{\Gamma}_\kappa$, 
we aim at the corresponding evolution equation with respect to
$t = \ln \kappa$ for the 2-point contributions alone,
\begin{equation}
\;\;\;\;\;\;\;\;\;\;\;
\frac{\partial}{\partial t}\, \overline{\Gamma}^{(2)}_\kappa
\;=\;
{\gamma}_\kappa^{(2)} 
\label{evoleq3}
\;.
\end{equation}
We emphasize that (\ref{evoleq3}) is still an {\it exact} equation:
no truncations have been imposed on the way from the
original evolution equation (\ref{evoleq2a}).
If we were to know $\gamma_\kappa^{(2)}$ exactly, then it
would be straightforward to solve for the evolution of 
$\overline{\Gamma}^{(2)}_\kappa$ with $\kappa$.
Unfortunately, the function $\gamma_\kappa^{(2)}$ on the right-hand side
is a tremendously complicated object, because it implicitly contains 
all sorts of contributions of higher order
in the gauge fields, which one would have to determine by
solving corresponding equations for $\overline{\Gamma}_\kappa^{(3)}$, 
$\overline{\Gamma}_\kappa^{(4)}$, and so forth. Fortunately, the
gauge symmetries of QCD allow to relate these higher-order
contributions among each other via the Slavnov-Taylor identities,
and it is possible, as we shall demonstrate, to obtain a closed
expression for $\gamma_\kappa^{(2)}$  without explicit
knowledge of the higher-order terms, but rather by their implicit
inclusion through the constraint equations that follow from first principles
\footnote{
This is analogous to the BBGKY hierarchy \cite{BBGKY} of
Green functions in field theory:
the $n$-point Green functions are
intimately coupled by an infinite set of equations of motion.
For example, the 1-point function (the mean field) is
determined by the Landau-Ginzburg equation, which contains
the 2-point function (the propagator). The 2-point
function itself is the solution of the Dyson-Schwinger equation,
which contains the 3-point and 4-point functions.
The 3-point and 4-point functions in turn are determined by even more
complicated equations that contain higher-order Green functions.
This scheme continues ad infinitum.
The hierarchy of the equations is exact, but in order to solve 
it approximately, it is usually truncated to a system
of equations involving only the 1- or 2-point functions.
To achieve self-consistency of the truncated set of equations
at, e.g.,  the $n=2$ level, the $n\ge3$ functions must be implicitly
included by additional constraint equations.
For instance, in QCD the Slavnov-Taylor identities
relate the 3-gluon vertex function to the propagator,
and can be used to eliminate the 3-point function.
We follow such a  path later in this paper.
}.
\bigskip

\subsubsection{Left-hand side of the evolution equation (\ref{evoleq3})}

Returning to (\ref{evoleq2a}),
we pick out from the series representation of $\overline{\Gamma}_\kappa[\overline{A}]$
in (\ref{Gamma30}) the contribution $\overline{\Gamma}_\kappa^{(2)}[\overline{A}]$
 that is quadratic in $\overline{A}$, and then consider $\overline{A}=0$,
\begin{equation}
\overline{\Gamma}_\kappa^{(2)}
\;\equiv\; 
\overline{\Gamma}_\kappa^{(2)}[\overline{A}=0]
\;=\; 
\left.
\frac{1}{2} \int \frac{d^4 q}{(2\pi)^4} 
\;\overline{A}_{a}^{\mu}(q)
\; \left(\overline{\Gamma}_\kappa^{(2)}(q,-q)\right)_{\mu\nu}^{ab}
\; \overline{A}_{b}^{\nu}(-q)
\;\right|_{\overline{A}=0}
\;.
\label{Gamma31}
\end{equation}
Now,
the two-point function under the integral on the right-hand side 
is related to the inverse gluon propagator by
$(\Delta_{\kappa})^{-1}_{\mu\nu} = \overline{\Gamma}_{\kappa ,\,\mu\nu}^{(2)} +
{\cal R}_{\kappa ,\,\mu\nu}$, since 
$\overline{\Gamma}_{\kappa ,\,\mu\nu}^{(2)} =
{\Gamma}_{\kappa ,\,\mu\nu}^{(2)} -  {\cal R}_{\kappa ,\,\mu\nu}$ 
and 
$\Gamma_{\kappa ,\,\mu\nu}^{(2)}= ({\cal G}^{(2)}_{\kappa})_{\mu\nu}^{-1}
= (\Delta_{\kappa})_{\mu\nu}^{-1}$.
We may, therefore, parametrize $\overline{\Gamma}_{\kappa ,\,\mu\nu}^{(2)}$
according to the most general tensor decomposition of the inverse
propagator $(\Delta_\kappa)^{-1}_{\mu\nu}$ that is compatible
with the constraining Ward identities for the class of axial gauges.
This requires two independent
scalar functions, $a_\kappa(q)$ and $b_\kappa(q)$, in which
the $q$-dependence can only
involve \cite{kummer} the two invariants $q^2/\Lambda^2$ and
$n^2 q^2/(n\cdot q)^2$. Introducing the variable
\begin{equation}
\chi \;\;\equiv\;\; \chi(n,q) \;\;=\;\; \frac{n^2\,q^2}{(n \cdot q)^2}
\label{chi}
\;,
\end{equation}
the dependence of $a_\kappa$ and $b_\kappa$ on $q$ and $n$ appears
as $a_\kappa(q)=a_\kappa(q^2,\chi)$, $b_\kappa(q)=b_\kappa(q^2,\chi)$.
Hence, we can  represent $\overline{\Gamma}_{\kappa\,\mu\nu}^{(2)}$ 
in the following form,
\begin{equation}
\; \left(\overline{\Gamma}_\kappa^{(2)}(q,-q)\right)_{\mu\nu}^{ab}
\;\;=\; \;
\delta^{ab} \left(\frac{}{}
a_\kappa (q^2,\chi) \;P_{\mu\nu}(q) \;+\; b_\kappa(q^2,\chi) \;Q_{\mu\nu}(q) 
\right)
\label{Gamma32}
\;,
\end{equation}
with the orthogonal projection operators
\footnote{
Here and in the following, negative powers of $n\cdot q$ are
understood in the principal value sense \cite{gaugereview}, which
ensures unitarity. Notice, that the last term $\propto n^{-2}$
in both $P_{\mu\nu}$ and in $Q_{\mu\nu}$, is actually $\propto (n\cdot q)^{-2}$,
as is evident from the definition of $\chi$, eq. (\ref{chi}).
}
\begin{eqnarray}
P_{\mu\nu}(q) &=&
\;g_{\mu\nu} \;+\;\frac{1}{1-\chi} \,
\left[
\;\;\chi\,\frac{q_\mu q_\nu}{q^2}
\,-\,\frac{n_\mu q_\nu + q_\mu n_\nu}{n\cdot q} 
\,+\,\chi\, \frac{n_\mu n_\nu}{n^2}
\right]
\label{proj1b}
\\
Q_{\mu\nu}(q) &=&
\;\;-\;\frac{1}{1-\chi} \;\;
\left[
\;\; \frac{q_\mu q_\nu}{q^2}
\,-\,\frac{n_\mu q_\nu + q_\mu n_\nu}{n\cdot q}
\,+\,\left( \chi\,-\,\frac{(1-\chi) n^2}{\xi q^2}\right) \, \frac{n_\mu n_\nu}{n^2}
\right]
\label{proj1a}
\;,
\end{eqnarray}
which have been constructed from the available vectors $q_\mu$, $n_\mu$
and from $g_{\mu}$ in the space $n_\mu P^{\mu\nu} = 0 = n_\mu Q^{\mu\nu}$.
In the absence of interactions, the bare parameters would be
$a_\kappa \rightarrow q^2+R_\kappa$ and $b_\kappa \rightarrow q^2+R_\kappa$.
In general, however, the scalar functions $a_\kappa$ and $b_\kappa$ in 
(\ref{Gamma32}) embody the full information about the running of
$\overline{\Gamma}_\kappa^{(2)}$ and, hence, of the gluon propagator 
which is determined by the inverse of $\overline{\Gamma}_\kappa^{(2)}$,
as we shall show below.
\bigskip

\subsubsection{Right-hand side of the evolution equation (\ref{evoleq3})}

Similar as above, we need to extract from ${\gamma}_\kappa[\overline{A}]$
in (\ref{evoleq2a}) the contribution ${\gamma}_\kappa^{(2)}[\overline{A}]$
 that is quadratic in $\overline{A}$ and then set $\overline{A}=0$.
We first notice that
\begin{equation}
\gamma_\kappa
\;\equiv\;
\gamma_\kappa [\overline{A} = 0]
\;=\;
\frac{1}{2} \int \frac{d^4 q}{(2\pi)^4 }  \frac{d^4 q'}{(2\pi)^4 }
\,\delta^4(q+q')\;
\left(
\frac{\partial}{\partial t} \,{\cal R}_\kappa(q^2)
\right)_{\mu\nu}^{ab}
\;\,
\left(\frac{}{}
\left[{\Gamma}_\kappa^{(2)}(q,q') \right]^{-1}\right)_{\mu\nu}^{ab}
\label{evoleq2c}
\end{equation}
where ${\cal R}_\kappa(q)$ is the Fourier transform of (\ref{Delta3a}),
\begin{equation}
{\cal R}_{\kappa\,,\mu\nu}^{\;\,ab}(q) \;=\;
{R}_{\kappa}(q^2) \;\delta^{ab}\,\left( g_{\mu\nu}\,-\,\frac{q_\mu q_\nu}{q^2}\right)
\;=\;
{R}_{\kappa}(q^2) \;\delta^{ab}\,\left( \,P_{\mu\nu}+Q_{\mu\nu}\right)
\label{Rkq}
\;.
\end{equation}
Next,
we decompose $\Gamma_\kappa^{(2)}$ in (\ref{evoleq2c}) into a kinetic term
($\Pi_\kappa^{(0)}$) and an interaction term ($\widehat{\Pi}_\kappa$),
\begin{equation}
(\Gamma_\kappa^{(2)})_{\mu\nu}^{ab}(q,q')
\;\;=\;\;\left(\frac{}{} \Pi_\kappa^{(0)} \;\;+\;\; \widehat{\Pi}_\kappa
\,\right)_{\mu\nu}^{ab}(q,q')
\;.
\label{Gamma15}
\end{equation}
From the relations (\ref{Gamma11a}), we infer
\begin{eqnarray}
\Pi_{\kappa ,\,\mu\nu}^{(0)\, ab} (q,q') &=&
\;\;\;\;\;\;
\left.
\frac{\delta^2 {\Gamma}_\kappa^{(0)}}
{\delta \overline{A}_a^\mu(q)\delta\overline{A}_b^\mu(q')}
\right|_{\overline{A}=0}
\;=\;
\;\;\;
2 i\;\delta^4(q+q')\;
\,\frac{\delta {\Gamma}_\kappa^{(0)}}{\delta \Delta_{\kappa\,ab}^{\;\mu\nu}(q)}
\nonumber\\
\widehat{\Pi}_{\kappa ,\,\mu\nu}^{\;\;\,ab} (q,q') &=&
\left.
\;\;\;\;\;\;
\frac{\delta^2 \widehat{\Gamma}_\kappa}
{\delta \overline{A}_a^\mu(q)\delta\overline{A}_b^\mu(q')}
\right|_{\overline{A}=0}
\;=\;
2i\;\delta^4(q+q')\;
\frac{\delta \widehat{\Gamma}_\kappa}{\delta \Delta_{\kappa\,ab}^{\;\mu\nu}(q)}
\label{Gamma16}
\;.
\end{eqnarray}
Applying these to the formulae (\ref{Gamma10} -- \ref{Gamma12}), 
after Fourier transformation to momentum space, we obtain for the kinetic term
\begin{eqnarray}
\Pi^{(0)\;ab}_{\kappa,\,\mu\nu}(q,-q)
& = &
\;\;\;\;\delta_{ab}\, \left( q^2 + R_\kappa(q^2)\right)\;
\left(\; g_{\mu\nu}\, -\,\frac{q_\mu q_\nu}{q^2}\,  +\; \frac{1}{\xi} \,
\frac{n_\mu n_\nu}{q^2}\right)
\label{Pi0}
\;,
\end{eqnarray}
while the interaction term gives
\begin{eqnarray}
\widehat{\Pi}^{\;\;ab}_{\kappa,\,\mu\nu}(q,-q)
&=&
\;\,\, \frac{g^2}{2}
\; \int \frac{d^4k}{(2\pi)^4}
\;\;W_{\;\;\mu\nu\lambda\sigma}^{(0),abcd}(q,k,-k,-q)\;
{\Delta}_\kappa^{\lambda\sigma\,,cd}(k) \;
\nonumber \\
& &
-\,\frac{i\,g^2}{2}
\; \int \frac{d^4k}{(2\pi)^4}
\;\;V_{\;\;\mu\lambda\sigma}^{(0),acd}(q,-k,-k')\;\;
{\Delta}_\kappa^{\lambda\lambda',\,cc'}(k)\;
{\Delta}_\kappa^{\sigma\sigma',\,dd'}(k')\;
{\cal V}_{\sigma'\lambda'\nu}^{d'c'b}(k',k,-q)
\nonumber \\
& &
+\;\frac{g^4}{6}
\; \int \frac{d^4k}{(2\pi)^4} \frac{d^4p}{(2\pi)^4}
\;\;{W}_{\;\;\mu\lambda\sigma\tau}^{(0),acde}(q,-k,-k',-p)\;\;
{\Delta}_\kappa^{\lambda\lambda',\,cc'}(k)\;
{\Delta}_\kappa^{\sigma\sigma',\,dd'}(k')\;
\nonumber \\
& &
\;\;\;\;\;\;\;\;\;\;\;\;\;\;\;\;\;\;\;\;\;\;\;\;\;
\;\;\;\;\;\;\;\;\;\;\;\;\;\;\;\;\;\;\;\;\;\;\;\;\;
\;\;\;\;\;\;
\times\;
{\Delta}_\kappa^{\tau\tau',\,ee'}(p)\;
{\cal W}_{\tau'\sigma'\lambda'\nu}^{e'd'c'b}(k,k', p,-q)
\nonumber \\
& &
+\;\frac{g^4}{24}
\; \int \frac{d^4k}{(2\pi)^4} \frac{d^4p}{(2\pi)^4}
\;\;W_{\;\;\mu\lambda\sigma\tau}^{(0),acde}(q,-k,-k',-p')\;\;
{\Delta}_\kappa^{\sigma\rho',\,df'}(k)\;
{\Delta}_\kappa^{\tau\rho'',\,ef''}(k')\;
\nonumber \\
& &
\;\;\;\;\;\;\;\;\;\;\;\;\;\;\;\;\;\;\;\;\;\;\;\;\;
\;\;\;\;\;\;
\times\;
{\cal V}_{\rho''\rho'\rho}^{f''f'f}(k,k', -p) \;\;
{\Delta}_\kappa^{\rho\lambda',\,fc'}(p)\;
{\Delta}_\kappa^{\lambda\sigma',\,cd'}(p')\;
\;\;{\cal V}_{\lambda'\sigma'}^{c'd'}(p, p', -q)
\label{Pi1}
\;,
\end{eqnarray}
with
$k' = q-k$ in the second term, $k' = q-k-p$ in the third term, and
$k' = q-k-p', \; p' = q-p$ in the last term.
Fig. 3 depicts diagramatically the inverse propagator 
$(\Gamma_\kappa^{(2)})_{\mu\nu}$, eq. (\ref{Gamma15}),
in terms of the contributions $\Pi_\kappa^{(0)}$, eq. (\ref{Gamma12a}),
and $\widehat{\Pi}_\kappa$, eq. (\ref{Gamma12}).
\bigskip

\vspace{-1.0cm}
\begin{figure}[htb]
\epsfxsize=450pt
\centerline{ \epsfbox{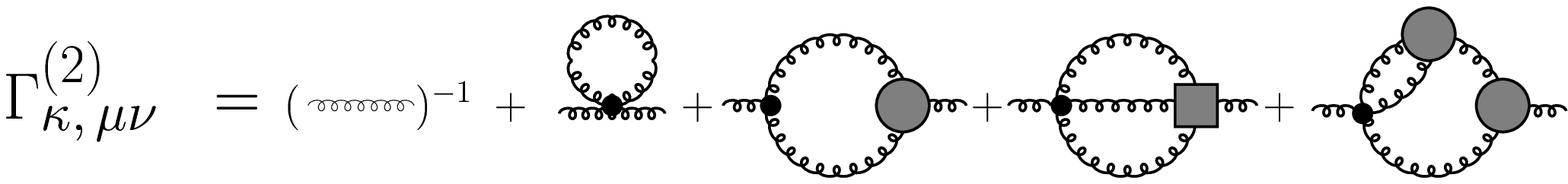} }
\vspace{-17.3cm}
\caption{
Diagrammatic representation of the inverse propagator 
$(\Gamma_\kappa^{(2)})_{\mu\nu} = (\Pi_\kappa^{(0)}+\widehat{\Pi}_\kappa)_{\mu\nu}$,
eqs. (\ref{Gamma15})-(\ref{Pi1}).
The first term is the contribution from the `kinetic part' $\Pi^{(0)}_\kappa$,
while the remaining terms arise from the `interaction
part' $\widehat{\Pi}_\kappa$.
The curly lines represent the {\it exact}
gluon propagator in the presence of the
infrared cut-off $\kappa$, the dots are {\it bare} 3-gluon
or 4-gluon vertices vertices, while
the shaded circles (boxes) are {\it exact} 3-gluon and 4-gluon vertices. 
\label{figure3}
}
\end{figure}
\bigskip
\bigskip

\noindent
Now let us define a partial derivative $\partial^\ast_t$
that acts only on the $t=\ln \kappa$ dependence of $\Pi^{(0)}_\kappa$, 
but not on $\widehat{\Pi}_\kappa$,
\begin{equation}
\partial^\ast_t \,\Gamma_\kappa^{(2)}\;\;:=\;\;
\partial^\ast_t \,\left(\Pi^{(0)}_\kappa+\widehat{\Pi}_\kappa\right) 
\;\;=\;\;\partial^\ast_t\,\Pi^{(0)}_\kappa
\;\;=\;\;\frac{\partial}{\partial t}\Pi^{(0)}_\kappa
\;,
\end{equation}
so that we may write the right-hand side of eq. (\ref{evoleq2a}) in
a form that is reminescent of the derivative of a one-loop expression,
but which is exact,
\begin{equation}
{\gamma}_\kappa 
\;\;=\;\; 
\frac{1}{2}
\int \frac{d^4 q}{(2\pi)^4 }\;
\;
\partial^\ast_t 
\,\left[\frac{}{}
\ln \left.{\Gamma}_\kappa^{(2)}\right._{\mu\,a}^{\mu\,a}(q,-q)
\right]
\;\;=\;\; 
\frac{1}{2} \,\mbox{Tr}\left[
\frac{\partial^\ast}{\partial t} 
\ln \left.{\Gamma}_\kappa^{(2)}\right. \right]
\;.
\label{evoleq2e}
\end{equation}
From this representation of $\gamma_\kappa$ we extract now the contribution
which corresponds to terms quadratic in the gluon fields, and therefore is
relevant for the evolution of the gluon propagator: We utilize
\begin{equation}
\mbox{Tr}\,[\;\partial_t^\ast \ln \Gamma_\kappa^{(2)}\;] 
\;=\; 
\mbox{Tr}\,[\;\ln \Pi_0\;]  
\;+\;\mbox{Tr}\,[\;\partial_t^\ast (\;\widehat{\Pi}_\kappa\; \Pi_\kappa^{(0)\;-1})]  
\;-\;\frac{1}{2} \mbox{Tr}\,[\;\partial_t^\ast (\;\widehat{\Pi}_\kappa\; 
\Pi_\kappa^{(0)\;-1}
\;\widehat{\Pi}_\kappa\; \Pi_\kappa^{(0)\;-1})\;] 
\;+\; \ldots
\label{expansion}
\;,
\end{equation}
where the dots refer to higher-order terms which 
are cubic and higher in the gluon fields and therefore
contribute only to the 3-point, 4-point functions, etc.,
but not to the gluon propagator or its inverse.
Now, the first term in (\ref{expansion}) amounts to an irrelevant
constant which may be dropped in view of (\ref{Pi0}), so that we 
finally arrive at 
\begin{eqnarray}
{\gamma}_\kappa^{(2)}  
& = &
\;\;\;\,
\frac{1}{2}\,
\int \frac{d^4 q}{(2\pi)^4 }\;
\;\partial^\ast_t 
\left[\frac{}{}
\widehat{\Pi}_{\kappa,\,\mu\nu}^{\;\;\;ab}(q,-q)\;\,(\Pi_{\kappa}^{-1})_{ba}^{\nu\mu}(-q,q)
\right]
\nonumber \\
& & -\; 
\frac{1}{4} \,
\int \frac{d^4 q}{(2\pi)^4 }\,\frac{d^4 p}{(2\pi)^4 }\;
\;\partial^\ast_t 
\left[\frac{}{}
\widehat{\Pi}_{\kappa ,\,\mu\lambda}^{\;\;\;ac}(q,-p)\,\;
(\Pi_{\kappa}^{-1})_{cd}^{\lambda\sigma}(-p,p)
\;\,\widehat{\Pi}_{\kappa,\,\sigma\tau}^{\;\;\;de}(p,-q)\,\;
(\Pi_{\kappa}^{-1})_{ea}^{\sigma\mu}(-q,q)
\right]
\;.
\label{gammaquad}
\end{eqnarray}
\bigskip

\subsubsection{The master equations for the gluon propagator}

Now we have collected all the ingredients for the evolution
equation (\ref{evoleq3}): $\overline{\Gamma}_\kappa^{(2)}$ 
appearing on the left-hand side, is given by 
(\ref{Gamma31} -- \ref{proj1b}), and ${\gamma}_\kappa^{(2)}$ 
on the right-hand side, is determined by 
(\ref{gammaquad}) together with (\ref{Pi0}) and (\ref{Pi1}).
In order to infer from this evolution equation two independent,
coupled scalar equations for the two unknown functions 
$a_\kappa$ and $b_\kappa$, we project
(\ref{evoleq3}) with $P_{\mu\nu}$ and  $Q_{\mu\nu}$,
given by (\ref{proj1a}) and (\ref{proj1b}), respectively.
Using $P_{\mu\lambda} P^{\lambda}_\nu = P_{\mu\nu}$, 
$Q_{\mu\lambda} Q^{\lambda}_\nu = Q_{\mu\nu}$, and 
$P_{\mu\lambda} Q^{\lambda}_\nu = 0$, as well as
$n^\mu (\Pi_\kappa^{(0)\;-1})_{\mu\nu}=0$,
$q^\mu (\widehat{\Pi}_\kappa)_{\mu\nu}=0$,
we obtain after  some algebra,
\begin{eqnarray}
\frac{\partial}{\partial t}\, a_\kappa(q^2,\chi)  &=&
\; \frac{1}{2}
\;\left(\;g_{\mu\nu}\;+\;
\frac{\chi}{1-\chi} \frac{n_\mu n_\nu}{n^2}\;\right)
\;\;\frac{\partial}{\partial t}\; \widehat{\Pi}_\kappa^{\mu\nu}(q,-q) 
\label{b1}
\\
\frac{\partial}{\partial t}\, b_\kappa(q^2,\chi)  &=&
\;\;- \;\frac{\chi}{1-\chi} \;
\frac{n_\mu n_\nu}{n^2}\;\;
\frac{\partial}{\partial t}\; \widehat{\Pi}_\kappa^{\mu\nu}(q,-q)
\label{a1}
\;.
\end{eqnarray}
We remind the reader of the complexity of these equations, which are
equivalent to (\ref{evoleq3}), and hence our comments 
after (\ref{evoleq3}) apply also here.
The key problem becomes clear in view of (\ref{Pi1}), which
shows that $\widehat{\Pi}_\kappa$ contains not only
the exact propagator $\Delta_\kappa$, but also the exact
3-gluon and 4-gluon vertex functions ${\cal V}$, respectively ${\cal W}$.
In principle, one would therefore have to solve even more
complicated equations for ${\cal V}$ and ${\cal W}$, and then
plug the solutions into $\widehat{\Pi}_\kappa$ of (\ref{Pi1}). 
Then (\ref{a1}) and (\ref{b1}) would contain on the
right-hand sides only the unknown $\Delta_\kappa$, the solution
of which we are after.
However, as we show in the next subsections, it is possible
to get rid of the explicit dependence on ${\cal V}$ and
${\cal W}$  by ({\it i}) eliminating the 4-gluon vertices
and ({\it ii}) expressing the 3-gluon vertices through the
propagator $\Delta_\kappa$ alone.
Then we can evaluate $\widehat{\Pi}_{\kappa}$, (\ref{a1}) and (\ref{b1}) 
serve to determine the functions $a_\kappa$ and $b_\kappa$
which, in turn, would give a unique solution to the
exact gluon propagator from (\ref{Gamma15}),
\begin{equation}
\Delta_{\kappa,\,\mu\nu}^{\;\;ab}(q)
\;=\;
\left({\Gamma}_\kappa^{(2)\;-1}\right)_{\mu\nu}^{ab}(q,-q)
\;\;=\;\;\left(\left[\Pi_\kappa^{(0)}\;+\widehat{\Pi}_\kappa 
\right]^{-1}\right)_{\mu\nu}^{ab}
\label{Gamma33}
\end{equation}
Decomposing the propagator analogous to (\ref{Gamma32}),
\begin{equation}
\Delta_{\kappa,\,\mu\nu}^{\;\;ab}(q)
\;=\;
\delta^{ab} \left(\frac{}{}
A_\kappa (q^2,\chi) \;S_{\mu\nu}(q) \;+\; B_\kappa(q^2,\chi) \;T_{\mu\nu}(q) 
\right)
\label{Gprop1}
\;,
\end{equation}
with the projectors
\begin{eqnarray}
S_{\mu\nu}(q) &=&
\;g_{\mu\nu} \;+\;\frac{1}{1-\chi} \,
\left[\;
\chi\;\left(1+\xi q^2\right) \, \frac{q_\mu q_\nu}{q^2}
\,-\,\frac{n_\mu q_\nu + q_\mu n_\nu}{n\cdot q} 
\,+\,\chi \, \frac{n_\mu n_\nu}{n^2}
\right]
\label{proj2b}
\\
T_{\mu\nu} (q) &=&
\;\;-\;\frac{1}{1-\chi} \;\;
\left[\;
\chi\;\left(1+\xi q^2\right) \, \frac{q_\mu q_\nu}{q^2}
\,-\,\frac{n_\mu q_\nu + q_\mu n_\nu}{n\cdot q}
\,+\, \frac{n_\mu n_\nu}{n^2}
\right]
\label{proj2a}
\;,
\end{eqnarray}
and inverting $\Pi_\kappa^{(0)}+\widehat{\Pi}_\kappa$ on the right-hand side
of (\ref{Gamma33}), one finds (c.f. Appendix D) that the functions 
$A_\kappa$ and $B_\kappa$ are related to  $a_\kappa$ and $b_\kappa$ by
\begin{equation}
A_\kappa (q^2,\chi) \; = \; \frac{1}{a_\kappa (q^2,\chi)}
\;,\;\;\;\;\;\;\;\;\;\;\;\;\;\;\;\;
B_\kappa (q^2,\chi) \; = \; \frac{\chi}{b_\kappa (q^2,\chi)}
\;.
\label{Gprop2}
\end{equation}
In the limit of vanishing coupling $g\rightarrow 0$, 
we have $a_\kappa \rightarrow q^2+R_\kappa$ and $b_\kappa \rightarrow q^2+R_\kappa$, so that the  {\it bare} propagator $\Delta_{\kappa}^{(0)}$, respectively its 
inverse $\Pi_\kappa^{(0)}$ are, 
\begin{eqnarray}
\Delta_{\kappa,\;\mu\nu}^{(0)}(q) 
\;&=&\;
\frac{S_{\mu\nu}\,+\,\chi\,T_{\mu\nu}}{q^2 +R_\kappa(q^2)}
\nonumber \\
\;&=&\;
\frac{1}{q^2 +R_\kappa(q^2)}\; 
\;
\left[
\;g_{\mu\nu} \,-\,\frac{n_\mu q_\nu + q_\mu n_\nu}{n\cdot q} 
\,+\,(n^2+\xi q^2)\,\frac{q_\mu q_\nu}{(n\cdot q)^2}
\right]
\label{proj3a}
\\
\nonumber \\
\Pi_{\kappa,\;\mu\nu}^{(0)}(q,-q) 
\;&=&\;
\left(q^2 +R_\kappa(q^2)\right)\;\,\left( \,P_{\mu\nu}\,+\, Q_{\mu\nu}\,\right)
\nonumber \\
\;&=&\;
\left(q^2 +R_\kappa(q^2)\right)\; 
\;
\left[
\;g_{\mu\nu} \,-\,\frac{q_\mu q_\nu}{q^2} 
\,+\,\left( n^2 q^2+ \frac{1}{\xi}\right)\frac{n_\mu n_\nu}{(n\cdot q)^2}
\right]
\label{proj3b}
\;,
\end{eqnarray}
and, since $\Delta_{\kappa}^{(0)} = \left(\Pi_{\kappa}^{(0)}\right)^{-1}$,  
the following inversion property holds: 
$
\Delta_{\kappa,\;\mu\lambda} 
\;(\, \Pi^{(0)}_{\kappa} +\widehat{\Pi}_\kappa\,)^{\lambda}_{\;\;\;\nu} 
= g_{\mu\nu}
$.
\bigskip
\bigskip

\subsection{The case  $\chi \;\rightarrow \;0$}
\smallskip

The system of evolution equations (\ref{b1}) and (\ref{a1}) 
for the functions $a_\kappa$, $b_\kappa$, and hence for $A_\kappa$,
$B_\kappa$, is still immensely difficult to solve, 
because, as is evident from (\ref{Pi1}),
the self-energy tensor $\widehat{\Pi}_\kappa$ contains products
of exact propagators $\Delta_\kappa$ (the solution of which
we do not know yet)  with  the exact 3-gluon and 4-gluon
vertex functions $\widehat{V}$ and $\widehat{W}$
(which are themselves unknown combinations of propagators).
However, we can make substantial progress, if
we can eliminate the {\it explicit} $\chi$-dependence, 
by considering  $\chi = n^2 q^2/(n\cdot q)^2 = 0$:
There are two possibilities to achieve this condition:
({\it i}) choosing $n^2 \rightarrow 0$, or ({\it ii}) 
considering $q^2/(n\cdot q)^2 \rightarrow 0$.
The first possibility corresponds to choosing, among all
the axial gauges with arbitrary $n^2$, the light-cone gauge with $n^2=0$.
The second possibility,
$q^2/(n\cdot q)^2 \rightarrow 0$, holds for any $n^2$, and 
corresponds to the {\it quasi-real} limit, by which we mean
the kinematic regime in which
the gluon energy $q_0$ is large as
compared to the virtual mass $\sqrt{q^2}$ so that the 
gluons are practically on-shell. Specifically, we require
for the gluon four-momentum $q_\mu=\left(q_0,{\bf q}_\perp, q_z\right)$
that
\begin{equation}
q_0^2 \;\;\approx \;\;q_z^2 \;\;\gg\;\; q^2_\perp \;\;\gg\; q^2
\;.
\label{mom1}
\end{equation}
This situation is typical for high-energy particle collisions
with (gluon) jet production, for example, hadronic collisions
with center-of-mass energy 
$E_{\rm cm} \,\lower3pt\hbox{$\buildrel > \over\sim$}\,100$ GeV,
where the gluon (and quark) fluctuations in the colliding hadrons
have highly boosted longitudinal momentum along the beam axis,
and comparably very small transverse momentum.
Bearing this physics picture in mind, it is then suggestive to
choose the  vector $n_\mu$ along the preferred longitudinal
$z$-direction that is dictated by the collision geometry, i.e.,
to choose $n_\mu$ in the $t$-$z$ plane, parametrized as
\begin{equation}
n_\mu\;=\;\left(u+v,{\bf 0}_\perp, u-v\right)
\;\;\;\;\;\;\;\;\;\;\;
n^2\;=\; 4\,u \,v
\;.
\label{mom2}
\end{equation}
The two assertions (\ref{mom1}) and (\ref{mom2}) imply 
\begin{equation}
n\cdot q \;\approx \; 2 \,v \, q_0 
\;\;\;\;\;\;\;\; \mbox{and} \;\;\;\;\;\;\;\;\;
\frac{q^2}{(n\cdot q)^2} \;\simeq \; 0 
\label{mom3}
\;.
\end{equation}
Consequently,  from (\ref{chi}) and (\ref{Gprop2}), we have 
for $q^2/(n\cdot q)^2 \rightarrow 0$ or $n^2 \rightarrow 0$
(assuming the functions $a_\kappa$ and $b_\kappa$ are finite for all $q$),
\begin{equation}
\chi \;\;\longrightarrow \;\;0
\;,\;\;\;\;\;\;\;\;\;\;\;\;
B_\kappa = \frac{\chi}{b_\kappa} \;\;\longrightarrow \;\;0
\label{limit1}
\;,
\end{equation}
so that we are left with only one unknown function
$A_\kappa = 1/a_\kappa$. We find that in this limit (\ref{b1}) and
(\ref{a1}) coincide, since
\begin{equation}
g_{\mu\nu}\widehat{\Pi}^{\mu\nu} \;\;=\;\; \widehat{\Pi}_\mu^{\;\,\mu} 
\;\;=\;\;
-\;3\;\frac{\chi}{1-\chi}\;\frac{n_\mu n_\nu}{n^2}\;\widehat{\Pi}^{\mu\nu}
\;,
\label{Pimumu}
\end{equation}
and therefore only the tensor structure $n_\mu n_\nu\widehat{\Pi}^{\mu\nu}$ appears
in both equations.
Using (\ref{Pimumu}) together with the definition of $\chi$, eq. (\ref{chi}),
and the expression (\ref{Pi1}) for $\widehat{\Pi}$,
we obtain the {\it master equation} for $a_\kappa$ in the limit (\ref{limit1}),
\begin{eqnarray}
\frac{\partial}{\partial t}\, b_\kappa(q^2,\chi)  
\;&=&\;
-\,\frac{\chi}{1-\chi} \;
\,\frac{n_\mu n_\nu}{n^2}\;\left( \frac{\partial}{\partial t}\; \widehat{\Pi}_\kappa^{\mu\nu}(q,-q)\right) 
\nonumber \\
\;.
\;&=&\;
-\,\left(\frac{q^2}{(n\cdot q)^2 - n^2 q^2}\right)
\;\;\left\{\frac{}{}
\;
\frac{1}{2} g^2 \;\int\frac{d^4k}{(2\pi)^4}
\; n^2 \; \left(\frac{\partial}{\partial t}
\Delta_\kappa^{\lambda\lambda}(k) \right)
\right.
\label{b2} \\
\;& &\;
\;\;\;\;\;\;\;\;\;\;\left. \frac{}{}
+\;\frac{i}{2} g^2 \;\int\frac{d^4k}{(2\pi)^4}
\; n\cdot (k - k')\; n^\nu\;
\left(\frac{}{}
\frac{\partial}{\partial t}
\left[
\Delta_\kappa^{\lambda\sigma}(k)\; \Delta_{\kappa ,\,\lambda}^{\;\;\;\;\,\sigma'}(k')\;
{\cal V}_{\sigma \sigma' \nu}(k',k,-q) 
\right]
\right)\,
\right\}
\nonumber
\;,
\end{eqnarray}
where $k' = q-k$, and we have utilized the form of the bare 3-gluon vertex
$V^{(0)}$, as given by (\ref{3vertex}).
Notice that (\ref{b2}) contains only the tadpole contribution
and the 3-gluon vertex contribution, as diagramatically
represented in Fig. 4: all the other 4-gluon terms
that are present in $\widehat{\Pi}_{\mu\nu}$ of (\ref{Pi1})
vanish identically upon contraction with $n_\mu$ and $n_\nu$,
because $\Delta_\kappa$ is orthogonal to $n$, which is a direct
consequence of the orthogonality of $\widehat{\Pi}$ with respect to
$q$ due to current conservation 
(both properties hold, of course, also for the bare functions
$\Delta_\kappa^{(0)}$ and $\Pi_\kappa^{(0)}$),
\begin{equation}
\label{inicond0}
n^\mu\,\Delta_{\kappa ,\;\mu\nu} \;\;=\;\;0\;\;= 
\;\;\Delta_{\kappa ,\;\mu\nu}\;n^\nu 
\;\;\;\;\;\;\;\;\;\;\;\;\;\;\;\;\;\;\;\;\;
q^\mu\,\widehat{\Pi}_{\kappa ,\;\mu\nu}\;=\;0\;=
\;\widehat{\Pi}_{\kappa ,\;\mu\nu}\;q^\nu
\;.
\end{equation}

\begin{figure}[htb]
\epsfxsize=450pt
\centerline{ \epsfbox{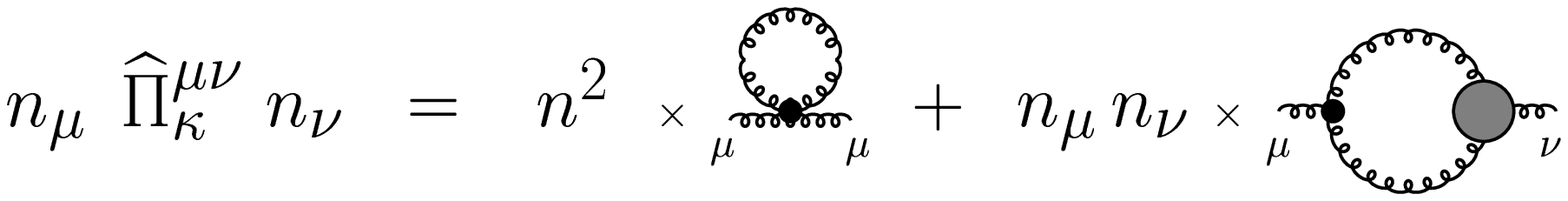} }
\vspace{-17.0cm}
\caption{
Diagrammatics of the contraction $n_\mu\widehat{\Pi}_{\kappa}^{\mu\nu} n_\nu$:
Only  the tadpole contribution (proportional to $n^2$)
and the 3-gluon vertex contribution (proportional to $n_\mu n_\nu$) survive.
All the  terms drop out upon contraction with $n_\mu$ and $n_\nu$,
\label{figure4}
}
\end{figure}
\bigskip
\bigskip

The initial conditions for the evolution equation (\ref{b2})
are dictated by asymptotic freedom in the ultraviolet limit
as $q\rightarrow \infty$, or more precisely, $q^2\rightarrow \Lambda^2$
with the normalization scale $\Lambda^2/\kappa^2 \rightarrow \infty$ 
[cf. (\ref{tdef})]:
\begin{equation}
a_\kappa(q^2,\chi) \;\;\stackrel{q^2\rightarrow\Lambda^2}{\longrightarrow}
\;\;q^2
\label{inicond1}
\;,
\end{equation}
which implies that the gluon propagator becomes the bare propagator
at the renormalization point,
\begin{equation}
\Delta_\kappa(q) \;\stackrel{q^2\rightarrow\Lambda^2}{\longrightarrow}
\;\;\Delta^{(0)}_{\kappa}(q)
\label{inicond2}
\;.
\end{equation}
As we move away from the asymptotic normalization scale $\Lambda$,
the  full gluon propagator propagator (\ref{Gprop1}) remains proportional
to its bare counterpart, modulo the function $A_\kappa = 1/a_\kappa$,
which encodes all effects of including softer and softer
gluon fluctuations in the evolution equation (\ref{b2}),
\begin{equation}
\Delta_{\kappa,\;\mu\nu}(q) 
\;\;=\;\;
 \frac{1}{a_\kappa(q^2,\chi)} \;\;S'_{\mu\nu}(q)
\;\;=\;\;
\left( \frac{q^2+R_\kappa(q^2)}{b_\kappa(q^2,\chi)}\right)
\;\; \Delta^{(0)}_{\kappa ,\;\mu\nu}(q) 
\label{proj3}
\;,
\end{equation}
with $S'_{\mu\nu}$ given by (\ref{proj2b}) at the value $\chi=0$, i.e.,
\begin{equation}
S'_{\mu\nu}(q) \;\;=\;\;
g_{\mu\nu} \,-\,\frac{n_\mu q_\nu + q_\mu n_\nu}{n\cdot q} 
\label{Tprime}
\;.
\end{equation}
Hence, the bare propagator and its inverse (taking now and henceforth 
$\xi \rightarrow  0$) reads
\begin{eqnarray}
\Delta_{\kappa,\;\mu\nu}^{(0)}(q) 
\;&=&\;
\frac{1}{q^2 +R_\kappa(q^2)}\; 
\left[
\;g_{\mu\nu} \,-\,\frac{n_\mu q_\nu + q_\mu n_\nu}{n\cdot q} 
\right]
\label{proj4a}
\\
\Pi_{\kappa,\;\mu\nu}^{(0)}(q,-q) 
\;&=&\;
\left(q^2 +R_\kappa(q^2)\right)\; 
\left[
\;g_{\mu\nu} \,-\,\frac{q_\mu q_\nu}{q^2} \right]
\label{proj4b}
\;,
\end{eqnarray}
and the  inversion property, noted after eq. (\ref{proj3b}), is modified for
$\xi \rightarrow 0$: $\;$
$
\Delta_{\kappa,\;\mu\lambda} 
\;(\, \Pi^{(0)}_{\kappa} +\widehat{\Pi}_\kappa\,)^{\lambda}_{\;\;\;\nu} 
= g_{\mu\nu}\,-\,n_\mu q_\nu/(n\cdot q)
$.
\bigskip

\subsection{Remarks}
\smallskip

Let us summarize the conceptual steps of the preceding subsections.
From the general form of the evolution equation (\ref{evoleq3}) for the 
quadratic (in the average gauge field) contributions of the 
average effective action,
we inferred a coupled set of equations (\ref{b1}) and (\ref{a1}) 
that determine the exact form of the gluon propagator via
(\ref{Gprop1}) and (\ref{Gprop2}) in terms
of the scalar functions $a_\kappa$ and $b_\kappa$.
In the case of $\chi = n^2 q^2/(n\cdot q)^2$, we could eliminate
the dependence on the function $b_\kappa$, and arrived at
the master equation (\ref{b2}) for  $a_\kappa$ alone, the solution of which
determines the full gluon propagator by simply
mutiplying the bare propagator with the single function
$a_\kappa$.
The presumption $\chi\rightarrow 0$ can be achieved
either by letting $n^2 \rightarrow 0$, or by
considering $q^2/(n\cdot q)^2 \rightarrow 0$.
The former possibility corresponds to going over to
the light-cone gauge, while the latter possibility is fulfilled 
in the kinematic regime (\ref{mom1}) of ``quasireal'' gluons.
In either case, we have the condition  (\ref{limit1}),
under which the master equation (\ref{b2}) is an exact
equation in the sense that it contains the full non-perturbative
evolution associated with the function $a_\kappa$ in general axial gauges
specified by the vector $n_\mu$ and the gauge parameter $\xi$.
\bigskip

\section{Solution  for the gluon propagator in the light-cone gauge }
\label{sec:section4}
\bigskip

Recall that (\ref{b2}) holds for the class of axial gauges
(\ref{gauge1}) in general, {\it viz.} for any choice of $n_\mu$
with finite $n^2$ and arbitrary gauge parameter $\xi$. 
For $n^2 \ne 0$,
the expression on the right-hand side of this equation
is then still very difficult to integrate, as has been discussed, 
e.g., in Ref. \cite{BBZ} for the case \footnote{This 
case would correspond to choosing $u = v = 1/2$ in eq. (\ref{mom2})}, 
$n_\mu = (1,0,0,0)$ $n^2=1$, and $\xi = 0$.
On the other hand, for $n^2 = 0$, which we will consider in the following, 
the right-hand side of (\ref{b2}) simplifies considerably, so that
an exact (numerical) integration is straightforward. Moreover,
we will show that it is even possible
to integrate  eq. (\ref{b2}) in closed form
by utilizing the methods of Ref.   \cite{alekseev1},
with the result being expressable in terms of elementary functions.
\bigskip

\subsection{Evolution of the renormalization function ${\cal Z}_\kappa$ for $n^2=0$}
\smallskip

The light-cone gauge can be specified by choosing, in the parametrization (\ref{mom2}),
the constant vector $n_\mu$, such that it is directed along the forward light-cone 
in the $t-z$ plane. Setting in (\ref{mom2}) $u=0$ and $v=1$, we have 
\begin{equation}
n_\mu\;=\;\left(1,{\bf 0}_\perp, -1\right)
\;\;\;\;\;\;\;\;\;\;\;
n^2\;=\; 0
\;.
\label{mom4}
\end{equation}
It follows then that $\chi = 0$, and if we introduce instead of
$a_\kappa$ the dimensionless renormalization function
\begin{equation}
{\cal Z}_\kappa(q^2) \;\equiv\;\frac{q^2}{a_\kappa(q^2,0)}
\;,
\label{zdef}
\end{equation}
with initial condition (\ref{inicond1}) at the normalization scale
$\Lambda^2 \gg \kappa^2$ in the ultraviolet:
\begin{equation}
{\cal Z}_\kappa(q^2) \;\;\stackrel{q^2\rightarrow \Lambda^2}{\longrightarrow}
\;\;\;1
\;,
\label{zinicond}
\end{equation}
then we may rewrite the evolution equation (\ref{b2}) as
\begin{eqnarray}
\frac{\partial}{\partial t}\, \frac{1}{{\cal Z}_\kappa(q^2)}  
\;&=&\;
\frac{\partial}{\partial t}\, \frac{a_\kappa(q^2,0)}{q^2}  
\nonumber \\
\;&=&\;
-\,\frac{n_\mu n_\nu}{(n\cdot q)^2} \;
\;\left( \frac{\partial}{\partial t}\; \widehat{\Pi}_\kappa^{\mu\nu}(q,-q)\right) 
\nonumber\\
\;&=&\;
-\,\frac{i}{2} \; g^2 \;\int\frac{d^4k}{(2\pi)^4}
\; \frac{n\cdot (k - k')}{(n\cdot q)^2}\; n^\nu\;
\left(\frac{}{}
\frac{\partial}{\partial t}
\left[
\Delta_\kappa^{\lambda\sigma}(k)\; \Delta_{\kappa ,\,\lambda}^{\;\;\;\;\,\sigma'}(k')\;
{\cal V}_{\sigma \sigma' \nu}(k',k,-q) 
\right]
\right)
\label{b3}
\;,
\end{eqnarray}
in which now only the 3-gluon contribution with the exact vertex function 
${\cal V}$ and exact propagators $\Delta_\kappa$ is present,
while the tadpole contribution, i.e., the first term on the right-hand side of
(\ref{b2}), vanishes since it is proportional to $n^2$.
The solution of (\ref{b3}) then determines the
full gluon propagator in terms of ${\cal Z}_\kappa$,
so that we have instead of (\ref{proj3}),
\begin{equation}
\Delta_{\kappa,\;\mu\nu}(q) 
\;\;=\;\;
{\cal Z}_\kappa(q^2)\;\; \Delta^{(0)}_{\kappa ,\;\mu\nu}(q) 
\label{proj4}
\;,
\end{equation}
with the  bare propagator $\Delta_\kappa^{(0)}$
given by (\ref{proj4a}).
\bigskip

\subsection{The spectral representation of propagator and vertex function}
\smallskip

The evolution equation  (\ref{b3}) still contains the unknown exact
3-gluon vertex function ${\cal V}$, which, as one would expect,
would have to be determined first, by solving a corresponding
evolution equation for ${\cal V}$, itself involving higher-order
vertex functions. Luckily, the gauge symmetry properties of QCD
imply the Slavnov-Taylor identities, which are the Ward identities
of QCD relating the vertex functions to the propagator. In general these
relations are non-trivial, however, in the class of axial gauges, 
the Slavnov-Taylor identities have a simple form. For example,
the 3-gluon vertex function ${\cal V}$ can be expressed in
terms of the propagator $\Delta_\kappa$ as 
\begin{equation}
q_\lambda \,\;{\cal V}_{\lambda\sigma\tau}(q,k,k') 
\;\;\Delta_\kappa^{\sigma\mu}(k)
\;\; \Delta_\kappa^{\tau\nu}(k')
\;\;=\;\;
 \Delta_\kappa^{\mu\nu}(k')\;\;-\; \;\Delta_\kappa^{\mu\nu}(k)
\;.
\label{ST1}
\end{equation}
where $(k' = q-k)$.
This Slavnov-Taylor identity suggests the following strategy:
({\it i})
construct an ansatz for ${\cal V}$,
in terms of $\Delta_\kappa$, such that (\ref{ST1}) is identically
satisfied, and, 
({\it ii}) insert this ansatz into the evolution equation
(\ref{b3}) for ${\cal Z}_\kappa$, upon which one obtains
a closed equation for the propagator $\Delta_\kappa$,
because of (\ref{proj4}) and (\ref{proj4a}). 
To do so, we adopt the elegant method of Delbourgo \cite{delbourgo2}
and represent the exact propagator in terms of its
spectral representation
\begin{equation}
\Delta_{\kappa ,\,\mu\nu}(q)\;\;=\;\;
S'_{\mu\nu}(q)\;\int d W^2 \;
\frac{\rho_\kappa(W^2)}{\left(q^2+R_\kappa(q^2)\right)-W^2}
\label{spectral}
\;,
\end{equation}
where $S'_{\mu\nu}(q)$ is defined by (\ref{Tprime}),
and the singularity at $W^2=q^2+R_\kappa$ in the denominator is 
to be evaluated with the usual $i\epsilon$ prescription.
The form (\ref{spectral}) includes the bare propagator (\ref{proj4a}), 
$\Delta_{\kappa ,\,\mu\nu}^{(0)} = S'_{\mu\nu}/(q^2+R_\kappa(q^2))$, 
upon setting $\rho_\kappa(W^2)= \delta(W^2)$. 
The physical interpretion of (\ref{spectral}) is very intuitive: It
expresses the  propagator for a gluon  with momentum $q$ and
subject to the infrared cut-off scale $\kappa$,
through the weighted {\it spectral density} $\rho_\kappa(W^2)$
which corresponds to the number density of virtual gluon
fluctuations with an effective mass $W$.
The case $\rho_\kappa(W^2) = \delta(W^2)$ corresponds then to a massless,
non-interacting on-shell gluon ($W=0$). 
This notion of the spectral density $\rho_\kappa$ is very reminescent of 
the gluon distribution function which is measured
in lepton-hadron or hadron-hadron collisions, and which
describes the substructure of a gluon in terms of virtual
fluctuations. We will return to this issue in the next Section.
\medskip

Inserting the spectral representation (\ref{spectral}) for $\Delta_\kappa$
into the Slavnov-Taylor identity (\ref{ST1}), one obtains an 
implicit equation for the 3-gluon vertex function ${\cal V}$ 
in terms of the spectral density $\rho_\kappa$.
Since $\Delta_\kappa$ and $\rho_\kappa$ are not known at this point,
we must make an ansatz for ${\cal V}$ that is compatible with the
Slavnov-Taylor identity. 
A possible form \cite{delbourgo2} that satisfies the identity (\ref{ST1}),
is the following spectral ansatz:
\begin{eqnarray}
& &
\;\Delta_\kappa^{\mu\lambda}(q)\;\Delta_\kappa^{\nu\sigma}(k)\;
 \Delta_\kappa^{\rho\tau}(k')
\;\; {\cal V}_{\lambda\sigma\tau}(q,k,k')
 \;\;=\;\;
\nonumber
\\
& & 
\;\;\;\;\;\;\;\;\;\;\;\;\;\;\;\;\;
=\;\frac{1}{3}\;\,\int dW^2
\;\rho_\kappa(W^2)\;
\frac{ 
S^{\prime\,\mu\lambda}(q)\; \;S^{\prime\,\nu\sigma}(k)\; \;S^{\prime\,\rho\tau}(k')
\;\;V^{(0)}_{\lambda\sigma\tau}(q,k,k')
}{
\left[\left(q^2+R_\kappa(q^2)\right)-W^2\right]
\left[\left(k^2+R_\kappa(k^2)\right)-W^2\right]
\left[\left(k^{'\,2}+R_\kappa(k^{'\,2})\right)-W^2\right]
}
\label{Vansatz}
\;.
\end{eqnarray}
The integrand on the left-hand side is symmetrical product of 
three propagators $S'_{\mu\nu}/[(p^2+R_\kappa(p^2))-W^2]$
and the bare 3-gluon vertex $V^{(0)}$,
weighted by the spectral density $\rho_\kappa(W^2)$.  Notice
that the combination of propagators and vertex function is just
what is required to solve the identity (\ref{ST1}), and moreover,
it respects Bose symmetry, because all three legs are represented
symmetrically.
Also, the appearance of the bare vertex
on the right-hand side of (\ref{Vansatz}) does {\it not} imply that
we are limiting ourselves to lowest order perturbation theory:
on the contrary, the propagators $\Delta_\kappa$ attached
to $V^{(0)}$ are the full propagators that embody
the dynamics from the (perturbative) ultraviolet regime all the way into
the (non-perturbative) infrared domain.
Nevertheless, (\ref{Vansatz}) is just an ansatz, and hardly unique:
one may think of constructing a different form that is also
compatible with (\ref{ST1}) but has a richer structure
\footnote{
Aitkinson {\it et al.} \cite{atkinson2} have conjectured that the form
(\ref{Vansatz}) does not necessarily comply with the Slavnov-Taylor
identity, because the index $\lambda$ of ${\cal V}_{\lambda\sigma\tau}$
is contracted with the $q$-propagator, so that it is not possible
to isolate a contraction of the vertex function with $q_\lambda$.
Instead a more complex ansatz is proposed in \cite{atkinson2}
which avoids this asymmetry.
However, in the light-cone gauge $n^2 =0$, the ansatz of Atkinson {\it et al.}
coincides with (\ref{Vansatz}) for $n^2 = 0$, so that one may
conclude that in the light-cone gauge these subtle ambiguities
are absent.
}.
\bigskip

\subsection{Solution for the spectral density $\rho_\kappa$ and the renormalization
function  ${\cal Z}_\kappa$}
\smallskip

Putting the pieces together, we first multiply 
(\ref{b3}) by $-(n\cdot q)^2 = - (n\cdot q) \,q_\nu\,n^\nu$, 
so that both sides of the equation are proportional to $n^\nu$.
Next,  we multiply both sides by $\Delta_{\kappa ,\,\mu\nu}(q)$,
in order to bring the right-hand side to the form 
$\Delta_\kappa\Delta_\kappa\Delta_\kappa \, {\cal V}$, as
required by (\ref{Vansatz}).
Finally, we insert the spectral representation (\ref{spectral}) 
and (\ref{Vansatz}) for
the propagators $\Delta_\kappa$, respectively  for
$\Delta_\kappa\Delta_\kappa\Delta_\kappa \, {\cal V}$.
As the result of these manipulations, we obtain the following equation,
which corresponds to (\ref{b3}):
\begin{eqnarray}
\frac{\partial}{\partial t}\, \frac{1}{{\cal Z}_\kappa(q^2)}  
\;&=&\;
\frac{\partial}{\partial t}\,
\int d W^2 \;\frac{\rho_\kappa(W^2)}{\left(q^2+R_\kappa(q^2)\right)-W^2\,+\,i\epsilon}
\;\;\;{\cal P}_\kappa(q^2,W^2)
\label{b4}
\;,
\end{eqnarray}
where
\begin{equation}
{\cal P}_\kappa(q^2,W^2)
\;\;=\;\;
\;q_\lambda \,S^{\prime\;\lambda \mu}(q)\;\widehat{\Pi}^{\;\prime}_{\kappa ,\,\mu\nu}(q^2,W^2)\,\;n^\nu
\;\;=\;\;
- \frac{q^2}{n\cdot q}\;\;n^\mu\;
\widehat{\Pi}^{\;\prime}_{\kappa ,\,\mu\nu}(q^2,W^2)\; n^\nu
\label{P}
\;,
\end{equation}
and
\begin{equation}
\widehat{\Pi}^{\;\prime}_{\kappa,\,\mu\nu}(q^2,W^2)
\;\;=\;\;
-\,\frac{i\,g^2}{2}
\; \int \frac{d^4k}{(2\pi)^4}
\;\;V_{\;\;\mu\lambda\sigma}^{(0),acd}(q,-k,-k')\;\;
S^{\prime\;\lambda\lambda',\,cc'}(k)\;
S^{\prime\;\sigma\sigma',\,dd'}(k')\;
V_{\;\;\sigma'\lambda'\nu}^{(0),d'c'b}(k',k,-q)
\label{b4a}
\end{equation}
is the  self-energy function (to order $g^2$) of an intermediate
virtual gluon with mass $W$.
The remarkable feature of this equation is that it is now {\it linear} 
in the spectral density $\rho_\kappa$ of the propagator, in contrast to
the previous equation (\ref{b3}) which involved a product of propagators.
After integration of (\ref{b4}) over $dt = d\kappa^2/(2\kappa^2)$
as defined by (\ref{tdef}), the formal solution for ${\cal Z}_\kappa^{-1}$ is:
\begin{equation}
\frac{1}{{\cal Z}_\kappa(q^2)}  
\;\;=\; \;
\frac{1}{{\cal Z}^{(0)}_\kappa(q^2)}\;\;+\;\;  
\int d W^2 \;\frac{\rho_\kappa(W^2)}{\left(q^2+R_\kappa(q^2)\right)-W^2\,+\,i\epsilon}
\;\;\;{\cal P}_\kappa(q^2,W^2)
\label{b5}
\;.
\end{equation}
Here the first term is determined by the initial condition (\ref{zinicond})
that ${\cal Z}_\kappa(q^2) = 1$ at the normalization point $\Lambda$.
As $q^2\rightarrow \Lambda^2$, 
the contribution ${\cal Z}_\kappa^{(0)\;-1}$ must reproduce the bare 
propagator with spectral density $\rho_\kappa(W^2)\rightarrow \delta(W^2)$
in the limit $g\rightarrow 0$ due to asymptotic freedom, i.e.,
\begin{equation}
\frac{1}{{\cal Z}^{(0)}_\kappa(q^2)}\;\;=\;\;  
q^2\;\int d W^2 \;\frac{\rho_\kappa(W^2)}{\left(q^2+R_\kappa(q^2)\right)-W^2\,+\,i\epsilon}
\label{b5a}
\;.
\end{equation}
What remains to be done is to compute the second term in (\ref{b5}). Thus,
we insert the explicit expressions for $S_{\mu\nu}^{\prime}$ of (\ref{b3}),
and $V^{(0)\;abc}_{\mu\nu\lambda}$ of (\ref{3vertex}) of Appendix B, into 
(\ref{b4a}) for $\Pi^{\;\prime}_\kappa(W^2,q)$, and after some algebra, we arrive at the 
following expression for ${\cal P}_\kappa$:
\begin{eqnarray}
{\cal P}_\kappa(q^2,W^2) \;&=&\;
- q^2\;
\;\frac{2i\,g^2\,C_G}{(2\pi)^4}
\int d^4k \;
\frac{n\cdot (k-k')\;\;\; n\cdot k'}
{\left[\left(k^2+R_\kappa(k^2)\right)-W^2\right] 
\left[\left(k^{'\,2}+R_\kappa(k^{'\,2})\right)-W^2\right]}
\;\equiv\;
- q^2\;
\;\frac{g^2\,C_G}{8\pi^4}
\;\; I_\kappa(q^2,W^2)
\label{b4b}
\;,
\end{eqnarray}
where $k' = q-k$ and the factor $C_G = N_c = 3$ results from the color trace 
$f^{acd}f^b_{\;\,cd} = \delta^{ab} C_G$.
We have abbreviated the integral (including a factor $1/i$)
as $I_\kappa(q^2,W^2)$ for later convenience.
Hence, (\ref{b5}) becomes:
\begin{equation}
\frac{1}{{\cal Z}_\kappa(q^2)}  
\;\;=\; \;
q^2 \;\int d W^2 \;\frac{\rho_\kappa(W^2)}{\left(q^2+R_\kappa(q^2)\right)-W^2\,+\,i\epsilon}
\;
\left[\frac{}{}
\;1\;\;-\;\; \frac{g^2\, C_G}{8\pi^4}\;I_\kappa(q^2,W^2)\;\right]
\;.
\label{b6}
\end{equation}
In order to evaluate  $I_\kappa(q^2,W^2)$, we must
now finally commit ourselves to a specific form of the infrared
regulator $R_\kappa(p^2)$.
In general, a closed analytic solution is not possible as long as $R_\kappa$
varies strongly with $p^2/\kappa^2$, so that a numerical
solution must be found on a computer. 
Specifically, we would like to use a slight generalization
of the form (\ref{Rk}) suggested in Sec. II,
\begin{equation}
R_\kappa(p^2)=
 p^2 \,\, \frac{\exp\left(-p^2/\kappa^2\right)}
{\exp\left(-p^2/\Lambda^2\right)  - \exp\left(-p^2/\kappa^2\right)}\;\; ,
\label{RkL}
\end{equation}
which includes an additional ultraviolet cut-off 
$\Lambda \gg\kappa$ and which 
contains (\ref{Rk}) for $\Lambda \rightarrow \infty$.
Such a form introduces a non-linear $p^2$-dependence in the denominators
$\left[\left(p^2+R_\kappa(p^2)\right)-W^2\right]^{-1}$ 
that appear in (\ref{b6}) and (\ref{b4b}), which discourage  an 
analytical evaluation.  We intend to
investigate solutions to (\ref{b6}) in the near future by integrating 
(\ref{b4b}) numerically, using the infrared regulator (\ref{RkL}).
\bigskip

\subsection{Asymptotic behavior of the gluon propagator}
\smallskip

Notwithstanding an exact numerical study of (\ref{b6}),
it is desirable to obtain  at least an approximate
analytical solution in the ultraviolet and the infrared limits.
This may elucidate 
the  behavior in these two extreme limits of  the 
gluon propagator $\Delta_\kappa = {\cal Z}_\kappa \Delta^{(0)}_\kappa$
within our specific approximate approach.
Furthermore, it may serve as a check for an exact numerical treatment.
In order to extract the behavior of ${\cal Z}_\kappa(q^2)$ for 
$q^2\rightarrow 0$ and $q^2\rightarrow \infty$, 
we note that the dominant contribution to 
$I_\kappa(q^2,W^2)$ of (\ref{b4b}) arises from fluctuations 
at small $k$ or  $k^{\prime}=q-k$; only the presence of the infrared 
regulator $R_\kappa$ prevents a divergence.
Hence, the integrand in (\ref{b4b}) is
substantially enhanced  in the infrared region, where
$k^2,k^{\prime\,2} \leq \kappa^2$, and where 
from (\ref{Rklim}), $R_\kappa \rightarrow \kappa^2$.
When on the other hand  $k^2,k^{\prime\,2} \gg \kappa^2$, the effect of the 
infrared regulator vanishes according to (\ref{Rklim}): $R_\kappa \rightarrow 0$.
Thus, we may replace $R_\kappa$ by
\begin{equation}
\;\;\;\;\;\;\;\;\;\;\;
R_\kappa(p^2) \;\;\longrightarrow\;\;\; \kappa^2
\;\;\;\;\;\;\;\;\;\;\;
(\;p \equiv q, k, k'\;)
\;,
\label{Rklim2}
\end{equation}
which  is independent of $p^2$, as desired, but
which has qualitatively the same effect as $R_\kappa(p^2)$ on the propagator, 
in both the infrared and the ultraviolet,
\begin{equation}
\frac{1}{p^2+R_\kappa(p^2) }\;\;\simeq\;\;
\frac{1}{p^2+\kappa^2 }\;\;\longrightarrow\;\;
\left\{
\begin{array}{l}
1/p^2 \;\;\;\;\;\;\mbox{for}\;\; p^2\rightarrow \infty
\\
1/\kappa^2 \;\;\;\;\;\;\mbox{for}\;\; p^2\rightarrow 0
\end{array}
\right.
\;.
\label{Rklim2a}
\end{equation}
Substituting (\ref{Rklim2}) in (\ref{b4b}), we obtain 
\begin{equation}
I_\kappa(q^2,W^2) \;\;\approx\;\;
\; \frac{1}{i}\int d^4k \;
\frac{n\cdot (k-k')\;\;\; n\cdot k'}
{\left[\left(k^2+\kappa^2\right)-W^2\right] 
\left[\left(k^{'\,2}+\kappa^2)\right)-W^2\right]}
\label{b6b}
\;,
\end{equation}
which can be evaluated exactly, 
by using the standard Feynman parametrization \cite{IZ},
and integrating over the momenta $k=q-k'$ in a space of $d = 2\omega$
dimensions,
\begin{eqnarray}
I_\kappa^{(\omega)}(q^2,W^2)  & = &
\;\frac{1}{i}\;\int d^{2\omega}k \;
\frac{n\cdot (k-k')\;\;\; n\cdot k'}
{\left[\left(k^2+\kappa^2\right)-W^2\right] 
\left[\left(k^{'\,2}+\kappa^2)\right)-W^2\right]}
\nonumber
\\
& = &
\;\frac{1}{i}\;
\int_0^1  dx\;\int d^{2\omega}k \;
\frac{
3 (n\cdot k)(n\cdot q)  - 2 (n\cdot k)^2 - (n\cdot q)^2 
}{
\left[k^2 - 2x k\cdot q + x q^2 + \kappa^2 - W^2\right]^{\,2}
}
\nonumber
\\
& = & 
\pi^\omega\;e^{-i\pi\omega} \;\Gamma(2-\omega)
\;\int_0^1 dx\; (1-x) (2x-1) 
\left(\frac{}{} \,x(1-x) q^2 + \kappa^2 -W^2 \,\right)^{\omega-2}
\;.
\end{eqnarray}
The remaining integral can be reduced to integrals of the type
$
\int_0^1 dx x^{u-1}(1-x)^{v-1} (x-y)^{-w}
$
which are integral representations of the hypergeometric
function $F(w, u; u+v; 1/y)
$,
so that the result for $I_\kappa$ can be cast in the following 
form \cite{alekseev1},
\begin{eqnarray}
I_\kappa^{(\omega)}(q^2,W^2)  
&=&
\pi^\omega\,e^{-i\pi\omega} \; (\kappa^2-W^2)^{\omega-2} \;\Gamma(2-\omega)\;
\;\;\times
\nonumber \\
& &
\;\;\;\;\;\;\;\;\;\;\;\;\;\;\;\;\;\;\;\;\;\;\;\;\;
\times\;\;
\left[\,\frac{1}{3} \, F\left(2-\omega ,\; 2 ; \;\frac{5}{2}; \;\frac{q^2}{4 (W^2-\kappa^2)} \right)
\;\;-\;\;
\frac{1}{2}\,
F\left(2-\omega ,\; 1 ; \;\frac{3}{2}; \;\frac{q^2}{4 (W^2-\kappa^2)} \right)
\right]\; .
\label{b6c}
\end{eqnarray}
The expression (\ref{b6b}) is singular in $d=4$ dimensions 
due to the pole of the Gamma function $\Gamma(2-\omega)$
which arises from the usual ultraviolet divergence of Feynman integrals of the 
type (\ref{b6b}).
If we were able to analytically compute of the original integral (\ref{b4b}) with
$R_\kappa$ given by (\ref{RkL}), instead of the approximate form (\ref{b6b}) with
$R_\kappa$ replaced by $\kappa^2$, this
divergence would be avoided due to the exponential suppression of momenta
$q > \Lambda$ in (\ref{RkL}).
The result (\ref{b6c}) of the approximate integral (\ref{b6b}) therefore has to be
regularized by hand, which we achieve by making a subtraction at 
some mass scale $\mu^2\ll \Lambda^2$, which we choose as $\mu^2 = \kappa^2$,
\begin{equation}
I_\kappa^{\rm (reg)}(q^2,W^2) \;=\;
\lim_{\omega\rightarrow 2} \;\left[\frac{}{} 
I_\kappa^{\rm (reg)}(q^2,W^2) \;-\; I_\kappa^{\rm (reg)}(\kappa^2,\kappa^2) \;\right]
\label{b6d}
\;.
\end{equation}
This regularized form $I^{\rm (reg)}(q^2,W^2)$ is then finite,
because, from the following property of the imaginary part of the
hypergeometric function $F(\alpha ,\beta ;\gamma ; x)$,
\begin{eqnarray}
F(\alpha ,\beta ;\gamma ; x+ i\epsilon)
&-&
F(\alpha ,\beta ;\gamma ; x- i\epsilon)
\;\; =  \;\;
\nonumber \\
& & 
\nonumber \\
&=& 
\frac{2\pi\,i\; \Gamma(\gamma)\;\theta(x-1)}{\Gamma(\alpha)\Gamma(\beta)
\Gamma(1+\gamma-\alpha-\beta)}
\;\,(x-1)^{\gamma-\alpha-\beta}\;
\; F(\gamma-\alpha,\gamma-\beta ;\,1+\gamma -\alpha-\beta ;\,1-x)
\;,
\end{eqnarray}
one readily infers that the factor $\Gamma(2-\omega)$
in (\ref{b6c}) cancels in the imaginary part of the regularized expression (\ref{b6d}),
while the real part is finite.
Hence, the limit $\omega \rightarrow 2$ is now well defined,
and (\ref{b6d}) can be evaluated  in terms of elementary functions,
by using some transformation properties \cite{lebedev} 
of the hypergeometric function. The result is:
\begin{equation}
I_\kappa^{\rm (reg)}(q^2,W^2) \;=\; \mbox{Re}\;I_\kappa^{\rm (reg)}(q^2,W^2) \;+\;
i\;\mbox{Im}\;I_\kappa^{\rm (reg)}(q^2,W^2) 
\label{Iresult}
\end{equation}
with the real part,
\begin{eqnarray}
\mbox{Re}\;I_\kappa^{\rm (reg)}(q^2,W^2) & =&
\,-\, \frac{1}{6} \left(1 - 4z \right)^{3/2}
\;\ln\left|\frac{1-\sqrt{1-  4z}}{1+\sqrt{1-4z}}\right| \;\theta(1-4z)
\nonumber \\
& &
\,-\, \frac{1}{3}\; \left(4z - 1\right)^{3/2}
\;\arctan\left(\frac{1}{\sqrt{4z-1}}\right) \;\theta(4z-1)
\nonumber \\
& &
\,+\, \frac{4}{3}\left(z-1\right)
\,+\; \frac{11}{6}\; \ln\left( \frac{z \,q^2}{\kappa^2} \right)
\,+\, \frac{\pi}{2\sqrt{3}}
\label{ReI}
\end{eqnarray}
the imaginary part,
\begin{equation}
\mbox{Im}\;I_\kappa^{\rm (reg)}(q^2,W^2) \;=\;
- \frac{\pi}{6}\;\left(1- 4z\right)^{3/2}\;\theta( 1- 4z)
\label{ImI}
\;,
\end{equation}
where
\begin{equation}
z \;\;=\; \frac{W^2-\kappa^2}{q^2}
\;\;\;\;\;\;\;\;\;\;\;\;\;\;\; (\;W^2 \ge \kappa^2\;)
\;.
\label{ca}
\end{equation}
Substituting (\ref{Iresult}) in  equation (\ref{b6}) for ${\cal Z}_\kappa^{-1}$,
we get
\begin{equation}
\frac{1}{{\cal Z}_\kappa(q^2)}  
\;\;=\; \;
q^2\; \int_0^\infty d W^2 \;
\frac{\rho_\kappa(W^2)}{\left(q^2+\kappa^2\right)-W^2\,+\,i\epsilon}
\;
\left\{\frac{}{}
\;1\;\;-\;\; \frac{g^2\, C_G}{8\pi^4}\;
I^{\rm (reg)}_\kappa(q^2,W^2)
\;\right\}
\;.
\label{b7}
\end{equation}
Upon taking the discontinuity at $q^2=W^2-\kappa^2$, using 
the principal-value prescription 
$(y\pm i \epsilon )^{-1} = P\,\left(\frac{1}{y}\right) \pm \pi i \delta(y)$, 
and calculating the imaginary part of eq. (\ref{b7}),
one arrives at the following integral equation for $\rho_\kappa$:
\begin{equation}
\rho_\kappa(q^2)\;\left[\;1+\frac{g^2\,C_G}{8\pi^2}\;
\mbox{Re} I^{\rm (reg)}_\kappa(q^2,q^2+\kappa^2)\;\right]
\;\;=\;\;
\frac{\delta(q^2)}{{\cal Z}_\kappa(q^2)}  
\;+\;
\frac{g^2\,C_G}{8\pi^3}\; \int_{\kappa^2}^{q^2/4+\kappa^2} d W^2 
\;\frac{\rho_\kappa(W^2)}{\left(q^2+\kappa^2\right)-W^2}
\;\,\mbox{Im} I^{\rm (reg)}_\kappa(q^2,W^2)
\label{b7a}
\;.
\end{equation}
For the case $g^2 = 0$, we recover, as anticipated,
the free solution for the spectral density, 
\begin{equation}
\rho_\kappa(q^2)\;\;\stackrel{g^2\rightarrow 0}{=}\;\delta(q^2)
\;\;\;\;\;\;\;\;\;\;\;\;\;\;
{\cal Z}_\kappa(q^2)\;\;\stackrel{g^2\rightarrow 0}{=}\;1
\label{b8a}
\;,
\end{equation}
which corresponds to a single bare on-shell gluon.
\smallskip

For the case $g^2 \ne 0$, 
we note that on the left-hand side of (\ref{b7a}),
$\mbox{Re} I_\kappa^{\rm (reg)}$
is to be evaluated from (\ref{ReI}) at $W^2=q^2+\kappa^2$,
i.e. $z = 1$,
while on the right-hand side of (\ref{b7a}) the $\theta$-function in
$\mbox{Im} I_\kappa^{\rm (reg)}$ from (\ref{ImI})
cuts off the upper integration limit at $z=1/4$, or, $W^2=q^2/4+\kappa^2$.
Furthermore, if we consider $q^2 \ge \kappa^2$
(keeping in mind to let $\kappa^2\rightarrow 0$ at the end),  
and subtract the `single-gluon' contribution (\ref{b8a}),
to define the `multi-gluon' contribution of virtual fluctuations,
\begin{equation}
\widehat{\rho}_\kappa(q^2)\;\;\equiv\;\;
{\rho}_\kappa(q^2)\;-\;\rho_\kappa^{(0)}(q^2)
\;,\;\;\;\;\;\;\;\;\;\;\;\;\;\;\;\;\;\;\;
{\rho}_\kappa^{(0)}(q^2)\;=\;\delta(q^2)
\label{rhotilde}
\;,
\end{equation}
we find after insertion of 
the expressions (\ref{ReI} -- \ref{ca}) into (\ref{b7a}),
\begin{equation}
\widehat{\rho}_\kappa(q^2)\;\;
\left[\, 1+\frac{11 g^2\,C_G}{48\pi^2}\;
\ln \left(\frac{q^2}{\kappa^2}\right)\,
\right]
\;\;=\;\;
-\,\frac{g^2\,C_G}{48\pi^2}
\;\;
\int_0^{q^2/4}\,dw^2
\frac{\left(1-\frac{4w^2}{q^2}\right)^{3/2}}{q^2-w^2\;\;\;}\;\;
\widehat{\rho}_\kappa(w^2)
\label{b8b}
\;,
\end{equation}
where we have shifted the variable of integration on the
right-hand side, $W^2 \rightarrow w^2 = W^2-\kappa^2$.
Notice the characteristic feature of the integral over $w^2$:
it  is dominated by the contributions from the region $w^2 \approx q^2$,
provided that $\rho_\kappa$ finite and well-behaved in that region.
From (\ref{b8b}), we now can extract the asymptotic behavior of $\rho_\kappa$
in the ultraviolet $q^2\rightarrow \infty$ and the infrared
$q^2\rightarrow 0$.
\begin{description}
\item[a)]
The ultraviolet limit $q^2\rightarrow \Lambda^2 $
($\Lambda^2\rightarrow \infty$):
In the large-$q^2$ limit, the logarithm in the brackets of the left-hand side
of (\ref{b8b}) dominates, so that approximately
\begin{equation}
q^2\;\widehat{\rho}_\kappa(q^2)\;\;
\left[\,\frac{11 g^2\,C_G}{48\pi^2}\;
\ln \left(\frac{q^2}{\kappa^2}\right)\,
\right]
\;\;\approx\;\;
-\,\frac{g^2\,C_G}{48\pi^2}
\;\;
\int_0^{q^2/4}\,d \ln w^2\;  
\;\;w^2\widehat{\rho}_\kappa(w^2)
\label{b9a}
\;.
\end{equation}
It is easy to see, that the form
\begin{equation}
\widehat{\rho}_\kappa(q^2) \;\;\approx \;\; \;  
\frac{1}{q^2}\;\;\;
\left(\,\frac{c_\infty}{\ln^2\left(q^2/\kappa^2\right)}\,\right)
\;, \;\;\;\;\;\;\;\;\;\;\;\;\;\;\;
c_\infty^{-1} \;=\; \,\frac{11 g^2 C_G}{48\pi^2}
\;,
\label{b9b}
\end{equation}
is a consistent  ultraviolet solution when substituted in (\ref{b9a}).
\item[b)]
The infrared limit $q^2\rightarrow 0$
($\kappa^2 \rightarrow 0$):
When $q^2\approx \kappa^2$ with $\kappa^2$ tending to $\kappa_{\rm PT}$, 
eq. (\ref{kappaPT}), we can drop the logarithm on the left-hand side
of (\ref{b8b}), so that,
\begin{equation}
q^2\;\widehat{\rho}_\kappa(q^2)\;\;
\;\;\approx\;\;
-\,\,\frac{g^2\,C_G}{48\pi^2}
\;\;
\int_0^{q^2/4}\,\frac{dw^2}{w^2}
\;\;\frac{\left(1-\frac{4w^2}{q^2}\right)}{1-\frac{w^2}{q^2}}^{3/2}
\;\;w^2 \widehat{\rho}_\kappa(w^2)
\label{b10a}
\;.
\end{equation}
An approximate solution  in this case is 
\begin{equation}
\widehat{\rho}_\kappa(q^2) \;\;\approx\;\;
\;  \frac{1}{q^2}\; \;\; \left(\, \frac{c_0\,\kappa^2}{q^2}\,\right)
\;, \;\;\;\;\;\;\;\;\;\;\;\;\;\;\;
c_0^{-1} \;=\; \frac{g^2 C_G}{48\pi^2}
\;,
\label{b10b}
\end{equation}
which is consistent with eq. (\ref{b9b}) in the infrared, when
$w^2\approx q^2 \rightarrow 0$.
\end{description}

\noindent
The actual gluon propagator $\Delta_{\kappa,\,\mu\nu}(q)$ is now
obtained by inserting the  spectral density (\ref{rhotilde})
into  the spectral representation (\ref{spectral}),
using the  expressions  for the ultraviolet limit and 
the infrared region,  (\ref{b9b}) and (\ref{b10b}), respectively:
\begin{eqnarray}
\Delta_{\kappa,\,\mu\nu}(q)  \;\; & \approx & \;\;
\frac{S^{\,\prime}_{\mu\nu}(q)}{q^2+\kappa^2}
\; \;
\left[ 1\;+\; \frac{11 g^2 C_G}{48\pi^2} \;\;\ln\left(\frac{q^2}{\kappa^2}\right)
\right]^{-1}
\;\;\;\;\;\;\;\;\;\;\;\;\;\mbox{for}\;\;q^2\gg \kappa^2 ,
\label{ZUsol} 
\\
\Delta_{\kappa,\,\mu\nu}(q)  \;\; & \approx & \;\;
\frac{S^{\,\prime}_{\mu\nu}(q)}{q^2+\kappa^2}
\; \;
\left[
1\;+\;\frac{g^2 C_G}{48\pi^2} \;\;\frac{q^2}{\kappa^2}
\right]^{-1}
\;\;\;\;\;\;\;\;\;\;\;\;\;\mbox{for}\;\;q^2\rightarrow 0 .
\label{ZIsol}
\end{eqnarray}
where $S^{\prime}_{\mu\nu}(q)$ is defined in (\ref{Tprime}).
In the ultraviolet limit $q^2\rightarrow\infty$, 
we recover the famous logarithmic dependence $\propto 1/q^2\ln(q^2)$,
while in the  infrared limit
$q^2\rightarrow 0$, the leading behavior is a power-law $\propto 1/q^4$.

The corresponding ultraviolet and infrared behavior 
of the renormalization function ${\cal Z}_\kappa(q^2)$ can be read off 
(\ref{ZUsol}) and (\ref{ZIsol}), 
by utilizing the relation between $\Delta_{\kappa,\,\mu\nu}$ and ${\cal Z}_\kappa$,
eqs. (\ref{proj4}) and (\ref{proj4a}).
These asymptotic results may be combined into a 
phenomenological, but hardly unique formula which interpolates smoothly
between the ultraviolet and the infrared limit:
\begin{equation}
\Delta_{\kappa,\;\mu\nu}(q) 
\;\;=\;\;
\frac{S^{\,\prime}_{\mu\nu}(q)}{q^2+\kappa^2}
\;\;\;
{\overline{\cal Z}}_\kappa(q^2)
\label{formula1}
\;,
\end{equation}
with
\begin{equation}
{\overline{\cal Z}}_\kappa(q^2)
\;\;=\;\;
(1-C(q^2))\;
\left[\,
1\;+\;\frac{g^2 C_G}{48\pi^2}\;\frac{q^2}{\kappa^2}
\right]^{-1}
\;\;+\;\;
C(q^2)\;
\left[\,
1\;+\; \frac{11g^2 C_G}{48\pi^2}\; \ln\left(\frac{q^2}{\kappa^2}\right)
\right]^{-1}
\label{formula2}
\;.
\end{equation}
Here the prefactors $1-C$ and $C$ interpolate between the
infrared and the ultraviolet limits, with $C$ tending to 1 as 
$q^2 \rightarrow \Lambda^2$ and approaching 0 as $q^2 \rightarrow 0$,
e.g. $C(q^2) = 2 [1-\Lambda^2/(q^2+\Lambda^2)]$.
\medskip

In Fig. 5a, we plot this form of  $\overline{{\cal Z}}_\kappa$
in comparison with the asymptotic results (\ref{b9b}) and (\ref{b10b}),
for different choices of $\kappa$. Fig. 5b shows the corresponding
gluon propagator $\Delta_\kappa$ in contrast to the free propagator
$\Delta_\kappa^{(0)}$.

\vspace{1.0cm}
\setcounter{figure}{4}
\begin{figure}[htb]
\begin{minipage}[t]{83mm}
\epsfxsize=150pt
\rightline{  \epsfbox{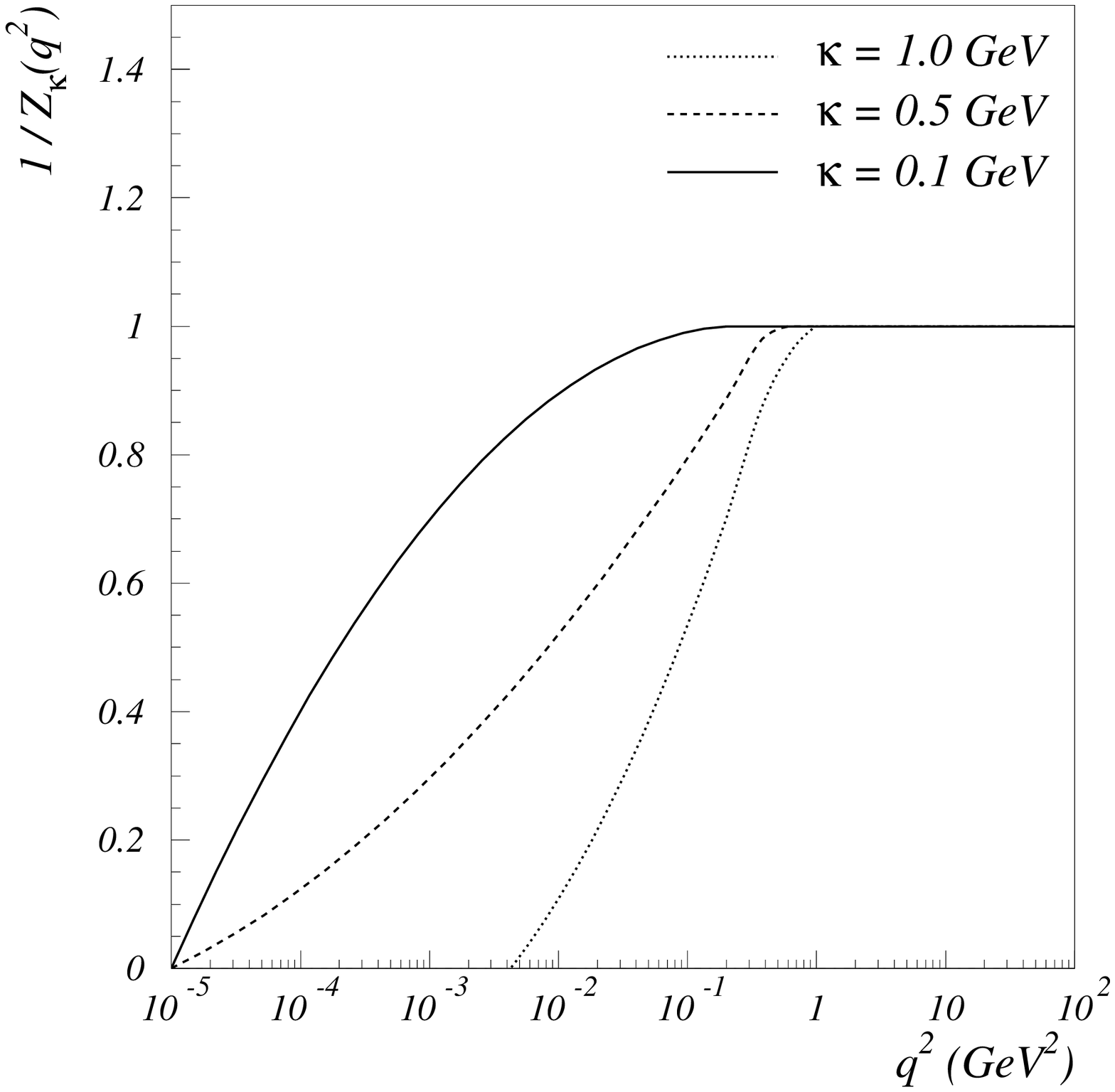} }
\vspace{-0.8cm}
\end{minipage}
\hspace{\fill}
\begin{minipage}[t]{75mm}
\epsfxsize=150pt
\centerline{ \epsfbox{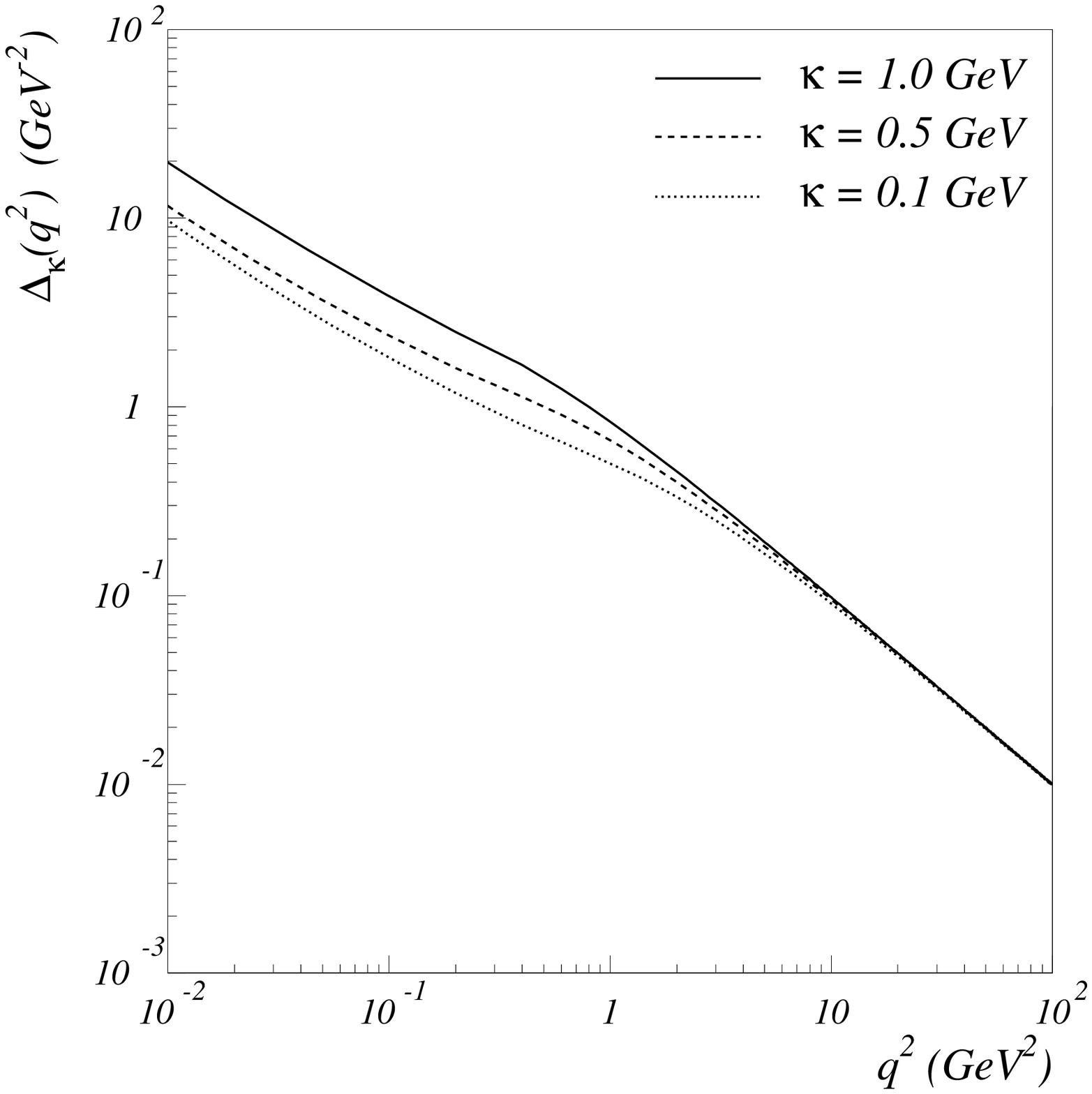}  }
\end{minipage}
\caption{
{\it Left panel:} The inverse renormalization function  $\overline{{\cal Z}}^{-1}_\kappa$,
eq. (\ref{formula2}) versus $q^2$ for different values of $\kappa$,
for different choices of $\kappa$. {\it Right panel:} the corresponding
gluon propagator $\Delta_\kappa$, eq. (\ref{formula1}),
 in contrast to the bare propagator $\Delta_\kappa^{(0)}$.
\label{figure5}
}
\end{figure}
\bigskip
\bigskip

\subsection{Remarks}

Let us summarize the strategy that has led to the main result of this
paper, namely the asymptotic
light-cone-gauge solutions of the renormalization function
${\cal Z}_\kappa$ for $q^2\rightarrow \infty$ and $q^2\rightarrow 0$,
eqs. (\ref{ZUsol}), (\ref{ZIsol}) and (\ref{formula2}). 
We derived an evolution equation (\ref{b3}) for ${\cal Z}_\kappa$ that
involves only the exact propagator $\Delta_{\kappa,\,\mu\nu}$
and the exact 3-gluon vertex ${\cal V}_{\mu\nu\lambda}$,
but no higher-order vertex functions. To obtain a closed
equation for the gluon propagator alone, the 3-gluon
vertex function was related to the propagator via the 
Slavnov-Taylor identity (\ref{ST1}) and constructed an {\it ansatz}
for   ${\cal V}_{\mu\nu\lambda}$, eq. (\ref{Vansatz}), which obeys 
the constraining Slavnov-Taylor identity. 
The necessity of making a particular (non-unique) ansatz
is clearly the weakest point in our approach,
yet it seems to be the only way to trade in the unknown 
${\cal V}_{\mu\nu\lambda}$ in order to  obtain a closed equation.
The resulting evolution equation (\ref{b4}) for ${\cal Z}_\kappa$ then
contains solely the gluon propagator in terms of its
spectral density $\rho_\kappa$, and thus expresses the
intimate relation between the renormalization function and
the full gluon propagtor (on the basis of the specific ansatz
for the 3-gluon vertex).
The final equation (\ref{b4}) for ${\cal Z}_\kappa$
could be solved analytically in terms of elementary
functions in the asymptotic ultraviolet $q^2\rightarrow\infty$
and the deep infrared $q^2\rightarrow 0$, provided
we approximate the infrared regulator $R_\kappa(q^2)$
by its asymptotic behavior in the limits $q^2\rightarrow \infty$
and $q^2\rightarrow 0$, respectively. 
\medskip

The  ultraviolet result (\ref{ZUsol}) for ${\cal Z}_\kappa$ 
is characterized by the logarithmic behavior 
consistent with asymptotic freedom, since 
the ratio of bare and renormalized coupling constants
$g_0^2/g^2 ={\cal Z}^{-1}_\kappa < 1$, corresponding 
anti-screening of the color charge,
i.e., the bare charge is larger than the renormalized one 
(as opposed to QED, where $e_0^2/e^2 > 1$, implying
screening of the electric charge due to virtual pair creation).
\medskip

The  infrared solution for ${\cal Z}_\kappa$, eq. (\ref{ZIsol}),
on the other hand, exposes a $1/q^2$ behavior, which would correspond 
to a linear static potential $V(r) \propto r$ as $r\rightarrow \infty$,
as expected for confinement in the long-wavelength limit (as opposed
to QED, where the infrared behavior is $\propto 1/q^2$, corresponding
to the classical Coulomb potential $V(r)\propto 1/r$).
Although the gluon propagator $\Delta_{\kappa,\,\mu\nu}$, 
and thus ${\cal Z}_\kappa$, is a gauge-dependent object,
its gauge-invariant physics content may be extracted by relating
it to the gauge-invariant Wilson loop \cite{wilson}.
\bigskip
\bigskip

\section{Phenomenological applications}
\bigskip

In this Section, we apply our results to the 
$\kappa$-dependent renormalization function
${\cal Z}_\kappa(q^2)$ to illustrate two important
phenomenological connections with experimentally measurable quantities,
namely the QCD running coupling $\alpha_s(q^2)$ and the gluon
distribution function $G(q)$.
First we infer from ${\cal Z}_\kappa$
the running of the coupling $\alpha_s(q^2)$, using standard
renormalization group arguments, and then we
relate ${\cal Z}_\kappa$ via the spectral density $\rho_\kappa$ of the gluon propagator,
to the gluon distribution function $G(q,\kappa)$ and its evolution
equation.

\subsection{Renormalization group equation and running coupling}

Recall  that the renormalized gluon propagator,
respectively the renormalized coupling satisfy, [cf. \ref{RG})], 
\begin{equation}
\Delta_{\kappa}(q^2) \;\;=\;\; 
{\cal Z}_\kappa(q^2) \; \Delta_{\kappa}^{(0)}(q^2)
\;\;\;\;\;\;\;\;\;\;\;\;\;\;
g(q^2) \;\;=\;\; {\cal Z}_\kappa(q^2)^{1/2} \;\; g_0
\label{RG1}
\;,
\end{equation}
where the scalar propagator function $\Delta_\kappa(q^2)$ is
related to $\Delta_{\kappa,\,\mu\nu}(q)$ by
\begin{equation}  
\Delta_{\kappa, \,\mu\nu}(q)\;\;\equiv\;\; \Delta_\kappa(q^2)\;\;S'_{\mu\nu}(q)
\label{RG2}
\end{equation}
with $S'_{\mu\nu}$ given by (\ref{Tprime}).
As in (\ref{RG0}), we specify the
initial conditions at the scale $\Lambda \gg 1$ GeV 
where we normalize the theory, such that
at $q^2 = \Lambda^2$, it coincides with the bare one,
\begin{equation}
{\cal Z}_\kappa(\Lambda^2) \;\;=\;\; 1 ,
\;\;\;\;\;\;\;\;\;\;\;\;\;
\Delta_{\kappa,\,\mu\nu}(\Lambda^2) \;\;=\;\; 
\Delta_{\kappa,\,\mu\nu}^{(0)}(\Lambda^2) ,
\;\;\;\;\;\;\;\;\;\;\;\;\;
g_0 \;\;=\;\; g(\Lambda^2)
\label{RG2a}
\; .
\end{equation}

In order to invoke the 
renormalization group formalism for the light-cone gauge representation of
$\Delta_{\kappa}(q)$, ${\cal Z}_\kappa(q^2)$,  and  $g(q^2)$,
it is convenient to introduce the dimensionless propagator function $D$ through 
\begin{equation}  
D_\kappa\left(\frac{q^2}{\Lambda^2}, \,g(\Lambda^2)\right) \;\;
\;\;\equiv \;\; q^2\;\Delta_{\kappa}(q^2)
\;,
\label{RG3}
\end{equation}  
How the physics  changes when we vary $\Lambda$ with ($\kappa$ fixed) is
described by the renormalization group equation for $D_\kappa$:
If we change the scale $\Lambda$, e.g. by $\Lambda^2 \rightarrow \lambda \; \Lambda^2$, 
then the renormalizability of the theory requires that this
is equivalent to a rescaling of $D_\kappa$ by the factor ${\cal Z}_\kappa$,
that is,
\begin{equation}
D_\kappa\left(\frac{q^2}{\Lambda^2}, \,g(\Lambda^2)\right) \;\;
= \;\;
{\cal Z}_\kappa\left(\frac{q^2}{\Lambda^2},\,g(\Lambda^2),\,  \lambda\right) \;\;
D_\kappa\left(\frac{q^2}{\lambda \Lambda^2}, g(\lambda \Lambda^2)\right)
\label{RG4}
\;,
\end{equation}
where we have written
\begin{equation}
{\cal Z}_\kappa(q^2) \;\;\equiv \;\;{\cal Z}_\kappa\left(\frac{q^2}{\Lambda^2},\,
g(\Lambda^2),\,\lambda\right)
\end{equation}
in order to expose the implicit $\lambda$-dependence in ${\cal Z}_\kappa$.
Now let us define the variable
\begin{equation}
\tau \; \equiv \;  -\,\ln\left(\frac{q^2}{\kappa^2}\right)
\;,
\end{equation}
and differentiate (\ref{RG4}) with respect to $\lambda$. Then, setting $\lambda =1$
yields the standard renormalization group equation:
\begin{equation}
\left[ \frac{}{}
\frac{\partial}{\partial \tau}\;+\; \beta_\kappa(g) \;+\; \eta_\kappa(\tau,g)
\;\right]\;
D_\kappa\left(\tau, g\right)
= \;\; 0
\;,
\label{RG5}
\end{equation}
where 
\begin{eqnarray}
\beta_\kappa(g) \; &\equiv &\;
\left. \frac{\partial}{\partial \lambda} \;g\left(\lambda \Lambda^2\right) \,\right|_{\lambda=1}
\label{RG6}
\;,
\\ & & \nonumber \\
\eta_\kappa(\tau, \,g) \; &\equiv &\; 
\left. \frac{\partial}{\partial \lambda}\; 
\ln {\cal Z}_\kappa\left(\frac{q^2}{\Lambda^2},\,g(\Lambda^2),\,\lambda\right)
\,\right|_{\lambda=1}
\label{RG7}
\;,
\end{eqnarray}
with $\beta_\kappa$ denoting the  {\it Callan-Szymanzik function} 
in the presence of the cut-off scale $\kappa$,
and $\eta_\kappa$ the  {\it anomalous dimension} $\eta_\kappa$, 
also being $\kappa$-dependent.
The solution (\ref{RG5}) to the renormalization group equation for $D_\kappa$
is obviously 
\begin{equation}
D_\kappa\left( \tau,\, g_0\right) \;\;=\;\;
D_\kappa\left( 0,\, g(-\tau)\right) \;\;
\exp\left[\,- \int_0^{\tau} d\tau^{\prime}\;\, 
\eta_\kappa\left(\tau^{\prime},\,g(-\tau^\prime)\right)\right]
\;,
\label{RG8}
\end{equation}
or,
\begin{equation}
D_\kappa\left( \frac{q^2}{\Lambda^2},\, g(\Lambda^2)\right) \;\;=\;\;
D_\kappa\left( \frac{q^2}{\kappa^2},\, g(q^2)\right) \;\;
\exp\left[\,- \int_{\kappa^2}^{q^2} \frac{d q^{\prime\;2}}{q^{\prime\;2}}\;\, 
\eta_\kappa\left(\frac{q^{\prime\;2}}{\kappa^2},\,g(q^{\prime\;2})\right)\right]
\;,
\label{RG8a}
\end{equation}
which shows, since $D_\kappa = q^2 \Delta_\kappa$, that the evolution 
of the gluon propagator is simply governed by the multiplicative factor 
${\cal Z}_\kappa$ involving the integrated anomalous dimension $\eta_\kappa$.
In view of (\ref{RG4}) we therefore can make the identification,
\begin{equation}
\ln\,{\cal Z}_\kappa\left(
\frac{q^2}{\kappa^2},\,g(q^2)\right)  \;\;=\;\;
-\frac{1}{2}\, \int_{\kappa^2}^{q^2} \frac{d q^{\prime\;2}}{q^{\prime\;2}}\;\, 
\eta_\kappa\left(\frac{q^{\prime\;2}}{\kappa^2},\,g(q^{\prime\;2})\right)
\label{RG7a}
\;.
\end{equation}
\medskip

In order to find the large-$q^2$ behaviour, we return to
the approximate solution $\overline{\cal Z}^{-1}_\kappa$ 
of (\ref{formula2}), and invert it by expanding in a power series in $g^2$,
\begin{equation}
{\overline{\cal Z}}_\kappa(q^2)
\;\;=\;\;
1\;-\; \frac{g^2}{(4\pi)^2}\;
\frac{11C_G}{3}\ln\left(\frac{q^2}{\kappa^2}\right)
\;\;+\;\; O\left(g^4\right)
\;,
\label{RG9}
\end{equation}
In the large-$q^2$ limit,  substitution of (\ref{RG9}) into (\ref{RG6}) then yields
the asymptotic behavior of the $\beta_\kappa$-function to order $O(g^3)$:
\begin{equation}
\beta_\kappa(g)\;\;=\;\; - \,\beta^{(0)}\,g^3\;\;+\;\; O\left(g^5\right)
\;\;\;\;\;\;\;\;\;\;\;\;\;\;\;\;\;\;\;\;\;
\beta^{(0)}\;=\; \frac{1}{(4\pi)^2}\frac{11 C_G}{3}
\;.
\label{RG10}
\end{equation}
The solution of (\ref{RG6}) together with (\ref{RG10}) then 
yields the (gauge-invariant) large-$q^2$ form of running coupling 
\begin{equation}
\overline{g}^2(q^2)\;\;=\;\;
\frac{g^2}{1\;+\frac{11 C_G}{3 (4\pi)^2}\,g^2\,\ln\left(q^2/\kappa^2\right)}
\;\;+\;\;O\left(g^4\right)
\;\; =\;\;
\frac{1}{1\;+\frac{11 C_G}{3 (4\pi)^2}\,\,\ln\left(q^2/\Lambda_{\rm QCD}^2\right)}
\;,
\label{RG11}
\end{equation}
with $\Lambda_{\rm QCD}^2=\kappa^2 \,\exp(-1/(\beta^{(0)} g^2)$ and 
$g = \overline{g}(\kappa^2)$.  Equivalent to (\ref{RG11}) is the 
running coupling $\alpha_s=\overline{g}^2/(4\pi)$ 
at 1-loop order,
\begin{equation}
\alpha_s^{(1)}(q^2)\;\;=\;\;
\frac{12\pi}{11 C_G\,\ln\left(q^2/\Lambda_{\rm QCD}^2\right)}
\;.
\label{RG12}
\end{equation}
Similarly, the solution  of (\ref{RG7})  in the large-$q^2$ limit gives the 
(gauge-dependent) anomalous dimension $\eta_\kappa(q^2)$ to order $O(g^2)$:
\begin{equation}
\eta(q^2)\;\;=\;\;
\frac{g^2}{(4\pi)^2}\frac{11 C_G}{6} \;\;+\;\;O\left(g^4\right)
\;.
\label{RG13}
\end{equation}
The large-$q^2$ estimates  (\ref{RG10} -- \ref{RG13}), resulting from
our approximate solution  $\overline{\cal Z}_\kappa$ of eq. (\ref{formula2}),
agree with the standard  results obtained within
perturbation theory for the pure gauge theory \cite{muta}.
\medskip

\subsection{Evolution of the gluon distribution function}

The gluonic substructure of a hadron can be measured in experiments, 
for instance in deep-inelastic lepton hadron scattering or high-energy 
hadronic collisions, through the gluon distribution function.
The gluon distribution function is defined \cite{collins82} as the density 
of gluon fluctuations inside a hadron,
that is, in terms of matrix elements in a hadron state of
specific operators that count the number of gluons carrying
a certain fraction $x$ of the hadron momentum $P$. The 
natural choice for such a number operator would be
${\cal A}_\mu{\cal A}^\mu$, however, in QCD this is not a gauge-invariant
object. Instead, one uses the gauge-invariant operator
${\cal F}_{\mu\nu}{\cal F}^{\mu\nu}$.
The precise definition of the gluon distribution function is most
conveniently expressed in the {\it infinite momentum frame}, 
in which the hadron moves in the $z-t$ plane along the light cone.
Employing  the standard light-cone representation of four-vectors,
\begin{eqnarray}
& &
v^\mu\;=\; \left( v^+, v^-, \vec{v}_\perp \right)
\;, \;\;\;\;\;\;\;
v^\pm \;=\; v_\mp \;=\; v^0\pm v^3
\;, \;\;\;\;\;\;\;
\vec{v}_\perp \;=\; \left(v^1,v^2\right)
\;, \;\;\;\;\;\;\;
v_\perp \;=\; \sqrt{\vec{v}_\perp^{\;2}}
\nonumber \\
& &
\;\;\;\;\;\;\;\;\;\;\;\;\;
v^2 \;=\; v^+v^- \;-\;v_\perp^2
\;,\;\;\;\;\;\;\;\;\;\;\;\;
v_\mu w^\mu \;=\; \frac{1}{2}\,
\left( v^+ w^- + v^- w^+ \right) \;-\;\vec{v}_\perp\cdot\vec{w}_\perp
\label{lcv1}
\;,
\end{eqnarray}
the gluon distribution function is then  the average number of gluons 
at light-cone time $r^+ =0$ in a hadron state $|P\rangle$ 
moving with momentum $P^+$, with the gluon fluctuations 
carrying a fraction $x = q^+/P^+$ in an interval $dx$ and
transverse momenta in a range $d^2 q_\perp$ \cite{collins82}:
\begin{equation}
G(x, {\bf q}_\perp)
\;\equiv\;
\frac{1}{xP^+}
\int d r^- d^2 r_\perp\;
e^{i (q^+ r^- -\vec q_\perp \cdot \vec r_\perp)} \;
\;\;
\langle\, P\, |
\,{\cal F}^{+\nu}(0,r^-,\vec r_\perp) \;\;{\cal E}(r^-,0)\;\;{\cal F}^+_{\;\;\nu}(0,0,\vec 0_\perp)\,
|\,P\, \rangle
\;.
\label{gluedist1}
\end{equation}
Here the path-ordering exponential
\begin{equation}
{\cal E}(r^-_2,r^-_1)\;\;\equiv\;\;
P\;\exp\left\{\,i g \,\int_{r^-_1}^{r^-_2}
dr^{\prime -} \,{\cal A}_a^+(0,r^{\prime -}, {\bf 0}_\perp)\;T^a
\right\}
\label{gluedist2}
\end{equation}
makes the definition (\ref{gluedist1}) with the
non-local operator ${\cal F}^{+\nu}(r^-_1) {\cal F}^+_\nu(r^-_2)$
fully gauge-invariant,
as it orders the gauge-field operators ${\cal A}_a^+T^a$
along the line-integral between $r^+_1$ and $r^+_2$.
Moreover, it  provides the link 
to compute the gluon distribution function in different gauges.
\smallskip

We adopt the general definition to our choice of light-cone gauge,
for which in terms of light-cone variables the choice
of the gauge vector $n_\mu$, eq. (\ref{mom4}) reads,
\begin{equation}
n_\mu\;\;=\;\; \left(n^+, n^-, \vec{n}_\perp\right) \;\;=\;\;
(0, 1, \vec{0}_\perp)
\label{gluedist3}
\end{equation}
so that the gauge constraint (\ref{gauge1}) becomes
\begin{equation}
n\cdot {\cal A} \;=\; {\cal A}^+ \;=\; {\cal A}_- \;=\; 0
\;.
\label{gluedist4}
\end{equation}
Thus, the factor  ${\cal E}$ in (\ref{gluedist1}) is equal to unity.
Futhermore, we note that specifically in the axial gauges (including the
light-cone gauge),\footnote{
The only non-vanishing components of the gauge-field tensor 
${\cal F}^{\mu\nu} = -{\cal F}^{\nu\mu}$ are
\begin{eqnarray}
& &\;\;\;\;\;\;\;
{\cal F}^{+-} \;=\; -\partial^+ {\cal A}^-
\;,\;\;\;\;\;\;
{\cal F}^{+\,i} \;=\; \partial^+ {\cal A}^i
\;,\;\;\;\;\;\;
{\cal F}^{-\,i} \;=\; \partial^- {\cal A}^i -\partial^i {\cal A}^-
-ig \left[{\cal A}^-,{\cal A}^i\right]
\;,\;\;\;\;\;\;
{\cal F}^{ij} \;=\; \partial^i {\cal A}^j -\partial^j {\cal A}^i
-ig \left[\overline{A}^i,{\cal A}^j\right]
\;.
\nonumber
\end{eqnarray}
}
\begin{equation}
{\cal F}^{+\nu}\,{\cal F}^{+}_\nu\;\;=\;\; 
(\partial^+ {\cal A}^i) \; (\partial^+ {\cal A}_i)
\label{gluedist5}
\;,
\end{equation}
where 
a summation over the transverse components $ i = 1,2$ is understood,
and $\partial^\pm = \partial/\partial r^\pm$.
This simple relation involves only the
transverse gauge fields ${\cal A}_i$,
which  has its physics origin in the fact that in the
axial gauges only the physical, transverse gluon degrees of freedom
propagate, while ${\cal A}^+$  vanishes and ${\cal A}^-$ is a pure
gauge which decouples.
As a consequence, the gauge-invariant definition (\ref{gluedist1})
of the gluon distribution takes the following form in the light-cone gauge:
\begin{equation}
G(x, {\bf q}_\perp)
\;\equiv\;
xP^+\; \int d r^- d^2 r_\perp \;
e^{i (q^+ r^- -\vec q_\perp \cdot \vec r_\perp)} \;
\;\;
\langle\, P\, |
\,{\cal A}^{i}(0,r^-,\vec r_\perp) \;{\cal A}_{i}(0,0,\vec 0_\perp)\,
|\,P\, \rangle
\;,
\label{gluedist6}
\end{equation}
summed over the transverse components $i=1,2$.
In order to extend this expression to accomodate
our scale-dependent formalism of Sec. IIB, in which the  gluon 2-point
functions carry an explicit $\kappa$-dependence due to the infrared
regulator ${\cal R}^{\mu\nu}_\kappa$, eq. (\ref{Delta3a}), 
we generalize (\ref{gluedist6}) by
\begin{equation}
{\cal A}^{i}{\cal A}_{i} \;\;\longrightarrow \;\;
{\cal A}_i\;\left(\delta^{ij}+\frac{{\cal R}_\kappa^{ij}(\partial^2)}{\partial^2}
\right)\;{\cal A}_{j} \;\;=\;\;
{\cal A}^{i}\;\left(1+\frac{R_\kappa(\partial^2)}{\partial^2}
\right)\;{\cal A}^{i} {\cal A}_i
\label{gluedist7}
\;,
\end{equation}
with $R_\kappa$ given by (\ref{Rk}) or (\ref{RkL}). Thus, the 
$\kappa$-dependent gluon distribution may be defined as
\begin{equation}
G_\kappa(x, {\bf q}_\perp, q^2)
\;\equiv\;
\frac{xP^+}{1+R_\kappa(q^2)/q^2}\,
\int d r^- d^2 r_\perp \;
e^{i (q^+ r^- -\vec q_\perp \cdot \vec r_\perp)} \;
\;\;
\langle\, P\, |
\,{\cal A}^{i}(0,r^-,\vec r_\perp) \;{\cal A}_{i}(0,0,\vec 0_\perp)\,
|\,P\, \rangle
\;,
\label{gluedist8}
\end{equation}
Now, as discussed in Appendix E, the expectation value of the gluon number 
operator ${\cal A}^i{\cal A}_i$
on the right-hand side is essentially the gluon spectral density
$\rho_\kappa$ that enters the spectral representation (\ref{spectral}) of the
gluon propagator. Precisely, it is the transverse spatial 
component $\rho_{\kappa\; i}^{(+)\;i} = \rho_\kappa^{1 1\;(+)}+\rho_\kappa^{2 2\;(+)}$, 
of  the causal correlation function 
\begin{equation}
\rho_{\kappa,\;\mu\nu}^{(+)}(q)
\;\;=\;\;
\int d^4 r \; e^{i q\cdot r}\;\;
\langle\, P\, | \,{\cal A}^\mu(r) \;{\cal A}^{\nu}(0)\, |\,P\, \rangle
\;.
\label{gluedist9}
\end{equation}
at $r^+ = 0$. Similarly, the anti-causal correlator is defined as
\begin{equation}
\rho_{\kappa,\;\mu\nu}^{(-)}(q)
\;\;=\;\;
-\;\int d^4 r \; e^{i q\cdot r}\;\;
\langle\, P\, | \,{\cal A}^\nu(0) \;{\cal A}^{\mu}(r)\, |\,P\, \rangle
\;.
\label{gluedist10}
\end{equation}
The spectral density is the sum of both contributions, 
\begin{equation}
\rho_{\kappa,\;\mu\nu}(q)
\;\;=\;\;
\frac{1}{2} \;\left[\rho_{\kappa,\;\mu\nu}^{(+)}(q)
\;\;+\;\;\rho_{\kappa,\;\mu\nu}^{(-)}(q)\right]
\;\;=\;\;
\rho_{\kappa,\;\mu\nu}^{(+)}(q)
\;,
\label{gluedist11}
\end{equation}
where the latter equality holds only if translational invariance is
preserved (in which case the crossing relation 
$\rho_{\kappa,\;\mu\nu}^{(+)} = - \rho_{\kappa,\;\nu\mu}^{(+)}
=\rho_{\kappa,\;\mu\nu}^{(-)}$ exists),
while it is invalid in physics situations where one encounters
a spatially inhomogenous medium. In the present context,
we are interested in the gluon distribution of a physical hadronic state
in free space, so that we may use (\ref{gluedist11}) to 
relate the spectral density  to  the gluon distribution. 
To do so, we first note that in the light-cone gauge, 
the tensor stucture of $\rho_{\kappa,\,\mu\nu}$ is identical to that
of the propagator [cf. Appendix E],
\begin{equation}
\rho_{\kappa,\;\mu\nu}(q)
\;\;=\;\;\rho_\kappa(q^2) 
\;\; \left(\, g_{\mu\nu} \,-\,\frac{n_\mu q_\nu + q_\mu n_\nu}{n\cdot q}\,\right)
\;\;=\;\;
\;\;\rho_\kappa(q^2) \;\; S^{\,\prime}_{\mu\nu}(q)
\;.
\label{gluedist12}
\end{equation}
Defining the density $\rho_\kappa(x,{\bf q}_\perp, q^2)$ through
\begin{equation}
\rho_\kappa(q^2) \;\; \equiv\;\; \int dx \,d^2 q_\perp \;
\rho_\kappa(x,{\bf q}_\perp, q^2)
\;,
\label{gluedist13}
\end{equation}
we see from  (\ref{gluedist8} -- \ref{gluedist13})
that the spectral density $\rho_\kappa(x,{\bf q}_\perp, q^2)$
can be identified with the gluon distribution (\ref{gluedist8}), 
\begin{equation}
\rho_\kappa(x,{\bf q}_\perp, q^2) \;\;=\;\; g_\kappa(x,{\bf q}_\perp, q^2)
\;,
\label{gluedist14}
\end{equation}
as one may intuitively expect, since the gluon distribution measures the density of gluonic
fluctuations which is nothing else but the `level density' of 
gluon states described by the spectral density.
Accordingly, the bare density $\rho_\kappa^{(0)}$, in the absence of interactions,
just
\begin{equation}
\rho^{(0)}_\kappa(x,{\bf q}_\perp, q^2) \;\;=\;\; g^{(0)}_\kappa(x,{\bf q}_\perp, q^2)
\;\;=\;\;\delta(1-x)\delta^2({\bf q}_\perp)\,\delta(q^2)
\;,
\label{gluedist14a}
\end{equation}
corresponds to single bare gluon carrying the full momentum fraction
$x = 1$. We remark that the density (\ref{gluedist14}) satisfies the
following sum rule \cite{west},
\begin{equation}
\int_0^{1} dx \,x\,
{\rho}_\kappa\left(x,{\bf q}_\perp,q^2\right) \;=\;1
\label{sumrule1}
\;.
\end{equation}
\medskip

As an immediate consequence of the  above identification of $\rho_\kappa$ 
with the gluon distribution $g_\kappa$,  the evolution
of the latter is governed again by the renormalization function ${\cal Z}_\kappa$:
Since the gluon propagator 
$\Delta_{\kappa,\,\mu\nu}(q) = {\cal Z}_\kappa (q^2) \Delta_{\kappa,\,\mu\nu}^{(0)}(q)$, 
we see from (\ref{spectral}) that also 
$\rho_{\kappa}(q^2) = {\cal Z}_\kappa(q^2) \rho_{\kappa}^{(0)}(q^2)$, 
To derive the precise form of the evolution equation for $g_\kappa$, let us 
consider the  transverse-momentum integrated  density,
\begin{equation}
\rho_\kappa(x,q^2)
\;\equiv\;\;\int d^2q_\perp \; \rho_\kappa(x,{\bf q}_\perp, q^2)
\;,
\label{gluedist14b}
\end{equation}
and introduce the $x$-moments
\begin{equation}
\widetilde{\rho}_\kappa(N,q^2) \;\;\equiv\; \;\int_0^1 dx \; x^{N-1}\; 
\rho_\kappa(x, q^2)
\;.
\label{gluedist15}
\end{equation}
The first moment is just
\begin{equation}
\rho_\kappa(q^2)\;\;=\;\; \widetilde{\rho}_\kappa(1,q^2)
\;\;=\;\; {\cal Z}_\kappa(q^2)\;\widetilde{\rho}^{(0)}_\kappa(1,q^2)
\;.
\label{gluedist16}
\end{equation}
Expressing ${\cal Z}_\kappa$ in terms of the anomalous dimension $\eta_\kappa$,
eq. (\ref{RG7a}),
\begin{equation}
{\cal Z}_\kappa(q^2) \;\;=\;\;
 -\frac{1}{2} \,\int_{\kappa^2}^{q^2} \frac{d q^{\prime\;2}}{q^{\prime\;2}}\;\,
\eta_\kappa\left(q^{\prime\;2},\,g(q^{\prime\;2})\right)
\label{gluedist17}
\;,
\end{equation}
the $N$-th moment generalization of  (\ref{gluedist16}) may be written
as
\begin{equation}
\widetilde{\rho}_\kappa(N,q^2)\;\;=\;\; 
\exp\left\{
 \,-\frac{1}{2} \int_{\kappa^2}^{q^2} \frac{d q^{\prime\;2}}{q^{\prime\;2}}\;\,
\eta_\kappa\left(N,q^{\prime\;2},\,g(q^{\prime\;2})\right)
\right\}
\;\; \widetilde{\rho}_\kappa^{(0)} (N,q^2)
\;\;\equiv\;\; 
{\cal Z}_\kappa(N,q^2)\; \widetilde{\rho}^{(0)}_\kappa(N,q^2)
\label{gluedist17a}
\;.
\end{equation}
The evolution  with  $q^2$ of the spectral density in $N$-space
is therefore governed by the evolution equation
\begin{equation}
q^2 \frac{\partial }{\partial q^2}\; \widetilde{\rho}_\kappa(N,q^2)\;\;=\;\; 
-\frac{1}{2}\,\eta_\kappa\left(N,q^{2},\,g(q^{2})\right)
\;\; \widetilde{\rho}_\kappa (N,q^2)
\label{gluedist18}
\;.
\end{equation}
If we define a probability distribution $P\left(x,g(q^{2})\right)$ as
\begin{equation}
\int_0^1 dx \;x^{N-1} \;P\left(N,g(q^{2})\right) \;\;\equiv\;\;
-\frac{1}{2}\,\eta_\kappa\left(N,q^{2},\,g(q^{2})\right)
\;,
\label{gluedist19}
\end{equation}
we can express (\ref{gluedist18}) as follows:
\begin{equation}
q^2 \frac{\partial }{\partial q^2}\; \rho_\kappa(x,q^2)\;\;=\;\; 
\int_x^1 \frac{dz}{z}\;P\left(x,g(q^2)\right)
\;\; \rho_\kappa (x,q^2)
\label{gluedist20}
\;.
\end{equation}
This evolution equation has the form of the DGLAP master equation
\cite{DGLAP}, however, with the essential difference that it contains
the non-perturbative infrared physics as well, while
the DGLAP equation corresponds to the perturbative limit of (\ref{gluedist20}).
This is easily realized by
expanding the probability function $P$ in power of $g^2$,
\begin{equation}
P\left(x,g(q^2)\right)
\;\;=\;\; \left(\frac{g^2(q^2)}{8\pi^2}\right)\; P^{(0)}(x)
\;+\; \left(\frac{g^2(q^2)}{8\pi^2}\right)^2\; P^{(1)}(x)
\;+\;\ldots
\label{gluedist21}
\;,
\end{equation}
and substituting in (\ref{gluedist20}). It is now evident that $P^{(0)}$
must coincide with the DGLAP probability for gluon splitting, $g\rightarrow gg$
\cite{DGLAP},
\begin{equation}
P^{(0)}(x) \;=\;
P^{\rm DGLAP}_{g\rightarrow gg}(x) \;=\;
2\;C_G\;\;\left(\,\frac{x}{1-x}_+\,+\,\frac{1-x}{x}\,+\,z (1-x)\,\right)
\label{gluedist22}
\;.
\end{equation}
Hence, one may regard \cite{curci}  $P\left(x,g(q^2)\right)$ as a generalization
of the DGLAP probability 
to all orders in $g^2/8\pi^2$, or $\alpha_s/2\pi$.
\medskip

The integral form of the evolution equation (\ref{gluedist20})
can now be expressed as
\begin{equation}
{\rho}_\kappa(x,q^2) \;=\;
{\cal Z}_\kappa(q^2)\; \rho_\kappa^{(0)}\left(x, q^2\right) 
\;+\;
\; {\cal Z}_\kappa(q^2)\;
\int_{\kappa^2}^{q^2}\frac{d q^{\prime\,2}}{q^{\prime\,2}}
\;
\int_0^1 \frac{dz}{x} \;
P\left(z, g(q^{\prime\,2})\right)\;
{\rho}_\kappa\left(\frac{x}{z},q^2\right)
\;{\cal Z}_\kappa^{-1}\left(\frac{q^{\prime\,2}}{z}\right)
\;,
\label{gluedist23}
\end{equation}
where $\rho_\kappa^{(0)}$ is defined in (\ref{gluedist14a}).
Multiplying by $x$ and integrating  over $x$ from 0 to $1$ yields on account
of  the sum rule (\ref{sumrule1}) an integral equation for ${\cal Z}_\kappa$
in terms of the probability $P$,
\begin{equation}
{\cal Z}^{-1}_\kappa(q^2) 
\;\;=\;\;
1\;\;+\;\;
\frac{}{}
\int_{\kappa^2}^{q^2}\frac{d q^{\prime\,2}}{q^{\prime\,2}}
\;
\int_0^1 dz \;
P\left(z,g(q^{\prime\,2})\right)\;
\;{\cal Z}^{-1}_{\kappa/\sqrt{z}}\left(q^2\right) 
\;.
\end{equation}
This equation is reminescent of (\ref{b6}) encountered in the
context of the evolution of the gluon propagator,
reflecting the universal role of the renormalization function 
${\cal Z}_\kappa$ in the light-cone gauge.
\bigskip

\section*{Acknowledgments}

This work was supported in part by the U.S. Department of Energy 
under contract DE-AC02-98CH10886.
\bigskip
\bigskip

\appendix

\section{Definitions and notation}
\label{sec:appa}

This Appendix gives a summary of the basic quantities encountered
in the paper, and the various notations used.
Throughout the paper pure $SU(3)_c$ Yang-Mills theory in
Minkowski space is considered, with $N_c=3$ colors
and the absence of quark degrees of freedom.
\medskip

Our convention for placing indices and labels are the following:
\begin{itemize}
\item
Lorentz vector indices $\mu,\nu,\ldots$ 
may be raised or lowered according to the Minkowski metric 
$g_{\mu\nu} = \mbox{diag}(1,-1,-1,-1)$, 
and the usual convention for summation over repeated indices is understood.
\item
Similarly, color indices $a,b,\ldots$ may be raised
or lowered according to the commutation rules of the $SU(3)$ generators,
eq. (\ref{A5}).
\item
All other labels that do not refer to internal degrees of
freedom, as e.g., $\Gamma_\kappa$ or $\Gamma^{(2)}$, are consistently 
placed either as subscripts or superscripts.
\end{itemize}
In order to avoid `inflationary labeling' with sub- or superscripts,
we often choose to suppress the color indices of vectors or tensors,
when the color labels corresponds to the Lorentz indices,
e.g., $\Gamma_{\mu\nu}^{ab}(q,q') \equiv \Gamma_{\mu\nu}$.
\smallskip

Furthermore, the following shorthand notations are employed:
\begin{eqnarray}
A\cdot B &\equiv& A_\mu g^{\mu \nu} B_\nu
\;,\;\;\;\;\;\;\;\;\;\;\;\;\;\;\;\;
\;\;\;\;\;\;\;\;\;
A\circ B \;\equiv\;  \int d^4x \,A_\mu(x)\, B^\mu(x)
\\
\left( A \,B \right) \cdot C &\equiv &
\,A^\mu B^\nu \;C_{\mu\nu} 
\;,\;\;\;\;\;\;\;\;\;\;\;\;\;\;\;\;
\left( A \, B \right)\circ C  \;\equiv \;
\int d^4x d^4 y \, \left(\frac{}{} A^\mu(x)\, B^\nu(y)\right)
\; C_{\mu\nu}(x,y)
\;.
\label{A1}
\end{eqnarray}
We use the symbol $Tr$ for the trace over discrete
indices,
\begin{equation}
\mbox{Tr}\left[ A B \right] 
\;=\;
\left\{
\begin{array}{l}
A_\mu^a(x) B_\nu^b(y)
g^{\mu\nu}\delta^{ab} \delta^4(x-y) 
\;\;\;\;\;\;\;\;\;\;\;\;
\mbox{in space-time}
\\  \\
A_\mu^a(k) B_\nu^b(k') g^{\mu\nu}\delta^{ab} \delta^4(k-k')
\;\;\;\;\;\;\;\;\;\;
\mbox{in momentum space}
\end{array}
\right.
\label{A2}
\;.
\end{equation}
Similarly, we use
the symbol $Sp$ for tracing over discrete indices as well as
integrating over continuous variables,
\begin{equation}
\mbox{Tr}\left[ A B \right] 
\;=\;
\left\{
\begin{array}{l}
\int d^4x d^4y \,\mbox{Tr} \left[A(x) B(y)\right]
\;\;\;\;\;\;\;\;\;\;\;\;\;\;\;\;\;
\mbox{in space-time}
\\  \\
\int
\frac{d^4k}{(2\pi)^4}\frac{d^4k'}{(2\pi)^4} \,\mbox{Tr}\left[A(k) B(k') \right]
\;\;\;\;\;\;\;\;\;\;\;\;
\mbox{in momentum space}
\end{array}
\right.
\label{A3}
\;.
\end{equation}
\bigskip

The {\it gauge field} is denoted by 
${\cal A}_\mu(x) \equiv T^a\,{\cal A}_\mu^a(x)$,
and the corresponding {\it gauge field tensor} and {\it covariant derivative}
are defined as
\begin{eqnarray}
{\cal F}_{\mu\nu}(x) &\equiv& T^a\,{\cal F}_{\mu\nu}^a(x)
\;=\; \frac{1}{(-i g)} \,\left[ D_\mu\,, D_\nu \right]
\nonumber \\
D_\mu(x) &\equiv& \partial_\mu \,-\,ig\,T^a\,{\cal A}_\mu^a(x)
\;=\; \partial_\mu \,-\,ig\,{\cal A}_\mu(x)
\;,
\label{A4}
\end{eqnarray}
or, explicitly in color components,
\begin{eqnarray}
{\cal F}_{\mu\nu}^a &=&
\partial_\mu {\cal A}_{\nu}^a \;-\; \partial_\nu {\cal A}_{\mu}^a 
\;+\; g\, f^{a}_{bc} \,{\cal A}_\mu^b {\cal A}_\nu^c
\nonumber \\
D_\mu^{ab} &=& \delta^{ab} \partial_\mu \,-\, 
g\,f^{ab}_c\,{\cal A}_\mu^c
\;.
\label{A5}
\end{eqnarray}
The derivative $\partial_\mu \equiv \partial /\partial x^\mu$
acts on the space-time argument $x^\mu = (x^0,\vec{x})$, and
the generators of the $SU(3)$ color group are the
traceless hermitian matrices $T_a$ with the structure constants $f^{abc}$,
as matrix elements ($a, b, \ldots$ running from $1$ to $N_c$) with
\begin{equation}
\mbox{Tr}\left(T^a, T^b\right) \;=\; N_c\;\delta^{ab}
\;,\;\;\;\;\;\;\;\;
\left[T^a,T^b\right] \;= \; + i\, f^{abc} T_c
\;,\;\;\;\;\;\;\;\;
\left(T^a\right)_{bc}
\;=\; - i\,f^{a}_{bc}\;=\; - i\,f^{abc}
\;. 
\label{A6}
\end{equation}
For example, the  {\it Yang-Mills action} reads then with these conventions:
\begin{eqnarray}
S_{\rm YM} &=&
-\;\frac{1}{4} 
\int d^4x
{\cal F}_{\mu\nu}^a(x) {\cal F}^{\mu\nu,\,a}(x)
\nonumber \\
&= & 
-\;\frac{1}{2} 
\;\int d^4x
\left\{
\frac{}{}
\left(\partial_\mu {\cal A}_{\nu}^a\right)^2 \;-\;
\left(\partial_\mu {\cal A}_{\nu}^a\right) 
\left(\partial^\nu {\cal A}^{\nu,\,a}\right)
\;+\;
g\, f_{abc} 
\left(\partial_\mu {\cal A}_\nu^a\right) 
{\cal A}^{\mu,\,b} {\cal A}^{\nu,\,c}
\;+\;
g^2\, f^{abc} f^{ab'c'}
{\cal A}_\mu^b {\cal A}_\nu^c {\cal A}^{\mu,\,b'} {\cal A}^{\nu,\,c'}
\right\}
\label{A7}
\end{eqnarray}
\bigskip

\section{Scale-dependent generating functionals and $n$-point functions}
\label{sec:appb}

Here we recollect the formulae  for the various functionals,
Green functions and vertex functions that we refer to in the paper.
We restrict ourselves to the case of non-covariant gauges and
focus our attention on the gauge field sector.
Our formulation is in complete analogy with the usual pathintegral formalism of QCD, except for the presence of the infrared scale $\kappa$ which
effectively truncates the theory to one which includes 
only field modes with momenta 
$\,\lower3pt\hbox{$\buildrel > \over\sim$}\,\kappa$.
In the limit $\kappa\rightarrow 0$ the full quantum theory is recovered,
whereas the opposite limit $\kappa \rightarrow \infty$ correponds to
the pure classical Yang-Mills theory.
\smallskip

The {\it scale-dependent vacuum persistance amplitude} 
$Z_\kappa [{\cal J}] = \langle\, 0 \,| \,0\,\rangle_{{\cal J},\,\kappa}$
in the presence of an external source ${\cal J}$ and the
infrared regulator $\Re_\kappa$ (with $\lim_{\kappa\rightarrow 0} \Re_\kappa = 0$) 
is defined as,
\begin{equation}
Z_\kappa[{\cal J}]\;=\;
{\cal N}'\; 
\;\int \,{\cal D}{\cal A} \; \mbox{det} \left(M\right)
\;\,\delta\left(F[{\cal A}]\right)
\;\;\exp \left[\frac{}{}\,i\, 
\left( S_{\rm YM}[{\cal A}] \;+ \;{\cal J} \circ {\cal A}\right)
\right]
\;\;\exp \left(\frac{}{}\,i\, \Re_\kappa[{\cal A}]\right)
\label{B1}
\;,
\end{equation}
and the expectation values of {\it time-ordered} products of field operators 
(in the presence of $\Re _\kappa$) are given by,
\begin{eqnarray}
& &
\langle \;{\cal A}_{\mu_1}^{a_1}(x_1) \ldots {\cal A}_{\mu_n}^{a_n}(x_n)\;\,\rangle_\kappa
\;\;\equiv\;\;
\langle\,0\,| \;\mbox{T}\,\left[ \,{\cal A}_{\mu_1}^{a_1}(x_1) \ldots {\cal A}_{\mu_n}^{a_n}(x_n) \,\right]\;|\,0\,\rangle_\kappa
\label{B1a}
\\
& & \nonumber
\\
& &
\;\;=\;\;
\frac{{\cal N}'}{Z_\kappa[0]}\;\;
\;\int \,{\cal D}{\cal A} \; \mbox{det} \left(M\right)
\;\,\delta\left(F[{\cal A}]\right)
\;\exp \left[\frac{}{}\,i\, 
\left( S_{\rm YM}[{\cal A}] \;+\;{\cal J} \circ {\cal A}\right)
\right]
\;\;\exp \left(\frac{}{}\,i\, \Re_\kappa[{\cal A}] \right)
\;\;\; \mbox{T} \left[
{\cal A}_{\mu_1}^{a_1}(x_1) \ldots {\cal A}_{\mu_n}^{a_n}(x_n)\,
\right]
\nonumber
\;.
\end{eqnarray}
Here
the functional integration is over all gauge field configurations
with the path-integral measure
${\cal D}{\cal A} \equiv \prod_x\prod_\mu\prod_a d {\cal A}_\mu^a(x)$,
and $S_{\rm YM}[{\cal A}] = - \frac{1}{4} \int d^4x \,{\cal F}_{\mu\nu}{\cal F}^{\mu\nu}$.
The determinant  $\mbox{det} ( M )$ is the Fadeev-Popov determinant
for the matrix
$M_{ab}(x,y)= \delta F^a_{{\cal A}}(x) / \delta \omega^b(y)$
with the gauge constraint for non-covariant gauges
$F^a [{\cal A}(x)] = n\cdot{\cal A}^a(x)= 0$ ($n^\mu$ being a constant
4-vector).
As discussed in Sect. 2, the factor  $\mbox{det} ( M ) \,\delta\left(F[{\cal A}]\right)$
can be converted into a ghost field contribution to the action
in the exponential of (\ref{B1}).
The great advantage of non-covariant gauges
is the decoupling of the ghost degrees of freedom from the gauge field,
so that (\ref{B1}) can be written as a sum of a ghost
contribution and a gauge field contribution,
\begin{equation}
Z_\kappa[{\cal J},\overline{\sigma},\sigma]\;\,=\;\,
Z_\kappa^{({\cal A})}[{\cal J}] \;\;+\;\; Z_\kappa^{(\eta)}[\overline{\sigma},\sigma]
\label{B2}
\;,
\end{equation}
where [ c.f. (\ref{Z2a})-(\ref{SG0}) ],
\begin{eqnarray}
Z_\kappa^{({\cal A})}[{\cal J}]
&=&
\int \,{\cal D}{\cal A} 
\;\exp\left\{ \;i \;\int d^4x \left( \frac{}{}
-\frac{1}{4} {\cal F}_{\mu\nu}^a{\cal F}^{\mu\nu}_a 
-\frac{1}{2\xi} \left(n^\mu {\cal A}^a_\mu\right)^2
+{\cal J}^a_\mu{\cal A}_a^\mu
\right)\;\right\}
\;\;\exp\left(\frac{}{} i\,\Re_\kappa[{\cal A}]\right)
\label{B2a}
\\
Z_\kappa^{(\eta)}[\overline{\sigma},\sigma]
&=&
\int\, {\cal D} {\cal A}
\;\exp\left\{ \;i \;\int d^4x \left( \frac{}{}
\overline{\eta}_a\;\left( \delta^{ab}\;n^\mu\,\partial_\mu\right)\;\eta_b
+\overline{\sigma}_a\eta^a+\sigma_a \overline{\eta}^a
\right)\;\right\}
\;\;\exp\left(\frac{}{} i\,\Re_\kappa[\overline{\eta},\eta]\right)
\label{B2b}
\;,
\end{eqnarray}
where $\Re_\kappa[{\cal A}]$ and 
$\Re_\kappa[\overline{\eta},\eta]$ are given by (\ref{Delta1}) and 
(\ref{Delta2}), respectively.
Concerning the dynamics of the gluon gauge fields, 
the ghost contribution amounts
to a constant term that factors out when generating the gluon Green functions
from (\ref{B2}) via repeated functional differentiation 
$
\left. Z_\kappa^{-1}[0]\; \delta^n Z_\kappa[{\cal J},\overline{\sigma},\sigma]
/\delta {\cal J}^n \right|_{{\cal J}=\overline{\sigma}=\sigma=0}
$.
For the same reason, the normalization ${\cal N}'$ in 
(\ref{B1}) is irrelevant. Hence we focus on the pure gauge field functional
$Z_\kappa^{({\cal A})}$, eq. (\ref{B2a}), and define for convenience
\begin{equation}
S_{\rm eff}[{\cal A},{\cal J}]
\;\;\equiv\;\;
\int d^4x\; \left(\frac{}{}
 -\frac{1}{4}\;{\cal F}_{\mu\nu}^a{\cal F}^{\mu\nu}_a
\;-\;\frac{1}{2\xi}\; \left(n^\mu {\cal A}^a_\mu\right)^2
\;+\;{\cal J}^a_\mu{\cal A}_a^\mu
\right)
\label{B3}
\;.
\end{equation}
\medskip

\subsection{The functional $Z_\kappa[{\cal J}]$}

\noindent
We write the gauge-field part of the scale-dependent vacuum persistence amplitude as,
\begin{equation}
Z^{({\cal A})}_\kappa[{\cal J}] \;=\;
\int \,{\cal D}{\cal A}
\;\exp \left[ i\, \frac{}{}\left( S_{\rm eff}[{\cal A},{\cal J}] 
+ \Re_\kappa[{\cal A}]\right) \;\right]
\label{B4}
\;.
\end{equation}
The gluon $n$-point Green functions $\widetilde{\cal G}_\kappa$
 (including both connected and disconnected parts) are then defined as
the expectation values of time-ordered ($T\{\ldots\}$) products of $n$ 
gauge fields in the presence of the infrared regulator $\Re_\kappa$,
\begin{eqnarray}
\left(\widetilde{\cal G}_\kappa^{(n)}(x_1,\ldots x_n)
\right)_{\mu_1\ldots\mu_n}^{a_1\ldots a_n}
& \equiv &
\;\;\langle\;\; {\cal A}_{\mu_1}^{a_1}(x_1) \ldots {\cal A}_{\mu_n}^{a_n}(x_n)
\;\;\rangle
\nonumber \\
& = &
\left.
\frac{(-i)^n}{Z^{({\cal A})}_\kappa[{\cal J}]} \;
\frac{ 
\delta^{n}  Z^{({\cal A})}_\kappa[{\cal J}] 
}
{
\delta {\cal J}_{a_n}^{\mu_n}(x_n) \delta {\cal J}_{a_{n-1}}^{\mu_{n-1}}(x_{n-1}) 
\ldots
\delta {\cal J}_{a_{1}}^{\mu_{1}}(x_{1}) 
}
\right|_{{\cal J}=0}
\nonumber \\
& = &
\left.
\frac{1}{Z^{({\cal A})}_\kappa[{\cal J}]} \;
\int \,{\cal D}{\cal A}
\;\exp \left[ i\, \frac{}{}\left( S_{\rm eff}[{\cal A},{\cal J}] + \Re_\kappa[{\cal A}]\right) \;\right]
\;\;T\left\{ {\cal A}_{\mu_1}^{a_1}(x_1) \ldots {\cal A}_{\mu_n}^{a_n}(x_n) \right\}
\right|_{{\cal J}=0}
\label{Zdef}
\;,
\end{eqnarray}
such that
the Volterra series representation of $Z$ reads
\begin{equation}
Z^{({\cal A})}_\kappa[{\cal J}] 
\;=\; \sum_{n=0}^\infty\;\frac{i^n}{n !}\;
\int d^4 x_n \ldots d^4 x_1\;
\left(\widetilde{\cal G}_\kappa^{(n)}(x_1,\ldots x_n)\right)_{\mu_1\ldots\mu_n}^{a_1\ldots a_n}
\;,
{\cal J}_{a_1}^{\mu_1}(x_1)\ldots {\cal J}_{a_n}^{\mu_n}(x_n)
\;.
\end{equation}
\medskip

\subsection{The functional $W_\kappa[{\cal J}]$}

\noindent
Corresponding to (\ref{B4}), we define 
the scale-dependent connected Green functional as,
\begin{equation}
W^{({\cal A})}_\kappa[{\cal J}] \;=\;
-\,i\,\ln Z^{({\cal A})}_\kappa[{\cal J}] \;=\;
-\,i\,\ln \left\{\;
\int \,{\cal D}{\cal A}
\;\exp \left[ i\, \frac{}{}\left( S_{\rm eff}[{\cal A},{\cal J}] + \Re_\kappa[{\cal A}]\right) \;\right]
\;\right\}
\label{B5}
\;.
\end{equation}
$W_\kappa$ generates  connected $n$-point Green functions ${\cal G}_\kappa$
in the presence of the infrared regulator $\Re_\kappa$,
\begin{eqnarray}
\left({\cal G}_\kappa^{(n)}(x_1,\ldots x_n)\right)_{\mu_1\ldots\mu_n}^{a_1\ldots a_n}
& \equiv &
\;\;\langle\;\; {\cal A}_{\mu_1}^{a_1}(x_1) \ldots {\cal A}_{\mu_n}^{a_n}(x_n)
\;\;\rangle^{(c)}
\nonumber \\
& = &
\left.
(-i)^{n-1}\;
\frac{ 
\delta^{n}  W^{({\cal A})}_\kappa[{\cal J}] 
}
{
\delta {\cal J}_{a_n}^{\mu_n}(x_n) \delta {\cal J}_{a_{n-1}}^{\mu_{n-1}}(x_{n-1}) 
\ldots
\delta {\cal J}_{a_{1}}^{\mu_{1}}(x_{1}) 
}
\right|_{{\cal J}=0}
\nonumber \\
& = &
\left.
(-i)^{n-1} \;
\frac{ 
\delta^{n-1} 
}
{
\delta {\cal J}_{a_n}^{\mu_n}(x_n) \delta {\cal J}_{a_{n-1}}^{\mu_{n-1}}(x_{n-1}) 
\ldots
\delta {\cal J}_{a_{2}}^{\mu_{2}}(x_{2}) 
}
\left(
\frac{(-i)}{Z^{({\cal A})}_\kappa[{\cal J}]} \;
\frac{ 
\delta Z^{({\cal A})}_\kappa[{\cal J}]
}
{
\delta {\cal J}_{a_{1}}^{\mu_{1}}(x_{1}) 
}
\right)
\right|_{{\cal J}=0}
\label{Wdef}
\;,
\end{eqnarray}
which generate the Volterra series
\begin{equation}
W^{({\cal A})}_\kappa[{\cal J}] 
\;=\; \sum_{n=0}^\infty\;\frac{i^{n-1}}{n !}\;
\int d^4 x_n \ldots d^4 x_1\;
\left({\cal G}_\kappa^{(n)}(x_1,\ldots x_n)\right)_{\mu_1\ldots\mu_n}^{a_1\ldots a_n}
\;,
{\cal J}_{a_1}^{\mu_1}(x_1)\ldots {\cal J}_{a_n}^{\mu_n}(x_n)
\;.
\label{B6}
\end{equation}
\medskip

\subsection{The effective action $\Gamma_\kappa[\overline{A}]$ and
average effective action $\overline{\Gamma}_\kappa[\overline{A}]$}

\noindent
The effective action is the generating functional for the 
proper vertex functions. It  is obtained as usual from Legendre
transformation of (\ref{B5}), by defining the average gauge field
$\overline{A}$ (as opposed the the gauge field ${\cal A}$ of the quantum
fluctuations),
$\overline{A}_\mu^a(x) \equiv \delta W_\kappa^{({\cal A})}
[{\cal J}]/\delta J_a^\mu(x) = \langle {\cal A}_\mu^a (x)\rangle$.
Then, the transformation of $W_\kappa$ yields
the scale-dependent vertex functional $\Gamma_\kappa$,
i.e., the effective action
in the presence of the infrared regulator $\Re_\kappa$,
\begin{equation}
\Gamma_\kappa[\overline{A}] \;=\;
W_\kappa^{({\cal A})}\;\;-\;\; {\cal J}\circ\overline{A}
\;=\;
-\,i\,\ln \left\{\;
\int \,{\cal D}{\cal A}
\;\exp \left[ i\, \frac{}{}\left( S_{\rm eff}[{\cal A},{\cal J}] 
+ \Re_\kappa[{\cal A}]\right) \; -\; {\cal J}\circ\overline{A}\right]
\;\right\}
\label{B7}
\;.
\end{equation}
One may think of (\ref{B7}) as a change of variables from $\left\{ {\cal J}(x)\right\}$
to $\left\{ \overline{A}(x)\right\}$, the latter being the natural
variable of the Legendre tranform $\Gamma_\kappa[\overline{A}]$. 
The derivative of $\Gamma_\kappa[\overline{A}]$ with respect to $\overline{A}$
gives the Legendre conjugate relation
$\delta \Gamma_\kappa[\overline{A}]/\delta \overline{A}_a^\mu(x) = - {\cal J}_\mu^a (x)$. 
Repeated functional derivatives of $\Gamma_\kappa[\overline{A}]$ 
generate the one-particle irreducible $n$-point functions,
or proper vertices, at the stationary point 
$\overline{A}= A_0$ that maximizes the effective action $\Gamma_\kappa[\overline{A}]$,
corresponding to vanishing sources ${\cal J}=0$:
\begin{equation}
\left.
\frac{\delta^{n}  \Gamma_\kappa[\overline{A}]}
{\delta \overline{A}_{a_n}^{\mu_n}(x_n)
\delta \overline{A}_{a_{n-1}}^{\mu_{n-1}}(x_{n-1})
\ldots
\delta \overline{A}_{a_{1}}^{\mu_{1}}(x_{1})}
\right|_{\overline{A}=A_0}
\;\;=\;\;
\left(\Gamma_\kappa^{(n)}(x_1,\ldots x_n)\right)_{\mu_1\ldots\mu_n}^{a_1\ldots a_n}
\;.
\label{B8}
\end{equation}
The series representation for $\Gamma_\kappa$ reads then:
\begin{equation}
\Gamma_\kappa[\overline{A}] 
\;=\; \sum_{n=0}^\infty\;\frac{1}{n !}\;
\int d^4 x_n \ldots d^4 x_1\;
\left(\Gamma_\kappa^{(n)}(x_1,\ldots x_n)\right)_{\mu_1\ldots\mu_n}^{a_1\ldots a_n}
\; \overline{A}_{a_1}^{\mu_1}(x_1)\ldots \overline{A}_{a_n}^{\mu_n}(x_n)
\;.
\end{equation}
\bigskip

\noindent
Finally, the  average effective action $\overline{\Gamma}_\kappa$
is defined as the effective action $\Gamma_\kappa$ of (\ref{B7}) minus
the infrared regulator $\Re_\kappa$ at $\overline{A}$,
\begin{equation}
\overline{\Gamma}_\kappa[\overline{A}] \;=\;
\Gamma_\kappa[\overline{A}] \;\,-\;\, \Re_\kappa[\overline{A}]
\label{effavGamma}
\;=\;
-\,i\,\ln \left\{\;
\int \,{\cal D}{\cal A}
\;\exp \left[ i\, \frac{}{}\left( S_{\rm eff}[{\cal A},{\cal J}] 
\; -\; {\cal J}\circ\overline{A}
\;+\; \Re_\kappa[{\cal A}]\right)  -  \Re_\kappa[\overline{A}]
\right]
\;\right\}
\label{B9}
\;.
\end{equation}
\medskip

\subsection{$n$-point Green functions and proper vertices for $n\le 4$}
\medskip
\noindent

Using the above definitions of the generating functional 
$Z_\kappa$, $W_\kappa$, $\Gamma_\kappa$ and $\overline{\Gamma}_\kappa$,
we list below the associated Green functions 
$\widetilde{\cal G}_\kappa^{(n)}$, ${\cal G}_\kappa^{(n)}$, 
$\Gamma_\kappa^{(n)}$, and $\overline{\Gamma}_\kappa^{(n)}$ for $ n = 1\ldots 4$.
\medskip

\noindent
The 1-point functions read:
\begin{eqnarray}
\left(\widetilde{\cal G}_\kappa^{(1)}(x)\right)_{\mu}^{a}
& = &
\langle\;\,{\cal A}_\mu^a(x)\;\,\rangle_\kappa \;=\;\overline{A}_\mu^a(x)
\nonumber \\
\left({\cal G}_\kappa^{(1)}(x)\right)_{\mu}^{a}
& = &
\langle\;\,{\cal A}_\mu^a(x)\;\,\rangle^{(c)}_\kappa \;=\;\overline{A}_\mu^a(x)
\nonumber \\
\left(\Gamma_\kappa^{(1)}(x)\right)_{\mu}^{a}
& = &
-{\cal J}_\mu^a(x)
\label{1pf} 
\;.
\end{eqnarray}
\medskip

\noindent
The 2-point functions are given by:
\begin{eqnarray}
\left(\widetilde{\cal G}_\kappa^{(2)}(x,y)\right)_{\mu\nu}^{ab}
& = &
\langle\;\,{\cal A}_\mu^a(x) {\cal A}_\nu^b(y) \;\,\rangle_\kappa
\;\;=\;\;
\Delta_{\kappa\;\mu\nu}^{ab}(x,y)\;+\;
\overline{A}_\mu^a(x) \overline{A}_\nu^b(y)
\nonumber \\
\left({\cal G}_\kappa^{(2)}(x,y)\right)_{\mu\nu}^{ab}
& = &
\langle\;\,{\cal A}_\mu^a(x) {\cal A}_\nu^b(y) \;\,\rangle^{(c)}_\kappa
\;\;=\;\;
\Delta_{\kappa\;\mu\nu}^{ab}(x,y)
\nonumber \\
\left(\Gamma_\kappa^{(2)}(x,y)\right)_{\mu\nu}^{ab}
& = &
\left[\; \Delta_{\kappa}^{-1}\;\right]_{\mu\nu}^{ab}(x,y)
\;,
\label{2pf}
\end{eqnarray}
where the exact gluon propagator ${\cal G}_\kappa^{(2)}$ 
and its inverse ${\Gamma}_\kappa^{(2)}$, are defined, respectively,  as 
\begin{equation}
\Delta_{\kappa\;\mu\nu}^{ab}(x,y)
\;\equiv\;
-i\,\frac{\delta}{\delta {\cal J}^\mu_a(x)}\,\langle \;{\cal A}_\nu^b(y)\;\rangle_\kappa
\;\;\;\;\;\;\;\;\;\;\;\;\;
\left[\; \Delta_{\kappa}^{-1}\;\right]_{\mu\nu}^{ab}(x,y)
\;=\;
i\,\frac{\delta}{\delta \langle\,{\cal A}^\mu_a(x)\,\rangle_\kappa}\,{\cal J}_\nu^b(y)
\;.
\label{Deltadef}
\end{equation}
\medskip

\noindent
For the 3-point functions one obtains:
\begin{eqnarray}
\left(\widetilde{\cal G}_\kappa^{(3)}(x,y,z)\right)_{\mu\nu\lambda}^{abc}
& = &
\langle\;\,{\cal A}_\mu^a(x) {\cal A}_\nu^b(y) {\cal A}_\lambda^c(z) \;\,\rangle_\kappa
\;\;=\;\;
(-i)\frac{\delta}{\delta {\cal J}_a^\mu(x)}\, \Delta_{\nu\lambda}^{bc}(y,z)
\;+\;
\left(\frac{}{}
\Delta_{\kappa\;\mu\nu}^{ab}(x,y)\,\langle {\cal A}_\lambda^c(z)\rangle_\kappa \;+\;
\right.
\nonumber \\
& & 
\;\;\;\;\;\;\;\;\;\;
+\;
\left. \frac{}{}
\Delta_{\kappa\;\nu\lambda}^{bc}(y,z)\,\langle {\cal A}_\mu^a(x)\rangle_\kappa \;+\;
\Delta_{\kappa\;\lambda\mu}^{ca}(z,x)\,\langle {\cal A}_\nu^b(y)\rangle_\kappa
\right)
 \;+\;
\langle {\cal A}_\mu^a(x)\rangle_\kappa
\langle {\cal A}_\nu^b(y)\rangle_\kappa
\langle {\cal A}_\lambda^c(z)\rangle_\kappa
\nonumber \\
\left({\cal G}_\kappa^{(3)}(x,y,z)\right)_{\mu\nu\lambda}^{abc}
& = &
\langle\;\,{\cal A}_\mu^a(x) {\cal A}_\nu^b(y) {\cal A}_\lambda^c(z) \;\,\rangle^{(c)}_\kappa \;\;=\;\;
(-i)\frac{\delta}{\delta {\cal J}_a^\mu(x)}\, \Delta_{\nu\lambda}^{bc}(y,z)
\nonumber \\
\left(\Gamma_\kappa^{(3)}(x,y,z)\right)_{\mu\nu\lambda}^{abc}
& = &
\;\;\;{\cal V}_{\mu\nu\lambda}^{abc}(x,y,z)
\;,
\label{3pf}
\end{eqnarray}
where the function ${\cal V}$
is the  exact proper 3-gluon vertex,
\begin{equation}
-i g\;{\cal V}^{abc}_{\lambda\mu\nu}(x,y,z)
\;\;=\;\;
-i g\;V^{\;\;abc}_{0\;\lambda\mu\nu}(x,y,z)
\;\;+\;\; O(g^3)
\;,
\end{equation}
which, to lowest order in the coupling constant $g$, reduces to the bare 
3-gluon vertex $V_{0}$,
\begin{eqnarray}
V_{0\;\lambda\mu\nu}^{\;\;abc}(x,y,z)
& =&
f^{abc} \;
\left\{
\frac{}{}
g_{\lambda\mu} (\partial_y - \partial_x)_\nu \,\delta^4(x,z) \delta^4(y,z)
\;+\;
g_{\mu\nu} (\partial_z - \partial_y)_\lambda \,\delta^4(y,x) \delta^4(z,x)
\right.
\nonumber \\
& &
\left.
\frac{}{}
\;\;\;\;\;\;\;\;\;\;\;
\;\;\;\;\;\;\;\;\;\;\;
\;+\;
g_{\nu\lambda} (\partial_x - \partial_z)_\mu \,\delta^4(x,y) \delta^4(z,y)
\right\}
\label{R3vertex}
\;.
\end{eqnarray}
In momentum space, it reads,
\begin{equation}
V_{0\;\lambda\mu\nu}^{\;\;abc}(k_1,k_2,k_3)
\;=\;
-\,i\,f^{abc} \,
\left\{
\frac{}{}
g_{\lambda\mu} (k_1 - k_2)_\nu \;+\;
g_{\mu\nu} (k_2 - k_3)_\lambda \;+\;
g_{\nu\lambda} (k_3 - k_2)_\mu \right\}
\label{3vertex}
\end{equation}
\medskip

\noindent
Finally, the 4-point functions have the following forms:
\begin{eqnarray}
\left(\widetilde{\cal G}_\kappa^{(4)}(x,y,z,w)\right)_{\mu\nu\lambda\sigma}^{abcd}
& = &
\langle\;\,{\cal A}_\mu^a(x) {\cal A}_\nu^b(y) {\cal A}_\lambda^c(z)
{\cal A}_\sigma^d(w) \;\,\rangle_\kappa
\;\;=\;\;
(-i)^2\frac{\delta^2}{\delta {\cal J}_a^\mu(x){\cal J}_b^\nu(y)}\,
 \Delta_{\lambda\sigma}^{cd}(z,w)
\;+
\nonumber \\
& & 
\;\;\;\;\;\;\;\;\;\;
\;+\;
(-i)\frac{\delta}{\delta {\cal J}_a^\mu(x)}\,
\left(\frac{}{}
\Delta_{\kappa\;\mu\nu}^{ab}(x,y)\,\langle {\cal A}_\lambda^c(z)\rangle_\kappa \;+\;
\Delta_{\kappa\;\nu\lambda}^{bc}(y,z)\,\langle {\cal A}_\mu^a(x)\rangle_\kappa \;+\;
\right.
\nonumber \\
& & 
\;\;\;\;\;\;\;\;\;\;
+\;
\left. \frac{}{}
\Delta_{\kappa\;\lambda\mu}^{ca}(z,x)\,\langle {\cal A}_\nu^b(y)\rangle_\kappa
\right)
 \;+\;
\langle {\cal A}_\mu^a(x)\rangle_\kappa
\langle {\cal A}_\nu^b(y)\rangle_\kappa
\langle {\cal A}_\lambda^c(z)\rangle_\kappa
\langle {\cal A}_\sigma^d(w)\rangle_\kappa
\nonumber \\
\left({\cal G}_\kappa^{(4)}(x,y,z,w)\right)_{\mu\nu\lambda\sigma}^{abcd}
& = &
\langle\;\,{\cal A}_\mu^a(x) {\cal A}_\nu^b(y) {\cal A}_\lambda^c(z)
{\cal A}_\sigma^d(w) \;\,\rangle^{(c)}_\kappa \;\;=\;\;
(-i)^2\frac{\delta^2}{\delta {\cal J}_a^\mu(x){\cal J}_b^\nu(y)}\,
 \Delta_{\lambda\sigma}^{cd}(z,w)
\nonumber \\
\left(\Gamma_\kappa^{(4)}(x,y,z,w)\right)_{\mu\nu\lambda\sigma}^{abcd}
& = &
\;\;\;{\cal W}_{\mu\nu\lambda\sigma}^{abcd}(x,y,z,w)
\;,
\label{4pf}
\end{eqnarray}
with the function ${\cal W}$
denoting the  exact proper 4-gluon vertex,
\begin{equation}
-g^2\; {\cal W}^{abcd}_{\lambda\mu\nu\sigma}(x,y,z,w)
\;\;=\;\;
-g^2\;W^{\;\;abcd}_{0\;\lambda\mu\nu\sigma}(x,y,z,w)
\;\;+\;\; O(g^4)
\;,
\end{equation}
which, to lowest order in the coupling, defines the usual bare 
4-gluon vertex $W_{0}$,
\begin{eqnarray}
W_{0\;\lambda\mu\nu\sigma}^{\;\;abcd}(x,y,z,w)
&=&
-\;
\left\{
\frac{}{}
\left( f^{ace}f^{bde} -  f^{ade}f^{cbe} \right) \, g_{\lambda\mu} g_{\nu\sigma}
\;+\;
\left( f^{abe}f^{cde} -  f^{ade}f^{bce} \right) \, g_{\lambda\nu} g_{\mu\sigma}
\right.
\nonumber\\
& &
\left.
\frac{}{}
\;\;\;\;\;\;\;\;\;\;\;
\;+\;
\left( f^{ace}f^{dbe} -  f^{abe}f^{cde} \right) \, g_{\lambda\sigma} g_{\nu\mu}
\right\}
\;
\delta^4(x,y) \delta^4(z,w) \delta^4(y,z)
\label{R4vertex}
\;.
\end{eqnarray}
In momentum space, it reads,
\begin{eqnarray}
W_{0\;\lambda\mu\nu\sigma}^{\;\;abcd}(k_1,k_2,k_3,k_4)
& =&
-\;
\left\{
\frac{}{}
\left( f^{ace}f^{bde} -  f^{ade}f^{cbe} \right) \, g_{\lambda\mu} g_{\nu\sigma}
\;+\;
\left( f^{abe}f^{cde} -  f^{ade}f^{bce} \right) \, g_{\lambda\nu} g_{\mu\sigma}
\right.
\nonumber \\
& &
\left.
\frac{}{}
\;\;\;\;\;\;\;\;\;\;\;\;\;\;\;\;\;\;\;\;\;\;
\;+\;
\left( f^{ace}f^{dbe} -  f^{abe}f^{cde} \right) \, g_{\lambda\sigma} g_{\nu\mu}
\right\}
\label{4vertex}
\;.
\end{eqnarray}
\medskip

\bigskip
\bigskip

\section{Fadeev-Popov determinant and decoupling of ghosts in the light-cone gauge}
\label{sec:appc}

In this Appendix the standard procedure of gauge field quantization
is applied to the class of non-covariant gauges (\ref{gauge1}), and it is shown that
ghost degrees of freedom are indeed absent, reducing the
general non-linear dynamics in of QCD essentially to a linear QED type 
dynamics.
We mention that
an alternative, non-standard  method was originally suggested and carried
out in detail by Kummer \cite{kummer}, which elegantly avoids
the ghosts altogether and instead introduces a Lagrange multiplier field
that carries the fictious degrees of freedom.
For an excellent review and bibliography, see Ref. \cite{gaugereview}.
Recall that under local gauge transformations
\begin{equation} 
g[\theta^a]\;\equiv\; \exp\left(  i g\,\theta^a(x)\, T^a\right)
\;,
\label{gtheta}
\end{equation} 
the gauge fields transform as
\begin{equation}
{\cal A}_\mu^a \;\longrightarrow\;
{\cal A}_\mu^{(\theta)\;a} \;=\;
g[\theta^a]\; {\cal A}_\mu^a \;g^{ -1}[\theta^a]
\;,
\end{equation}
implying that 
${\cal F}_{\mu\nu}^a {\cal F}_{\mu\nu}^a \;=\;
{\cal F}_{\mu\nu}^{(\theta)\;a} {\cal F}_{\mu\nu}^{(\theta)\;a}$,
and thereby ensuring the gauge invariance of the Yang-Mills action 
$S_{\rm YM}[{\cal A}] = - \frac{1}{4} \int d^4x \,{\cal F}_{\mu\nu}{\cal F}^{\mu\nu}$.
However, a source term of the form ${\cal J}\circ {\cal A}$
is not gauge invariant under the transformations (\ref{gtheta}).
Consequently, the functional
\begin{equation}
Z_\kappa^{(\rm naive)}
\;=\; \int \,{\cal D}{\cal A} \;
\;\exp \left[\frac{}{}\,i\, 
\left( S_{\rm YM}[{\cal A}] \;+ \;{\cal J} \circ {\cal A}\right)
\right]
\;\;\exp \left(\frac{}{}\,i\, \Re_\kappa[{\cal A}]\right)
\label{Znaive}
\end{equation}
is also not a gauge invariant quantity.
As is well known, this can be remedied by applying the formal Fadeev-Popov 
\cite{FP} procedure and integrate in the path-integral
$Z_\kappa$ over all possible gauge transformations $g(\theta^a)$ subject to
the linear subsidiary condition 
\begin{equation}
\phi^a [{\cal A}_\mu^{(\theta)}] \;\equiv\;
n^\mu\,{\cal A}_\mu^{(\theta)\;a}(x) \;-\; \beta^a(x)
\;\stackrel{!}{=}\;0
\label{phiA}
\end{equation}
with normalized space-like vector $n^\mu$ and $\beta^a(x)$ an arbitrary
weight function.
The Fadeev-Popov trick to implement  
the constraint (\ref{phiA}) in the non-invariant functional $Z_\kappa^{(\rm naive)}$
by multiplying with
\begin{equation}
1\;\,=\;\,
\int {\cal D} \theta\;\prod_a\delta\left(\phi^a [{\cal A}_\mu^{(\theta)}]\right)
\; \mbox{det}(\, M\,)
\;,
\end{equation}
where the determinant is the Jacobian for the change of variables
$\phi^a\rightarrow \theta^a$,
\begin{equation}
\mbox{det}\,\left( M_{ab} \right)
\;=\; \mbox{det}\left(\frac{\delta \phi^a[{\cal A}_\mu^{(\theta)}]
}{
\delta \theta^b}\right)_{\phi^a [{\cal A}_\mu^{(\theta)}] = 0}
\;=\;
\left\{
\int {\cal D} \theta\;\prod_a\delta\left(\phi^a [{\cal A}_\mu^{(\theta)}]\right)
\right\}^{-1}
\;.
\label{detF}
\end{equation}
Following this procedure one arrives at 
\begin{equation}
Z_\kappa \;=\;
\int \,{\cal D}{\cal A} \; \mbox{det} (\,M\,)\;
\prod_a \delta \left( \phi^a[{\cal A}_\mu]\right)
\;\exp \left[\frac{}{}\,i\, 
\left( S_{\rm YM}[{\cal A}] \;+ \;{\cal J} \circ {\cal A}\right)
\right]
\;\;\exp \left(\frac{}{}\,i\, \Re_\kappa[{\cal A}]\right)
\label{Znew}
\;,
\end{equation}
which is now a gauge invariant expression due to the proper account 
of the subsidiary condition (\ref{phiA}) that guarantees the correct 
transformation properties of the gauge fields in the presence of the 
sources ${\cal J}$.

To obtain the final form of $Z_\kappa$ as quoted in (\ref{Z2}), one
carries out the functional integration over the arbitrary funtions $\beta^a(x)$
introduced in (\ref{phiA}), by choosing, e.g., a Gaussian weight functional
\begin{equation}
w[\beta^a]\;=\;
\exp\left\{- \frac{i}{2\xi}
\,\int d^4x \,\left[\beta^a(x)\right]^2
\right\}
\;,
\end{equation}
with the real-valued parameter $\xi$, upon which
the Fadeev-Popov determinant $\mbox{det}( M )$ can be rewritten 
in a more suitable way:
\begin{equation}
\mbox{det} ( \,M\,)\;=\;
\int {\cal D}\beta \;\prod_a\;\exp\left\{-\frac{i}{2\xi}
\,\int d^4x \,\left[\beta^a(x)\right]^2 
\right\}
\;\delta\left(
n^\mu\,{\cal A}_\mu^{(\theta)\;a}(x) \;-\; \beta^a(x)
\right)
\;.
\label{detFnew}
\end{equation}
In order to calculate the determinant, it is sufficient to integrate over 
$\theta^a$ in a small vicinity where the argument of the $\delta$-function 
passes through zero at given ${\cal A}^{(\theta)\;a}$ and $\beta^a$.
For infinitesimal gauge transformations
\begin{equation}
g[\theta^a] \;\longrightarrow\; 
\delta g[\theta^a] \;=\; 1\;+\;ig\, \theta^a(x) \,T^a
\;,
\end{equation}
the gauge fields transform as
\begin{equation}
{\cal A}_\mu^{a} \;\longrightarrow\;
{\cal A}_\mu^{a} \;+\; \delta {\cal A}_\mu^{a} 
\;\;,\;\;\;\;\;\;\;\;\;\;\;\;\;\;
\delta {\cal A}_\mu^{a} 
\;=\; g f^{a}_{bc} \theta^b{\cal A}_\mu^c \;+\;\partial_\mu \theta^a
\;,
\end{equation}
so that one obtains
\begin{eqnarray}
\;\delta\left( n^\mu\,{\cal A}_\mu^{(\theta)\;a}(x) \;-\; \beta^a(x) \right)
&=&
\;\delta\left( n^\mu\,{\cal A}_\mu^{(\theta)\;a}(x) 
\;+\; g f^{a}_{bc}\theta^b\,n^\mu {\cal A}_\mu^{(\theta)\;c}\;+\; 
n^\mu \partial_\mu \,\theta^a \;-\;\beta^a \right)
\nonumber \\
&=&
\;\delta\left(
g f^{a}_{bc}\theta^b\,\beta^c\;+\; n^\mu \partial_\mu \,\theta^a
\right)
\;,
\end{eqnarray}
because $n^\mu{\cal A}_\mu^{(\theta)\;a} = \beta^a$.
This latter expression is evidently independent of
${\cal A}_\mu^a$.
Therefore, when substituted into (\ref{detFnew}) and the
integrations carried out,
\begin{equation}
\mbox{det}\left( M\right)\;=\;
\mbox{det}\left(\frac{}{}
 \delta^{ac}\;n^\mu \;[\delta_a^b \partial_\mu \,+\, 
g\,f^{cb}_d\,{\cal A}_\mu^d]\right)
\;\;=\;\;\mbox{det}\left( \delta^{ab}\;n\cdot\partial \right)
\;,
\end{equation}
one sees that $\mbox{det}( M )$ is
also independent of the gauge fields, and hence can be pulled out of
the path-integral $Z_\kappa$ and absorbed in the overall normalization.
The final result is then:
\begin{equation}
Z_\kappa \;=\;
{\cal N}\;\int \,{\cal D}{\cal A} \;
\;
\exp \left\{i \left( \frac{}{} S_{\rm YM} \left[{\cal A} \right]
\;+\; S_{\rm fix}^{(\xi)}\left[n\cdot{\cal A}\right]
\;+\;{\cal J}\circ {\cal A} \;+\; \Re_\kappa[{\cal A}]
\right)
\right\}
\;,
\end{equation}
where, from (\ref{detFnew}),
\begin{equation}
S_{\rm fix}^{(\xi)}\left[n\cdot{\cal A}\right]
\;\equiv\;
\exp\left\{- \frac{i}{2\xi}
\,\int d^4x \,
\left[n\cdot {\cal A}^a(x)\right]^2\right\}
\;.
\end{equation}
In conclusion,
the property of gauge field independence of the Fadeev-Popov determinant
proves that there are indeed
no ghost fields coupling to the gluon fields, hence the formulation is
ghost-free.
\bigskip
\bigskip

\newpage

\section{Gluon propagator and polarization tensor in the axial gauges,
and in the light-cone gauge}
\bigskip

\subsection{The general case}

In order to find the explicit form
of the gluon propagator $\Delta_\kappa = {\cal G}_\kappa^{(2)}$, we  
first evaluate its inverse, ${\Gamma}_\kappa^{(2)}= ({\cal G}_\kappa^{(2)})^{-1}$,
from the second functional derivative of $\Gamma_\kappa$ with respect
to $\overline{A}$, and then invert it. With the conventions of
Appendix B, we have,
for the inverse of the {\it exact} gluon propagator $\Delta_\kappa$,
\begin{equation}
\left(\Delta_{\kappa}\right)^{-1}_{\mu\nu} (x,y) \;=\; \Gamma^{(2)}_{\kappa\;\mu\nu} \;\;=\;\;
\left.
\frac{ \delta^2 \Gamma_\kappa[\overline{A}]}{\delta \overline{A}^\nu(y)\delta \overline{A}^\mu(x)} 
\right|_{\overline{A}=0}
\;\;=\;\;
\left.
\frac{2i\, \delta \Gamma_\kappa[\overline{A}=0]}{\delta \Delta_{\nu\mu}(y,x)} 
\right.
\;,
\end{equation}
where the explicit form of the effective action $\Gamma_\kappa[\overline{A}]$ is given by
(\ref{Gamma11a}) and (\ref{Gamma11b}) together with the expressions 
(\ref{Gamma11})-(\ref{Gamma12}).
\medskip
The {\it exact} propagator $\Delta_\kappa$ 
is related to the {\it bare} propagator $\Delta_{\kappa}^{(0)}$ 
and the proper self-energy tensor $\Pi_\kappa$
through the Dyson-Schwinger equation
\begin{equation}
\left( \Delta_\kappa^{-1}\right)_{\mu\nu}
\;=\;
\left(\Delta^{(0)\;\;-1}_{\kappa}\right)_{\mu\nu} \;+ \;\widehat{\Pi}_{\kappa, \,\mu\nu}
\;\;=\;\;\Pi^{(0)}_{\kappa,\,\mu\nu} \;+ \;\widehat{\Pi}_{\kappa,\,\mu\nu}
\;,
\label{D1}
\end{equation}
where, in the class of axial gauges, the propagator is transverse to
the gauge vector $n_\mu$, 
\begin{equation}
n_\mu \,\Delta_\kappa^{\mu\nu}  \;=\; 0 \;=\; \Delta_\kappa^{\mu\nu}\, n_\nu
\label{transverse1}
\;,
\end{equation}
while the polarization tensor is strictly transverse with respect to
the external momentum $q_\mu$ (the conjugate of $\partial_\mu$),
\begin{equation}
\partial_\mu \,\widehat{\Pi}_\kappa^{\mu\nu}  \;=\; 0 \;=\; \partial_\nu \,\widehat{\Pi}_\kappa^{\mu\nu}
\label{transverse2}
\;,
\end{equation}
and both are symmetric under interchange of indices and arguments,
\begin{equation}
\Delta_\kappa^{\mu\nu} \;=\; \Delta_\kappa^{\nu\mu}
\;,\;\;\;\;\;\;\;\;\;\;\;\;\;\;\;\;
\Pi_\kappa^{\mu\nu} \;=\; \Pi_\kappa^{\nu\mu}
\label{transverse3}
\;.
\end{equation}
\smallskip

In order to  infer the general form of the exact propagator $\Delta_{\kappa}$, 
we apply (\ref{2point}) to (\ref{Gamma11} -- \ref{Gamma12}),
and carry out the Fourier transformation to momentum space.
Then one observes that the axial-gauge representation of the
inverse gluon  propagator (\ref{D1}), $\Delta_\kappa^{-1}=\Pi^{(0)}_\kappa+\widehat{\Pi}_\kappa$,
can be decomposed into two independent Lorentz tensor components \cite{kummer},
\begin{equation}
\; \left(\Delta_\kappa^{-1}\right)_{\mu\nu}^{ab}(q)
\;\;=\; \;
\delta^{ab} \left(\frac{}{}
a_\kappa (q^2,\chi) \;P_{\mu\nu}(q) \;+\; b_\kappa(q^2,\chi) \;Q_{\mu\nu}(q)
\right)
\label{X3}
\;,
\end{equation}
with  the projectors  $P_{\mu\nu}= P_{\nu\mu}$ and 
$Q_{\mu\nu}=Q_{\nu\mu}$, 
\begin{eqnarray}
P_{\mu\nu}(q) &=&
\;g_{\mu\nu} \;+\;\frac{1}{1-\chi} \,
\left[
\;\;\chi\,\frac{q_\mu q_\nu}{q^2}
\,-\,\frac{n_\mu q_\nu + q_\mu n_\nu}{n\cdot q}
\,+\,\chi\, \frac{n_\mu n_\nu}{n^2}
\right]
\label{X5}
\\
Q_{\mu\nu}(q) &=&
\;\;-\;\frac{1}{1-\chi} \;\;
\left[
\;\; \frac{q_\mu q_\nu}{q^2}
\,-\,\frac{n_\mu q_\nu + q_\mu n_\nu}{n\cdot q}
\,+\,\left( \chi\,-\,\frac{(1-\chi) n^2}{\xi q^2}\right) \, \frac{n_\mu n_\nu}{n^2}
\right]
\label{X4}
\;,
\end{eqnarray}
which are  orthogonal (in the space
transverse to $q$) and obey the relations 
$P_{\mu\lambda}P^\lambda_\nu =P_{\mu\nu}$, 
$Q_{\mu\lambda}T^\lambda_\nu =Q_{\mu\nu}$, 
$P_{\mu\lambda}Q^\lambda_\nu =0$. 
The invariant functions, $a_\kappa$ and $b_\kappa$ depend in general on
$q^2$ and on the variable 
\begin{equation}
\chi \;\;\equiv\;\; \chi(n,q) \;\;=\;\; \frac{n^2\,q^2}{(n \cdot q)^2}
\label{X2}
\;,
\end{equation}
because the inverse propagator (\ref{X5}) requires a scaling invariance under the
change $n_\mu \rightarrow \lambda \,n_\mu$. 
In a similar way, one may parametrize the propagator itself as
\begin{equation}
\Delta_{\kappa,\,\mu\nu}^{\;\;ab}(q)
\;=\;
\delta^{ab} \left(\frac{}{}
A_\kappa (q^2,\chi) \;S_{\mu\nu}(q) \;+\; B_\kappa(q^2,\chi) \;T_{\mu\nu}(q)
\right)
\label{X1}
\;,
\end{equation}
with  different projection operators $S_{\mu\nu}= S_{\nu\mu}$ and 
$T_{\mu\nu}=T_{\nu\mu}$, 
\begin{eqnarray}
S_{\mu\nu}(q) &=&
\;g_{\mu\nu} \;+\;\frac{1}{1-\chi} \,
\left[\;
\chi\;\left(1+\xi q^2\right) \, \frac{q_\mu q_\nu}{q^2}
\,-\,\frac{n_\mu q_\nu + q_\mu n_\nu}{n\cdot q}
\,+\,\chi \, \frac{n_\mu n_\nu}{n^2}
\right]
\label{X10b}
\\
T_{\mu\nu} (q) &=&
\;\;-\;\frac{1}{1-\chi} \;\;
\left[\;
\chi\;\left(1+\xi q^2\right) \, \frac{q_\mu q_\nu}{q^2}
\,-\,\frac{n_\mu q_\nu + q_\mu n_\nu}{n\cdot q}
\,+\, \frac{n_\mu n_\nu}{n^2}
\right]
\label{X10a}
\;,
\end{eqnarray}
which are again orthogonal (but now in the space transverse to $q$),
satisfying
$S_{\mu\lambda}S^\lambda_\nu =S_{\mu\nu}$, 
$T_{\mu\lambda}T^\lambda_\nu =T_{\mu\nu}$, 
$S_{\mu\lambda}T^\lambda_\nu =0$, and moreover 
$n^\mu S_{\mu\nu} = 0 = n^\mu T_{\mu\nu}$.
Using (\ref{X5})-(\ref{X1}) in
\begin{equation}
\Delta_{\kappa,\,\mu\lambda}(q)\;\Delta^{\;\;\lambda}_{\kappa ,\,\nu}(q)
\;\;\stackrel{!}{ = }\;\; g_{\mu\nu}
\label{X6}
\;,
\end{equation}
it is straightforward to derive
\begin{equation}
A_\kappa (q^2,\chi) \;\;=\;\; \;\;\frac{1}{a_\kappa (q^2,\chi)}
\;,\;\;\;\;\;\;\;\;\;\;\;\;\;\;\;\;
B_\kappa (q^2,\chi) \;\;=\;\; \chi\;\;\frac{1}{b_\kappa (q^2,\chi)}
\;.
\label{X7}
\end{equation}
The bare propagator $\Delta_\kappa^{(0)}$ corresponds to $\widehat{\Pi}=0$
in (\ref{D1}) and (\ref{X3}), which yields 
$a_\kappa = b_\kappa = q^2 +R_\kappa(q)^2$, and so,
\begin{eqnarray}
\Delta_{\kappa,\;\mu\nu}^{(0)\;ab}(q)
\;&=&\;
\frac{\delta^{ab}}{q^2 +R_\kappa(q^2)}
\;\;\left[\,S_{\mu\nu}\,+\,\chi\,T_{\mu\nu}\,\right]
\nonumber \\
\;&=&\;
\frac{\delta^{ab}}{q^2 +R_\kappa(q^2)}\;
\;
\left[
\;g_{\mu\nu} \,-\,\frac{n_\mu q_\nu + q_\mu n_\nu}{n\cdot q}
\,+\,(n^2+\xi q^2)\,\frac{q_\mu q_\nu}{(n\cdot q)^2}
\right]
\label{X8}
\;,
\end{eqnarray}
while the inverse $\left(\Delta_\kappa^{(0)}\right)^{-1} = \Pi^{(0)}_\kappa$ reads 
\begin{eqnarray}
\left(\Delta_\kappa^{(0)\;\;-1}\right)^{ab}_{\mu\nu}(q) 
\; &=& \;\delta^{ab}\; \left( q^2+R_\kappa(q^2)\right) \; 
\;\left[\; P_{\mu\nu}\;+\; Q_{\mu\nu}\;\right]
\nonumber \\
\;& = & \;
\delta^{ab}\; \left( q^2+R_\kappa(q^2)\right) \; 
\;
\left[
\;g_{\mu\nu} \,-\,\frac{q_\mu q_\nu}{q^2}
\,+\,\left(n^2 q^2 + \frac{1}{\xi}\right)\;\frac{n_\mu n_\nu}{(n\cdot q)^2}
\right]
\label{X9}
\end{eqnarray}

In order to derive expressions for
the exact propagator $\Delta_\kappa$ and its inverse
$\Delta^{-1}_\kappa$, in correspondence to
(\ref{X8}) and (\ref{X9}), it is useful inspect in more detail the
structure of the polarization tensor $\widehat{\Pi}_\kappa$, and
which of its contributions are dominant.
Let us define the dimensionless scalar functions
\begin{equation}
\Pi_\kappa^{(1)}(q) \;\;\equiv \;\;
1\;-\;\frac{a_\kappa(q^2,\chi)}{q^2+ R_\kappa(q^2)}
\;\;\;\;\;\;\;\;\;\;\;\;\;\;
\Pi_\kappa^{(2)}(q) \;\;\equiv \;\;
\frac{a_\kappa(q^2,\chi)-b_\kappa(q^2,\chi)}{q^2+ R_\kappa(q^2)}
\;,
\end{equation}
and rewrite eq. (\ref{X3}), on account of (\ref{D1}) as
\begin{eqnarray}
\widehat{\Pi}_{\kappa,\,\mu\nu}(q,-q) \;& = & \;
\left(\Delta_{\kappa}^{-1}\;-\; 
\Delta_{\kappa}^{(0)\;-1}\right)_{\mu\nu}
\;\;=\;\;
\left(\Delta_{\kappa}^{-1}\;-\; \Pi_{\kappa ,\,\mu\nu }^{(0)}\right)
\nonumber \\
& &
\nonumber \\
& =& 
\;
\left(\;g_{\mu\nu} -\frac{q_\mu q_\nu}{q^2}\right) \,\Pi^{(1)}_\kappa
\;+\;
\left(\frac{q_\mu q_\nu}{q^2}
-
\frac{n_\mu q_\nu + q_\mu n_\nu}{n\cdot q}
\,+ \,(n^2 q^2+{\xi}^{-1}) \,
\frac{q^2 n_\mu n_\nu}{(n\cdot q)^2}\right) \,\Pi^{(2)}_\kappa 
\label{X11}
\;,
\end{eqnarray}
which implies the relations
\begin{equation}
\Pi^{(1)}_\kappa \;\;=\;\; \frac{1}{2} \left(g_{\mu\nu} + (n^2 q^2+{\xi}^{-1}) \,
\frac{n_\mu n_\nu}{(n\cdot q)^2} \right)\,
\widehat{\Pi}_\kappa^{\mu\nu}
\;\;\;\;\;\;\;\;\;\;\;\;\;\;\;\;\;\;\;
\Pi^{(2)}_\kappa \;\;=\;\;
3 \,\Pi^{(1)}_\kappa \;-\; \widehat{\Pi}_{\kappa ,\,\mu}^{\;\;\;\;\mu}
\;.
\label{X12}
\end{equation}
In view of (\ref{X11}), one realizes that $\widehat{\Pi}_\kappa$ consists
of a covariant piece $\propto \Pi^{(1)}_\kappa$ plus a non-covariant piece
$\propto \Pi^{(2)}_\kappa$. Furthermore, comparing with (\ref{X9}),
in is obvious that it is solely the covariant contribution that survives
in the limit of vanishing coupling $g = 0$, because then $a_\kappa = b_\kappa =
q^2 + R_\kappa$, so that $\Pi_\kappa^{(2)} = 0$.
\bigskip

\noindent
From (\ref{X9}) and (\ref{X11}), we read off the inverse of the exact gluon propagator
(going over to $\xi \rightarrow  0$), 
\begin{equation}
\left(\Delta_\kappa^{-1}\right)_{\mu\nu}^{ab}(q)
\;=\;
\delta^{ab}\;\left( q^2\;+\;R_\kappa(q^2)\right)\;
\left\{\frac{}{}
\left(g_{\mu\nu} -\frac{q_\mu q_\nu}{q^2}\right) \,\left(1-\Pi_\kappa^{(1)}\right)
\;+\;
\left(\frac{q_\mu q_\nu}{q^2}
-
\frac{n_\mu q_\nu + q_\mu n_\nu}{n\cdot q}
+
\frac{n^2 q^2}{(n\cdot q)^2}\; \frac{n_\mu n_\nu}{n^2}\right) \,\Pi_\kappa^{(2)}
\right\}
\label{X13}
\;.
\end{equation}
and the actual  gluon propagator is readily 
obtained by inverting (\ref{X13}),
\begin{equation}
\Delta_{\kappa\;\mu\nu}^{\;\;ab} (q)
\;=\;
\;\frac{\delta^{ab}}{q^2\,+\,R_\kappa(q^2)}\;\,
\;\left(\frac{1}{1 - \Pi_\kappa^{(1)}}\right)
\;
\left\{g_{\mu\nu}
\;-\; \frac{n_\mu q_\nu + q_\mu n_\nu}{n\cdot q} 
\;+\;
\frac{q_\mu q_\nu}{q^2}
\;\left(
\frac{n^2 q^2}{(n\cdot q)^2}
\;
\frac{\Pi_\kappa^{(2)}}{1 - (\Pi_\kappa^{(1)}-\Pi_\kappa^{(2)})}\right)
\right\}
\label{X14}
\;.
\end{equation}
\smallskip

\subsection{The  case  $\chi\rightarrow 0$}

Inspection of the
expressions (\ref{X13}) and (\ref{X14}) exhibit  the relative importance of the
contributions $\propto \Pi_\kappa^{(1)}$ and $\Pi_\kappa^{(2)}$: 
If the terms involving $\Pi_\kappa^{(2)}$ could be droppped, then both (\ref{X13}) and
(\ref{X14}) would become simply the bare counterparts (\ref{X8})
and (\ref{X9}) for $\xi \rightarrow 0$, modulo the factors
$1-\Pi_\kappa^{(1)}$, respectively $1/(1-\Pi_\kappa^{(1)})$.
Now, there is no immediate argument why  $\Pi^{(2)}_\kappa$ itself should
be negligable as compared to $\Pi_\kappa^{(1)}$, so that the
only way the $\Pi^{(2)}_\kappa$-term in the propagator (\ref{X14}) 
could be small or even vanishing, is when
\begin{equation}
\chi \;\;\frac{n^2 q^2}{(n\cdot q)^2}\;\;\longrightarrow \;\;0
\;,
\end{equation}
which implies
\begin{equation}
\frac{q^2}{(n\cdot q)^2} \;\longrightarrow \;0
\;\;\;\;\;\;\;\;\;\;\;\;\;\;\;
\mbox{or}
\;\;\;\;\;\;\;\;\;\;\;\;\;\;\;
n^2 \;\longrightarrow \;0
\;.
\end{equation}
The first condition corresponds to very large momentum component
along the direction of $n$, for example, if $n$ is chosen along the
$z$-axis, then $q_z \rightarrow \infty$ would do the job.
The second condition, on the other hand, corresponds to 
picking, out of the  class of axial gauges, specifically the light-cone gauge.
Under either of these conditions, one arrives the very simple forms
for (\ref{X13}) and (\ref{X14}):
\begin{equation}
\left(\Delta_{\kappa}^{-1}\right)_{\mu\nu}^{ab}(q)
\;=\;
\delta^{ab}\;\left( q^2\;+\;R_\kappa(q^2)\right)\;
\,\left(1-\Pi_1\right) \;
\left(\;g_{\mu\nu} -\frac{q_\mu q_\nu}{q^2}\right) 
\label{X15}
\;.
\end{equation}
\begin{equation}
\Delta_{\kappa\;\mu\nu}^{\;\;\,ab} (q)\;=\; 
\;\frac{\delta^{ab}}{q^2\,+\,R_\kappa(q^2)}\;\,
\;\left(\frac{1}{1 - \Pi_1}\right)
\;
\left(\;g_{\mu\nu} 
\;-\; \frac{n_\mu q_\nu + q_\mu n_\nu}{n\cdot q} 
\right)
\label{X16}
\;,
\end{equation}
One sees that now the effect of gluon self-interactions are encoded
multiplicatively, so that we can express
both $\Delta_\kappa$ and $\Delta_\kappa^{-1}$
as the bare counterparts $\Delta_{\kappa}^{(0)}$,
respectively  $\Delta_\kappa^{(0)\;-1}$,
modulo a scalar renormalization function
${\cal Z}_\kappa(q^2,\chi) |_{\chi\rightarrow 0} \equiv {\cal Z}_\kappa(q^2)$,
\begin{eqnarray} 
\left(\Delta_\kappa^{-1}\right)_{\mu\nu}^{ab}(q) 
\;
&=&
\;\;\;
\frac{1}{{\cal Z}_\kappa(q^2)}\;\;
\left(\Delta_{\kappa}^{(0),\;-1}\right)_{\mu\nu}^{ab}(q) 
\;\;\;= \;\;\; 
\delta^{ab}\;\left( \frac{q^2\;+\;R_\kappa(q^2)}
{{\cal Z}_\kappa(q^2)}
\right)
\; \;
\left(\;
g_{\mu\nu} -\frac{q_\mu q_\nu}{q^2}
\right)
\label{X18}
\\
& & \nonumber \\
& & \nonumber \\
\;\;\Delta_{\kappa,\;\mu\nu}^{\;\;ab}(q) \; \;\;&=&
\;\;\;\;\;
{\cal Z}_\kappa(q^2) \;\;\;\left(\Delta_\kappa^{(0)}\right)_{\mu\nu}^{ab}(q) 
\;\;\;\;\;\;= \;\;\; \;\; \delta^{ab}\;\left(\frac{
{\cal Z}_\kappa(q^2)
}{q^2+R_\kappa(q^2)}
\right)\;\;
\left(\; g_{\mu\nu} \;-\; \frac{n_\mu q_\nu + q_\mu n_\nu}{n\cdot q} 
\right)
\label{X19}
\;,
\end{eqnarray} 
where the renormalization function ${\cal Z}_\kappa(q^2)$ is related to the gluon
self-energy $\Pi_\kappa^{(1)}$ by
\begin{equation}
{\cal Z}_\kappa(q^2) \;=\; \frac{1}{1\,-\,\Pi_\kappa^{(1)}(q)}
\label{X20}
\;,
\end{equation}
with initial  condition at $q^2 = \Lambda^2\rightarrow \infty$,
\begin{equation}
\left. {\cal Z}_\kappa(q^2)\right|_{q^2=\Lambda^2} \;\;=\;\; 1
\label{X21}
\;,
\end{equation}
so to ensure that the full propagator equals the bare one in
the asymptotic freedom regime when ${\Pi}^{(1)}_\kappa, \Pi^{(2)}_\kappa \rightarrow 0$.
\bigskip
\bigskip

\newpage

\section{Spectral representation of the gluon propagator  in the axial gauges}
\bigskip

In this Appendix we discuss in more detail the relation between
the gluon propagator $\Delta_\kappa$ and its spectral density $\rho_\kappa$,
as introduced in Sec. 3, eq. (\ref{spectral}).
Recall, that the gluon propagator is formally defined,
according to (\ref{B1a}), (\ref{Wdef}) and (\ref{2pf}), 
as the connected 2-point Green function in the presence of
the infrared cut-off $\kappa$,
involving the {\it time-ordered} product of two gauge fields at 
space-time points $x$ and $y$
\footnote{We suppress here the superscript $(\rm c)$ for `connected'}:
\begin{equation}
\Delta_{\kappa,\,\mu\nu}^{\;\;\;ab}(x,y) \;\;=\;\;
\langle\,0\,|\;\mbox{T}\;\left[\,{\cal A}_\mu^a(x) {\cal A}_\nu^b(y)\,\right]\,|\,0\,\rangle_\kappa
\label{E1}
\;.
\end{equation}
Analogously we define now the  gluon correlation function
as the non-time ordered 2-point function
which describes the correlation between two gluon fields 
at $x$ and $y$, irrespective of their time history and spatial origin:
\begin{equation}
\rho_{\kappa,\,\mu\nu}^{\;\;\;ab}(x,y) \;\;=\;\;
\langle\,0\,|\;\left[\,{\cal A}_\mu^a(x) {\cal A}_\nu^b(y)\,\right]\,|\,0\,\rangle_\kappa
\label{E2}
\;.
\end{equation}
In momentum space, we write (\ref{E1}) and (\ref{E2}) as,
\begin{eqnarray}
\Delta_{\kappa,\;\mu\nu}(q)\;&=&\;
\int d^4 x \;e^{i\,q\cdot x}\;\; \Delta_{\kappa,\;\mu\nu}(x,0)
\label{E3}
\\
\rho_{\kappa,\;\mu\nu}(q)\;&=&\;
\int d^4 x \;e^{i\,q\cdot x}\;\; \rho_{\kappa,\;\mu\nu}(x,0)
\label{E4}
\;.
\end{eqnarray}
We recall that 
both the propagator and the correlator depend on $q$ and $n$, more precisely
on $q^2$ and $n\cdot q$.
Let us now focus on the correlation function $\rho_{\kappa,\,\mu\nu}$ and
then work our way back to the propagator $\Delta_{\kappa,\,\mu\nu}$.
Following  \cite{west}, we take the commutator in (\ref{E2}) apart and define
(suppressing the color indices, as they are in paralell with the Lorentz indices),
\begin{eqnarray}
\rho_{\kappa,\,\mu\nu}^{(+)}(q) \;&\equiv&\;
\int d^4x \; e^{i\,q\cdot x}\;
\langle\,0\,|\;{\cal A}_\mu(x) {\cal A}_\nu(0)\,|\,0\,\rangle_\kappa
\label{E5}
\\
\rho_{\kappa,\,\mu\nu}^{(-)}(q) \;&\equiv&\;
-\,\int d^4x \; e^{i\,q\cdot x}\;
\langle\,0\,|\;{\cal A}_\nu(0) {\cal A}_\mu(x)\,|\,0\,\rangle_\kappa
\label{E6}
\;.
\end{eqnarray}
Hence,
\begin{equation}
\rho_{\kappa,\,\mu\nu}(q) \;\;=\;\;
\rho_{\kappa,\,\mu\nu}^{(+)}(q) \;+\; \rho_{\kappa,\,\mu\nu}^{(-)}(q) 
\;\;=\;\;
\int d^4x \; e^{i\,q\cdot x}\;
\langle\,0\,|\;\left[\,{\cal A}_\mu^a(x) {\cal A}_\nu^b(0)\,\right]\,|\,0\,\rangle_\kappa
\;,
\label{E7}
\end{equation}
and we have the following properties,
\begin{equation}
\rho_{\kappa,\,\mu\nu}(q) \;\;=\;\; - \rho_{\kappa,\,\mu\nu}(-q)
\label{E8}
\end{equation}
\begin{equation}
\rho_{\kappa,\,\mu\nu}^{(-)}(q) \;\;=\;\; - \rho_{\kappa,\,\mu\nu}^{(+)}(-q)
\label{E9}
\;.
\end{equation}
\medskip

Now, recall that in the axial gauges  the $q$-dependence of both the 
propagator $\Delta_\kappa$ and the correlator $\rho_\kappa$ can enter 
only in terms of the two invariants $q^2$ and $(n\cdot q)^2$.
It is therefore useful to introduce a notation for the decomposition of
an arbitrary four-vector $v^\mu$ into its longitudinal ($v^\mu_L$)
and transverse components ($v^\mu_T$) with respect to the gauge vector $n^\mu$:
\begin{equation}
v^\mu_L \;\;=\;\; \left(n\cdot v\right)\;\;n^\mu
\;,\;\;\;\;\;\;\;\;\;\;\;\;\;
v^\mu_T \;\;=\;\; v^\mu\;-\;v^\mu_L
\label{v1}
\end{equation}
with
$
v_L^2 = \left(n\cdot v\right)^2\;n^2
$,
$
v_T^2 = v^2 - v_L^2
$,
and
$
n\cdot v_L = n\cdot v_T= 0
$.
Thus, the $q$-argument in $\rho_\kappa$, for instance, reads with this notation,
\begin{equation}
\rho_{\kappa,\,\mu\nu}(q) \;\;=\;\; \rho_{\kappa,\,\mu\nu}(q^2,q_L^2)
\label{v2}
\;.
\end{equation}
\medskip

In order to derive the relation between the time-ordered
product of gauge fields (\ref{E1}) in the propagator $\Delta_\kappa$ 
and the non-time-ordered product
(\ref{E2}) in the correlation function $\rho_\kappa$, 
we proceed now as follows.
Let $\left\{ | \,N\,\rangle \right\}_{{\cal H}_G}$ denote a complete set of
states which spans the Hilbert space  ${\cal H}_G$ of all possible
gluon configurations (one of which is the vacuum state $| \,0\,\rangle$).
Inserting then this complete set of gluon states into
(\ref{E5}) gives
\begin{equation}
\rho_{\kappa,\;\mu\nu}^{(+)}(q)\;\;=\;\;
\sum_n\;\,
\langle\;0\;|\;{\cal A}_\mu\;|\;N\;\rangle_\kappa\;
\langle\;N\;|\; {\cal A}_\nu\;|\;0\;\rangle_\kappa
\;\;(2\pi)^4\;\delta^4(q- p_N)
\;,
\label{E10}
\end{equation}
and require $q_0 = p_0 \ge 0$. 
Inserting (\ref{E10}) into (\ref{E5}) and inverting the Fourier transform,
one readily finds,
\begin{eqnarray}
\langle\,0\,|\;{\cal A}_\mu(x) {\cal A}_\nu(0)\,|\,0\,\rangle_\kappa
\;&=&\;
\int \frac{ d^4 q }{(2\pi)^4}\;\,
e^{- i\,q\cdot x}\;\,\theta(q_0)\;\,\rho_{\kappa,\,\mu\nu}(q^2,q_L^2)
\nonumber \\
\;&=&\;
\int \frac{ d^4 q }{(2\pi)^4}\;\,
e^{- i\,q\cdot x}\;\,\theta(q_0)\;\,
\int_{-\infty}^\infty  dy_L \;\, e^{i\,q_L\cdot y_L} 
\;\,\widetilde{\rho}_{\kappa,\,\mu\nu}(q^2,y_L)
\label{E11}
\;,
\end{eqnarray}
where $\widetilde{\rho}_\kappa(q^2,y_L)$ is the longitudinal transform of (\ref{v2}). 
Introducing the advanced and retarded functions,
$\Delta_\kappa^{(+)}$ and $\Delta_\kappa^{(-)}$, respectively, 
\begin{equation}
\Delta^{(\pm)}_{\kappa}(x)\;\;=\;\;
\pm\;\int \frac{ d^4 q }{(2\pi)^4}\;\,
e^{- i\,q\cdot x}\;\,\theta(q_0)\;\,\delta(q^2-\kappa^2)
\; ,
\label{advret}
\end{equation}
one can express (\ref{E11}) as
\begin{equation}
\langle\,0\,|\;{\cal A}_\mu(x) {\cal A}_\nu(0)\,|\,0\,\rangle_\kappa
\;\;=\;\;
\int_0^\infty dq^2 \int_{- \infty}^\infty dy_L\;\,
\widetilde{\rho}_{\kappa,\,\mu\nu}(q^2,y_L)
\;\, \Delta^{(+)}_{\kappa}(x-y_L)
\; .
\label{E12}
\end{equation}
Similarly, from the crossing relations (\ref{E8}) and (\ref{E9}),
one obtains for the reversed product of the gauge fields,
\begin{equation}
\langle\,0\,|\;{\cal A}_\nu(0) {\cal A}_\mu(x)\,|\,0\,\rangle_\kappa
\;\;=\;\;
\int_0^\infty dq^2 \int_{- \infty}^\infty dy_L\;\,
\widetilde{\rho}_{\kappa,\,\mu\nu}(q^2,y_L)
\;\, \Delta^{(-)}_{\kappa}(-x+y_L)
\; .
\label{E12a}
\end{equation}
With the above relations we can now express the
time-ordered product of the gauge fields, which determines
the propagator $\Delta_\kappa$ via (\ref{E1}), as 
\footnote{
To be precise,
here the indices $i,j=1,2,3$ should be restricted to the spatial components
of the gauge fields ${\cal A}$ and of $\rho_\kappa$, because 
in the coordinate representation, the 
tensor structure of $\rho_{\kappa,\,\mu\nu}$ (cf. (\ref{E18} below)
leads to space-time derivatives $\partial_\mu$ acting on the
$\Delta_\kappa^{(\pm)}$ functions, which causes the
time-ordering operation does not commute with the time derivatives
arising from the time components of $\rho_{\kappa,\,\mu\nu}$.
}
\begin{equation}
\langle\,0\,|\;\mbox{T}\;\left[\,{\cal A}_\mu^a(x) {\cal A}_\nu^b(0)\,\right]\,|\,0\,\rangle_\kappa
\;\;=\;\;
\int_0^\infty dq^2 \int_{- \infty}^\infty dy_L\;\,
\widetilde{\rho}_{\kappa,\,\mu\nu}(q^2,y_L)
\;\left[\frac{}{}
\;\theta(x_0)\;\Delta_\kappa^{(+)}(x-y_L) \;+\;
\;\theta(-x_0)\;\Delta_\kappa^{(-)}(x-y_L) \;\right]
\label{E13}
\;.
\end{equation}
If the gauge vector is chosen to be space-like or light-like,  
i.e., $n^2 \le 0$, then causality allows us to replace 
$\theta(x_0)$ by $\theta(x_0-y_L)$, in which case
(\ref{E13}) together with (\ref{E1}) yields
\begin{equation}
\Delta_{\kappa,\mu\nu}(x,0)
\;\;=\;\;
\int_0^\infty dq^2 \int_{- \infty}^\infty dy_L\;\,
\widetilde{\rho}_{\kappa,\,\mu\nu}(q^2,y_L)
\;\;\Delta_\kappa^{(F)}(x-y_L)
\label{E14}
\end{equation}
where $\Delta_\kappa^{(F)}$ is the standard Feynman function,
\begin{equation}
\Delta_\kappa^{(F)}(x)
\;\;=\;\;
\theta(x_0)\;\Delta_\kappa^{(+)}(x) \;+\;
\;\theta(-x_0)\;\Delta_\kappa^{(-)}(x) 
\label{E15}
\;.
\end{equation}
In momentum space, eq. (\ref{E14}) reduces to the well-known
spectral (or Lehmann) representation,
\begin{equation}
\Delta_{\kappa,\,\mu\nu}(q)
\;\;=\;\;
\int_0^\infty dq^{\prime\;2} \;
\frac{
{\rho}_{\kappa,\,\mu\nu}(q^2,q_L^2)}{q^2 - q^{\prime\;2}}
\;.
\label{E16}
\end{equation}
If we decompose the tensor structure of $\Delta_{\kappa,\,\mu\nu}$ as  
in Sec. 3, eq. (\ref{Gprop1}), or Appendix D, eq. (\ref{X1}),
\begin{equation}
\Delta_{\kappa,\,\mu\nu}(q)
\;=\;
A_\kappa (q^2,\chi) \;S_{\mu\nu}(q) \;+\; B_\kappa(q^2,\chi) \;T_{\mu\nu}(q)
\label{E17}
\;,
\end{equation}
with  $\chi = n^2 q^2/(n\cdot q)^2$ as before,
and  analogously, 
for  the correlation function $\rho_{\kappa,\,\mu\nu}$,
\begin{equation}
\rho_{\kappa,\;\mu\nu}(q)\;\;=\;\;
\rho^A_\kappa (q^2,\chi) \;S_{\mu\nu}(q) \;+\; \rho^B_\kappa(q^2,\chi) \;T_{\mu\nu}(q)
\label{E18}
\;,
\end{equation}
then eq. (\ref{E16}) may be written as
(noting that $\chi = n^2 q^2/(n\cdot q)^2 = n^2 q^2 / q_L^2$, i.e. $\chi \propto q_L^{-2}$),
\begin{equation}
\Delta_{\kappa,\,\mu\nu}(q)
\;\;=\;\;
S_{\mu\nu}(q) \;\;
\int_0^\infty dq^{\prime\;2} \;
\frac{
{\rho}^A_{\kappa}(q^2,\chi)}{q^2 - q^{\prime\;2}}
\;\;+\;\;
T_{\mu\nu}(q) \;\;
\int_0^\infty dq^{\prime\;2} \;
\frac{
{\rho}^B_{\kappa}(q^2,\chi)}{q^2 - q^{\prime\;2}}
\;.
\label{E19}
\end{equation}
In the case $\chi \rightarrow 0$, corresponding to $q_L^2 \rightarrow \infty$,
the second term in (\ref{E19}) tends to zero, as discussed in Appendix D.
Thus, with 
\begin{equation}
\rho^A_\kappa(q^2,\chi) \;\;\stackrel{\chi \rightarrow 0}{\longrightarrow}
\;\; \rho_\kappa(q^2)
\;\;\;\;\;\;\;\;\;\;\;\;\;\;\;\;\;\;
\rho^B_\kappa(q^2,\chi) \;\;\stackrel{\chi \rightarrow 0}{\longrightarrow}
\;\; 0
\;.
\label{E20}
\end{equation}
we are left with
\begin{equation}
\Delta_{\kappa ,\,\mu\nu}(q)\;\;=\;\;
S'_{\mu\nu}(q)\;\int d q^{\prime\;2} \;
\frac{\rho_\kappa(q^{\prime\;2})}{q^2-q^{\prime \;2}}
\label{E21}
\;,
\end{equation}
which is precisely eq. (\ref{spectral}) with 
the substitution $q^2 \rightarrow q^2 + R_\kappa$ in the denominator.
\medskip

The spectral representation (\ref{E19}) or (\ref{E21}) has a rather
intuitive  physics interpretation:
The propagator for a gluon of momentum $q$ 
is a sum over all intermediate virtual gluon states
of momentum $q^\prime$ with virtuality $q^{\prime \;2}$ and
with level density  $\rho_\kappa(q')$,
and  weighted with the
phase-space factor $1/(q^2 -q^{\prime\;2}$.
For a bare  gluon with momentum $q\rightarrow\infty$
in the ultraviolet, no virtual fluctuations would be present,
so that $\rho_\kappa(q^{\prime \;2}) = \delta(q^{\prime\;2})$.
On the other extreme, an  infrared gluon with momentum $q\rightarrow 0$
is dressed by a dense cloud of soft virtual  fluctuations,
so that $\rho_\kappa(q^{\prime \;2})$ can be a very broad distribution.

\end{document}